\documentclass[pdflatex,sn-basic,Numbered]{sn-jnl}

\usepackage{graphicx}%
\usepackage{multirow}%
\usepackage{amsmath,amssymb,amsfonts}%
\usepackage{amsthm}%
\usepackage{mathrsfs}%
\usepackage[title]{appendix}%
\usepackage{xcolor}%
\usepackage{textcomp}%
\usepackage{manyfoot}%
\usepackage{booktabs}%
\usepackage{algorithm}%
\usepackage{algorithmicx}%
\usepackage{algpseudocode}%
\usepackage{listings}%
\usepackage{bm}%
\usepackage{bbm}%
\usepackage{mathtools}%
\usepackage{physics}%
\usepackage{newtxtext, newtxmath}
\usepackage{braket}%
\usepackage{microtype}

\theoremstyle{thmstyleone}%
\theoremstyle{thmstyletwo}%

\theoremstyle{thmstylethree}%

\usepackage{cleveref}

\newcommand{\ee}{\text{e}}
\newcommand{\ii}{\text{i}}
\newcommand{\A}{\text{a}}
\newcommand{\B}{\text{b}}
\newcommand{\C}{\text{c}}

\newcommand{\SetAncilla}{\mathcal{A}}
\newcommand{\SetRelevant}{\mathcal{R}}

\newcommand{\R}{\boldsymbol{R}}
\newcommand{\K}{\boldsymbol{K}}
\newcommand{\QmOp}[1]{\hat{#1}}
\newcommand{\QmUnity}{\mathbbm{1}}
\newcommand{\Vect}[1]{\bm{#1}}

\begin{document}

\title[Multi-Photon Quantum Rabi Models with Center-of-Mass Motion]{Multi-Photon Quantum Rabi Models with Center-of-Mass Motion}


\author*[1]{\fnm{Sabrina} \sur{Hartmann}}\email{sabrina.hartmann@uni-ulm.de}

\author[1]{\fnm{Nikolija} \sur{Mom\v{c}ilovi\'{c}}}

\author[2]{\fnm{Alexander} \sur{Friedrich}}\email{alexander.friedrich@dlr.de}

\affil*[1]{\orgdiv{Institut f{\"u}r Quantenphysik and Center for Integrated Quantum Science and Technology (IQ\textsuperscript{ST})}, \orgname{Universit{\"a}t Ulm}, \orgaddress{\street{Albert-Einstein-Allee~11}, \city{Ulm}, \postcode{D-89069},  \country{Germany}}}

\affil[2]{\orgdiv{Institute of Quantum Technologies}, \orgname{German Aerospace Center (DLR)}, \orgaddress{\street{Wilhelm-Runge-Stra\ss e~10}, \city{Ulm}, \postcode{D-89081 Ulm}, \country{Germany}}}


\abstract{The precise control of light-matter interaction serves as a fundamental cornerstone for diverse quantum technologies, spanning from quantum computing to advanced atom optics. Typical experiments often feature multi-mode fields and require atoms with intricate internal structures to facilitate e.g., multi-photon interactions or cooling. Moreover, the development of high-precision cold-atom sensors necessitates the use of entangled atoms. We thus introduce a rigorous, second-quantized framework for describing multi-$\Lambda$-atoms situated within a cavity, encompassing their center-of-mass motion and interaction with a two-mode electromagnetic field. A key feature of our approach resides in the systematic application of a Hamiltonian averaging theory to the atomic field operators, which enables the derivation of a compact and physically insightful effective Hamiltonian governing the system's time evolution. This effective Hamiltonian not only naturally incorporates the well-known AC-Stark and Bloch-Siegert shifts, notably augmented by contributions from all involved ancillary states, but also distinctly includes effects arising from counter-rotating terms, thereby offering a more complete description than conventional Rotating Wave Approximation (RWA)-based models. A significant finding of our investigation is the emergence of a particle-particle interaction mediated by ancillary states, stemming directly from the inherent combined quantum nature of the coupled electromagnetic field and atomic system. We illustrate the practical implications of our model through a detailed analysis of atomic Raman diffraction, explicitly determining the time evolution for a typical initial state and demonstrating the resulting Rabi oscillations. Furthermore, our theoretical results confirm that both the atomic and optical Hong-Ou-Mandel effect can be observed within this Raman configuration, utilizing a specific two-particle input state, thereby extending the utility of Raman diffraction for quantum interference phenomena while also providing a fully microscopic model of the effect.}




\maketitle

\section{Introduction}
The original atom-field interaction model \cite{Dirac1927,Fermi1932}, introduced by Dirac in 1927 and refined by Fermi in 1932, provided the first consistent description of atomic absorption and emission processes of light accounting e.g., for the Doppler effect \cite{Fermi1932}. These pioneering studies laid the groundwork for the modern fields of quantum and matter-wave optics. In the simplest cases, the atom-field interaction becomes a two-level atom~\cite{Rabi37} in a circularly polarized and classical electromagnetic field. On the other hand, the expression “Rabi model” is nowadays used to describe a variety of analogous systems, all of which have been inspired by the original problems \cite{Larson}. For example, a two-level system coupled to a quantum harmonic oscillator is often referred to as the “quantum Rabi model”. The semiclassical \cite{BlochSiegertRabiProblem,Shirley,Grifoni_Hänggi,Yan} as well as the quantized version of the Rabi problem \cite{Irish_AA,Zueco,Braak} have been extensively studied in the last decades and several extensions like coupling to two photons or two modes \cite{Zhang_2modes,Dossa,Wu_2Mode} of the electromagnetic field, multiple photons \cite{Gardas} or two qubits \cite{Chilingaryan} have been made. A second-quantized approach for the atomic and the electromagnetic field is presented in Ref.~\cite{Baldiotti_Molina}.

In the context of the so-called Rotating Wave Approximation (RWA), the Hamiltonian is divided into a fast (“counter-rotating”) and slow (“co-rotating”) rotating part. The counter-rotating terms are eliminated under the assumption of weak coupling and near-resonance to reduce the quantized Rabi model to a simpler version, usually referred to as the Jaynes-Cummings model \cite{Jaynes_Cummings}. 

The Rabi model and its variants represent the most straightforward approach describing light-matter interaction and are used in many fields of physics, such as solid state physics \cite{Wallraff_App,Peter_App,Hennessy,Tomka_App} and quantum optics \cite{Deutsch_App,Rabl_App}, specifically cavity and circuit QED \cite{Raimond_App,Blais_App,Niemczyk_App}.

The quantum Rabi model, long considered as not analytically solvable, was solved exactly in 2011 by Braak \cite{Braak}. However, the analytical solution is expressed in terms of transcendental functions and thus, approximations for the Rabi model are still valuable. Several techniques have been developed to find approximations in the regime where the RWA is not valid. Methods such as adiabatic approximation \cite{Irish_AA,Rossatto_AA}, generalized adiabatic elimination \cite{Li_GAA}, generalized rotating wave approximation \cite{Irish_GRWA,Zhang} or intermediate rotating wave approximation \cite{WANG2015779} have been developed for the strong to ultra-strong coupling regime, while an approximation for large detuning is presented in Ref.~\cite{Zueco}.

The Rabi model considers only a single, first-quantized two-level atom and one optical mode, which is the simplest case. Although there are several extensions of the Rabi model, they are often limited to just one or two atoms approximated as a two-level system. However, many applications such as stimulated Raman adiabatic passage (STIRAP) \cite{Bergmann_stirap,Falci_stirap,Novikov_stirap} and atomic Raman diffraction \cite{Moler,Hartmann_Regimes,Hartmann_Third} require an extension to a $\Lambda$-system \cite{Fewell,Ture_Lambda}. For the description of BECs \cite{Dalfovo_BEC,Haine_BEC,becker2018space,SzigetiBEC} or entanglement between atoms \cite{Luecke2011,hosten_measurement_2016,PhysRevLett.116.093602,LangeEntanglement,ReviewPezze,SzigetiReview,PhysRevLett.127.140402,colombo2022time,luo2024}, a tool to measure beyond the standard quantum limit (SQL) \cite{PhysRevA.47.5138,PhysRevA.47.3554}, the single-atom description may not be sufficient. Thus, a many-particle description is crucial for modeling state-of-the-art quantum sensors that operate below the SQL \cite{Szigeti2014,PhysRevX.15.011029,greve2022entanglement}. 

Therefore, we present a many-body description in which atoms in a cavity are considered as a multi-$\Lambda$-system coupled to two optical modes. The optical modes couple ground and excited states to one of $N$ ancilla states. This is a more realistic approach to light-matter interactions, given that lasers typically address the entire manifold rather than a single ancilla state \cite{kumar2025}. This results in minor energy shifts that are not captured by the common $\Lambda$-system.

Even a single $\Lambda$-system is not analytically solvable. Thus, we reduce the complexity of the system by time averaging the atomic field operator, which allows us to retain the counter-rotating terms and decouple the dynamics of the ancilla states from those of the ground and excited states in the resulting effective Hamiltonian. 

The structure of the present article is as follows: In \Cref{sec:MultiLambdaModel} we introduce the Hamiltonian for a multi-level atom interacting with a electromagnetic field, and subsequently reduce the interaction to the aforementioned two-mode and multi-$\Lambda$ scheme. The objective of this study is to obtain the time-evolution of the atomic field operator in \Cref{sec:DeterminationHeff}, which is governed by the effective Hamiltonian. The corresponding calculations are quite extensive and have been relegated to the appendix. First, we transform the Hamiltonian in Appendix~\ref{app:LambdaSystemIP} into the interaction picture. In Appendix~\ref{app:DerivationHeff}, we develop a method for temporal averaging of an operator and subsequently apply it to the multi-$\Lambda$ scheme. The interaction picture is then reversed, and the resulting expression is rearranged for the purpose of enhanced physical interpretation and subsequently discussed in \Cref{sec:BuildingBlocks}. In \Cref{sec:EnergyShifts} we introduce the AC-Stark and Bloch-Siegert shifts, which are then employed in \Cref{sec:IdentificationCouplings} to define the (self-) couplings of the effective Hamiltonian. Moreover, we introduce in \Cref{sec:MatrixRepresentation} a matrix formalism, which we utilize together with the Heisenberg equation of motion in \Cref{sec:Heisenberg} to facilitate the comparison between our result and the semi-classical model. Finally, we consider in \Cref{sec:Application} some specific applications. Therefore, we derive in \Cref{sec:DerivationU} the time evolution operator of the reduced 2-by-2 Hamiltonian. This is outlined in more detail in Appendix~\ref{app:V}. Subsequently, we apply in \Cref{sec:ApplicationSP} the time-evolution operator to a generic optical and a single-particle atomic state and identify the beam splitter operator, which describes this diffraction process. The corresponding evaluation of the Hamiltonian’s center-of-mass component is presented in Appendix~\ref{app:EvalCOM}. We complete our analysis with the consideration of two-particle atomic states in \Cref{sec:HongMandel}. We demonstrate that, by selecting a specific two-particle atomic input state, Raman diffraction can be utilized to observe the Hong-Ou-Mandel effect in both the atomic and optical system.

\section{Two-mode field and atoms in a cavity}
\label{sec:MultiLambdaModel}
In this paper, we present a non-relativistic, completely second-quantized \cite{Martin_MP,Kadanoff_MP} description of atoms in a cavity interacting with an electromagnetic two-mode field. We begin with a generic atomic dipole with $N$ levels and an atomic center-of-mass Hamiltonian in \Cref{sec:GenericAtom}. Thereafter, we introduce a generic electromagnetic field in \Cref{sec:GenericLightField}. In \Cref{sec:GenericIA} we discuss the interaction of the atoms with a two-mode electromagnetic field. Subsequently, in \Cref{sec:LambdaModel}, we show how the model can be reduced to one ground and one excited state coupled to an ancillary state manifold. To this end, we transform in Appendix~\ref{app:LambdaSystemIP} the initial $N$ level system coupled to the electromagnetic field into an interaction picture. Subsequently we apply the method of time averaging at the level of the atomic field operators, as outlined in Appendix~\ref{app:DerivationHeff}, to our system. This procedure enables the expression of the time evolution of the atomic field operator in terms of an effective Hamilton operator.

It should be noted that throughout this article, first-quantized operators are denoted with script letters, while second-quantized operators are indicated with a hat. Vectors are presented in bold.

\subsection{Generic atomic dipole operators and atomic Hamiltonian}
\label{sec:GenericAtom}
A large fraction of today’s cold atom experiments are conducted with atomic gases instead of individual atoms. These gases are inherently many-particle systems that do not preserve particle number when coupled to an environment, and thus are most conveniently described within the framework of second quantization \cite{PhysRevA.50.2207}. The recently published study in Ref.~\cite{PhysRevA.110.L041301} considers a closely related problem that is, in principle, contained in our model due to the expansion to a multilevel atom and a two-mode electromagnetic field. Furthermore, our approach differs from that taken in the aforementioned study in that we do not require the conservation of the particle number. Here, we limit our consideration to indistinguishable atoms, i.e., a bosonic system with internal states $\ket{\mathscr{l}}$ labeled by $\mathscr{l}$. Accordingly, we introduce bosonic non-relativistic atomic field operators for each internal state, which, in position representation, take the form
\begin{equation}
		\QmOp{\psi}_\mathscr{l}(\R)= \sum_\xi \braket{\R \vert \phi_\xi} \hat{a}_{\mathscr{l}\xi} =  \sum_{\xi} \phi_\xi(\R) \hat{a}_{\mathscr{l}\xi},
		\label{eq:FieldOperator}
\end{equation}
expanded into a complete and orthonormal basis $\ket{\mathscr{l}, \phi_\xi}$ with the single-particle wavefunction $\phi_\xi(\R)$, where $\xi$ denotes the external state of the atom\footnote{In principle, one can also pick a Fourier-integral style decomposition for the field operator as usually done in free space.}. Consequently, the single-particle wave functions obey the completeness relation $\sum_\xi \phi_\xi(\R) \phi_\xi^\ast(\R^\prime) = \delta(\R-\R^\prime)$ and $\int \dd^3 R ~\phi_\xi^\ast (\R) \phi_\chi^{\phantom{\ast}}(\R)=\delta_{\xi\chi}$. Again, the index $\mathscr{l}$ is used to account for the internal state, e.g., a hyperfine state in a hydrogen-like atom, while $\xi$ describes the external degrees of freedom in that state. Therefore, the atomic field operator $\QmOp{\psi}_\mathscr{l}(\R)$ annihilates a single particle in internal state $\ket{\mathscr{l}}$ at position $\R$ while $\QmOp{\psi}^\dagger_\mathscr{l}(\R)$ creates a particle with the same internal state and position. The bosonic annihilation and creation operators, $\hat{a}_{\mathscr{l}\xi} $ and $\hat{a}_{\mathscr{l}\xi}^\dagger$, satisfy the bosonic commutation relations, i.~e., $[\hat{a}_{\mathscr{l} \xi},\hat{a}_{\mathscr{k} \chi}^\dagger]=\delta_{\mathscr{l}\mathscr{k}} \delta_{\xi \chi}$. Hence, in terms of the atomic field operators we have the relations 
\begin{align}
		[\QmOp{\psi}_\mathscr{l}(\R),\QmOp{\psi}_\mathscr{k}^\dagger(\R^\prime)]=\delta_{\mathscr{l} \mathscr{k}}\delta(\R-\R^\prime),
	\text{~}
		[\QmOp{\psi}_\mathscr{l}(\R),\QmOp{\psi}_\mathscr{k}(\R^\prime)]= 0  \text{~and~} 
		[\QmOp{\psi}_\mathscr{l}^\dagger(\R),\QmOp{\psi}_\mathscr{k}^\dagger (\R^\prime)]=0
		\label{eq:commutation-relations-matter-field}
\end{align}
with Kronecker delta $\delta_{\mathscr{l} \mathscr{k}}$ and Dirac's delta distribution $\delta(\R-\R^\prime)$.

With these field operators we can define the (atomic) dipole operator density
\begin{equation}
    \QmOp{\Vect{\wp}}(\Vect{R}) = \sum_{\mathscr{l},\mathscr{k}} \left(\QmOp{\psi}_\mathscr{l}^\dagger(\R) \Vect{\wp}_{\mathscr{l} \mathscr{k}} \QmOp{\psi}_\mathscr{k}(\R)+ \text{h.c.} \right),
    \label{eq:electric-dipole-density-definition}
\end{equation}
which describes transitions between the internal states $\ket{\mathscr{l}}$ and $\ket{\mathscr{k}}$ with transition amplitudes $\Vect{\wp}_{\mathscr{l} \mathscr{k}}$. Note that the diagonal elements of the dipole operator density vanish~\cite{Steck2022QuantumAtomOptics} due to the parity symmetry properties of the dipole coupling, i.e. $\Vect{\wp}_{\mathscr{l}\mathscr{l}}=0$, and $\Vect{\wp}_{\mathscr{l} \mathscr{k}}=\Vect{\wp}_{\mathscr{k} \mathscr{l}}^\ast$.

The atomic Hamiltonian takes the form
\begin{equation}
    \hat{H}_\text{A} = 
        \QmOp{H}_\text{ext} + \QmOp{H}_\text{int}
    = \int \dd^3 \R \sum_{\mathscr{l}} \QmOp{\psi}_\mathscr{l}^\dagger(\R) \bigl[ \mathcal{H}_\text{COM}^{(\mathscr{l})} + \hbar \omega_\mathscr{l}  \bigr] \QmOp{\psi}_\mathscr{l}(\R)
    \label{eq:AtomHamiltonianIP}
\end{equation}
and is comprised of two distinct parts: first, an external part, denoted as $\hat{H}_\text{ext}$, which contains a generic single-particle center-of-mass component $\mathcal{H}_\text{COM}^{(\mathscr{l})}$, which might e.g., take the form $\mathcal{H}_\text{COM}^{(\mathscr{l})}=\Vect{P}^2/(2M_\mathscr{l})+V_\mathscr{l}(\Vect{R})$. Second, an internal energy part $\hat{H}_\text{int}$, which consists of a frequency $\omega_\mathscr{l}$ associated with the internal state $\ket{\mathscr{l}}$. A summation over all considered internal states $\ket{\mathscr{l}}$ results in the full atomic Hamiltonian. For our treatment, we assume that our atomic sample is dilute and thus, we neglect particle-particle interactions. In principle, they can be incorporated and carried through the derivation with some minor adjustments. However, the central focus of this article does not lie in particle-particle interactions.

\subsection{Quantized electromagnetic field and free-field Hamiltonian}
\label{sec:GenericLightField}
A generic quantized electromagnetic field in a cavity can be described by the field operators
\begin{equation}
    \begin{aligned}
    \label{eq:mode-decomposed-field-operators-cavity}
    \QmOp{\Vect{E}}(\R) &= \frac{\ii}{\sqrt{2}} \sum_{\alpha}\mathcal{E}_{\alpha}\left(\QmOp{\alpha} \Vect{u}_{\alpha}(\R)-\QmOp{\alpha}^\dagger \Vect{u}_{\alpha}^\ast(\R)\right)\\
    \QmOp{\Vect{B}}(\R) &= \frac{1}{\sqrt{2}}\sum_{\alpha}\mathcal{B}_{\alpha}\left(\QmOp{\alpha}~\nabla \times \Vect{u}_{\alpha}(\R)+\QmOp{\alpha}^\dagger ~\nabla \times \Vect{u}_{\alpha}^\ast(\R)\right)
\end{aligned}
\end{equation}
for the electric field and the magnetic induction in the Schrödinger picture. The summation is performed over all modes via the multi-index $\alpha$, jointly accounting for polarization $\sigma_\alpha$ and mode number $m_\alpha$ via $\alpha=(m_\alpha \sigma_\alpha)$ so that non-separable cavity geometries as cylindrical setups~\cite{Kakazu1996}, are also included in Eq.~\eqref{eq:mode-decomposed-field-operators-cavity}. The field amplitudes are given by $\mathcal{E}_{\alpha}=\sqrt{\hbar \Omega_\alpha/\epsilon_0}$ and $\mathcal{B}_\alpha =\sqrt{\hbar/(\epsilon_0\Omega_\alpha)}$ while $\Vect{u}_{\alpha}(\R)$ denotes the corresponding (vector) mode function obeying the orthonormality relation 
$\int_{V}~\mathrm{d}^3 \R ~\Vect{u}^{\dagger}_{\alpha}(\R) \Vect{u}_{{\alpha^{\prime}}}(\R)= \delta_{\alpha {\alpha^{\prime}}}$. The electromagnetic creation operator $\QmOp{\alpha}^\dagger$ and the corresponding annihilation operator $\QmOp{\alpha}$ of the mode $\alpha$ satisfy the bosonic commutation relations, i.~e. $[\hat{\alpha}, \hat{\alpha}'^\dagger]=\delta_{\alpha \alpha'}$. 

In the absence of charges and currents the Hamiltonian for the electromagnetic field is then described as
\begin{equation}
    \QmOp{H}_\mathrm{L} = \sum_{\alpha} \hbar \Omega_{\alpha} \left(\QmOp{n}_{\alpha}+\frac{1}{2}\right).
    \label{eq:H_L}
\end{equation}
Here the number operator $\QmOp{n}_{\alpha}$ of the mode with frequency $\Omega_\alpha$ is defined as $ \QmOp{n}_\alpha=  \QmOp{\alpha}^\dagger\QmOp{\alpha}$ and accounts for the number of photons in the Fock state $\ket{n_\alpha}$ via $\QmOp{n}_\alpha\ket{n_\alpha} = n_\alpha \ket{n_\alpha}$.

\subsection{Interaction between atom and light field}
\label{sec:GenericIA}
In order to describe the interaction between light and matter, we approximate the atoms as dipoles using the dipole operator density $\QmOp{\Vect{\wp}}(\Vect{R})$ and consider the electric field $\hat{\Vect{E}}$ at the center-of-mass position of the atom $\R$, leading to
\begin{equation}
    \QmOp{H}_{\text{DP}}=-\int \dd^3 \R~ \QmOp{\Vect{\wp}}(\Vect{R}) \cdot  \hat{\Vect{E}}(\R).
    \label{eq:DP}
\end{equation}

Using the definitions of the dipole operator density and the electric field operator, given in \Cref{eq:electric-dipole-density-definition} and \Cref{eq:mode-decomposed-field-operators-cavity} respectively, we obtain for the interaction Hamiltonian
\begin{equation}
\label{eq:dipole-coupling-density}
    \QmOp{H}_{\text{DP}} = \hbar
       \sum_{\substack{\alpha\\ \mathscr{l},\mathscr{k}}} \int \dd^3 \R~\left[\QmOp{\psi}_\mathscr{l}^\dagger(\R) \Omega_{\mathscr{l}\mathscr{k}\alpha}(\R) \QmOp{\alpha}  \QmOp{\psi}_\mathscr{k}(\R)
       + \QmOp{\psi}_\mathscr{k}^\dagger(\R) \Lambda^*_{\mathscr{l}\mathscr{k}\alpha}(\R)  \QmOp{\alpha} \QmOp{\psi}_\mathscr{l}(\R) \right]-\text{\text{h.c.}},
\end{equation}
where we used the definition of the complex scalar product $\Vect{w}\cdot \Vect{v} = \Vect{w}^\dagger \Vect{v}$ with the properties $\Vect{w} \cdot \Vect{v} = (\Vect{w}^\ast \cdot \Vect{v}^\ast)^\ast$ and $\Vect{z}^\ast \cdot \Vect{w} = (\Vect{z}\cdot \Vect{w}^\ast)^\ast$. Moreover we defined the co-rotating and counter-rotating coupling strengths $\Omega_{\mathscr{l}\mathscr{k}\alpha}(\R)=-\ii \mathcal{E}_{\alpha}\Vect{\wp}_{\mathscr{l}\mathscr{k}}\cdot \Vect{u}_{\alpha}(\R)/(\sqrt{2}\hbar)$  and \mbox{$ \Lambda^*_{\mathscr{l}\mathscr{k}\alpha}(\R)=-\ii \mathcal{E}_{\alpha}\Vect{\wp}_{\mathscr{l}k}^\ast\cdot \Vect{u}_{\alpha}(\R)/(\sqrt{2}\hbar)$}, respectively.

Henceforth, our focus will be on the electric dipole transition. However, the case of a magnetic dipole transition can be straightforwardly obtained by replacing $-\QmOp{\wp}(\Vect{R})\cdot \QmOp{E}(\Vect{R})\mapsto \QmOp{\mu}(\Vect{R})\cdot \QmOp{B}(\Vect{R})$ in Eq.~\eqref{eq:dipole-coupling-density}. In this context, $\QmOp{\mu}(\Vect{R})$ is the magnetic dipole density defined in full analogy to Eq.~\eqref{eq:electric-dipole-density-definition}. Subsequently, the same approach is taken as outlined in the following sections. In principle, it is also possible to treat an electric and magnetic dipole transitions simultaneously along these lines.

\subsection{Atom-light coupling for a $\Lambda$-like system}
\label{sec:LambdaModel}
\begin{figure}[tb]
    \centering
    \includegraphics[width=1.0\textwidth]{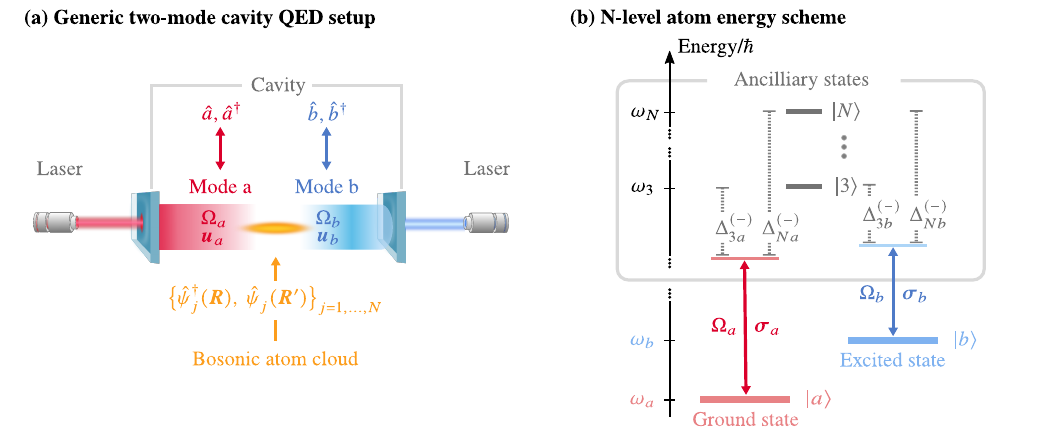}
    \caption{{\bfseries Panel (a)}: Schematic depiction of our generic two-mode cavity setup with $N$-level bosonic atoms inside the cavity. The externally pumped modes (e.g. via two lasers) have frequencies $\Omega_\A$ and $\Omega_\B$ and are described by the mode functions $\Vect{u}_\A$ and $\Vect{u}_\B$. The quantized nature of light is taken into account by the creation and annihilation operators, $\{\QmOp{\A},\QmOp{\A}^\dagger\}$ and $\{\QmOp{\B},\QmOp{\B}^\dagger\}$, for the respective modes.
    The internal states of the atoms $\ket{\mathscr{l}}$ are associated with the internal energy $\hbar \omega_\mathscr{l}$ with $\mathscr{l}=1,2,3,\hdots,N$. In second quantization the atomic cloud is modeled by the atomic field operators $\{\QmOp{\psi}_\mathscr{l},\QmOp{\psi}_\mathscr{l}^\dagger\}$ obeying the bosonic commutation relations $[\QmOp{\psi}_\mathscr{l}(\R),\QmOp{\psi}_\mathscr{k}^\dagger(\R^\prime)]=\delta_{\mathscr{l}\mathscr{k}} \delta(\R-\R^\prime)$,
    $[\QmOp{\psi}_\mathscr{l}(\R),\QmOp{\psi}_\mathscr{k}(\R^\prime)]= 0$
    and \mbox{$[\QmOp{\psi}_\mathscr{l}^\dagger(\R),\QmOp{\psi}_\mathscr{k}^\dagger (\R^\prime)]=0$}.
    {\bfseries Panel (b):}
    Generalized $\Lambda$-level scheme of atoms in the cavity. For simplicity's sake the internal states of the atoms are associated with the orbitals/states $\ket{\mathscr{l}}$ where we keep an implicit identification with the field operators $\ket{\mathscr{l}}\leftrightarrow\{\QmOp{\psi}_\mathscr{l},\QmOp{\psi}_\mathscr{l}^\dagger\}$ in mind. Moreover, we assume that a direct transition between the states $\ket{1}\equiv\ket{\A}$ and $\ket{2}\equiv\ket{\B}$ is forbidden by atomic selection rules. The ground state is designated as $\ket{\A}$ and couples to the ancilla states $\ket{3}$ to $\ket{N}$ via the electric field of the $\A$-mode in electric dipole approximation with frequency $\Omega_{\A}$ and polarization $\Vect{\sigma}_\A$. Conversely, the excited state $\ket{\B}$ couples to the ancilla states via the $\B$-mode of the electric field with frequency $\Omega_{\B}$ and polarization $\Vect{\sigma}_\B$. The modes are strongly detuned from single-photon resonance (but not from two-photon resonance) via the detunings $\big\{\Delta_{j\A}^{(-)}\big\}_{j=3,\hdots,N}$ for the transition from the state $\ket{\A}$ and by $\big\{\Delta_{j\B}^{(-)}\big\}_{j=3,\hdots,N}$ in case of state $\ket{\B}$. The relative size and distance of the atomic frequencies or rather the energy levels is not depicted to scale as indicated by the interrupted frequency axis.}
    \label{fig:Atom}
\end{figure}

In practice, $\Lambda$- or ladder-like systems are of significant interest as they form the theoretical backbone of quantum and atom optics experiments. As an example we select a~$\Lambda$-like system and an electric dipole transition. In order to make our model a bit more realistic than a simple three-level system, we couple the ground state
$\ket{1}$ and the excited state $\ket{2}$ not only to one ancillary state, but to a whole manifold of states. This assumption allows our model to account for a multi-$\Lambda$ coupling via an ancillary state manifold with different coupling strengths. As illustrated in \Cref{fig:Atom}, the model is composed of the $N-2$ ancilla states $\ket{j}$ with $j \in \mathcal{A} = \lbrace 3,...,N \rbrace$. The detuning is given by
\begin{equation}
    \Delta^{(\pm)}_{j \alpha} =\Omega_\alpha \pm \omega_{j\alpha},
    \label{eq:Detuning}
\end{equation}
with $\omega_{j\alpha}= \omega_j - \omega_\alpha$, where $\alpha \in \mathcal{R} = \{\A, \B\}$. Here, we made the replacements $\ket{1}\equiv\ket{\A}$, $\ket{2}\equiv\ket{\B}$. Due to selection rules, each of these ancilla states is coupled to the ground state $\ket{\A}$ only via the electric field of frequency $\Omega_\A$ and polarization $\sigma_\A$, and to the excited state $\ket{\B}$ via an electric field of frequency $\Omega_\B$ and polarization $\sigma_\B$. This allows us to reduce our notation by one index, as the used frequency and addressed ground state can be labeled with the same index. This leads to the following replacements: $ \Omega_{\mathscr{l} \alpha \alpha} \rightarrow \Omega_{\mathscr{l}\alpha}$ and $\Lambda_{\mathscr{l} \alpha \alpha} \rightarrow \Lambda_{\mathscr{l}\alpha}$ in $\QmOp{H}_\text{DP}$ in \Cref{eq:dipole-coupling-density}.
After retaining only the polarization allowed transitions in  $\QmOp{H}_\text{DP}$ we thus end up with the interaction Hamiltonian
\begin{equation}
        \hat{H}_\text{AL}= \hbar \int \dd^3 \R \sum_{\alpha=\A,\B} \sum_{j=3}^N \left[\QmOp{\psi}_j^\dagger(\R)  \Omega_{j\alpha}(\R)\hat{\alpha} \QmOp{\psi}_{\alpha}(\R)  
    + \QmOp{\psi}_{\alpha}^\dagger(\R)  \Lambda^*_{j\alpha}(\R)\hat{\alpha} \QmOp{\psi}_{j}(\R)  \right] - \text{\text{h.c.}}
    \label{eq:InteractionHamiltonianSP}
\end{equation}
in the Schrödinger picture. 

\section{Effective Hamiltonian and matrix representation}
\label{sec:DeterminationHeff}
The objective of this study is to determine the effective Hamilton operator, which governs the time evolution of the atomic field operator. To this end, we derive in Appendix~\ref{app:LambdaSystemIP} and Appendix~\ref{app:DerivationHeff} an expression for the effective Hamiltonian in the Schrödinger picture. The resulting expression is rather complicated. Accordingly, we separate it into logically distinct contributions in \Cref{sec:BuildingBlocks}. Subsequently, we identify in \Cref{sec:EnergyShifts} the energy shifts and in \Cref{sec:IdentificationCouplings} the couplings, which determine the effective Hamiltonian. In \Cref{sec:MatrixRepresentation} we introduce a matrix representation, which transforms the effective Hamiltonian into a more intuitive form, thereby facilitating its interpretation. We close this section, by presenting the Heisenberg equations of motion in \Cref{sec:Heisenberg}, which are governed by the effective Hamiltonian. A related approach is presented in \cite{Barajas2025}, which is based on the averaging of the Hamilton operator. 

\subsection{Building blocks of the the effective Hamiltonian}
\label{sec:BuildingBlocks}
The time evolution of the atomic field operator is governed by the effective Hamiltonian. To derive the effective Hamiltonian, we transform the system Hamiltonian into an interaction picture with respect to $\hat{H}_\mathrm{int}$ and $\hat{H}_\mathrm{L}$ in Appendix~\ref{app:LambdaSystemIP}. The Hamiltonian can then be decomposed into a time-independent part and a second contribution with time dependencies in its exponentials, belonging to a class of so-called harmonic Hamiltonians. 

Subsequently, in Appendix~\ref{app:DerivationHeff}, we transfer the method of time-averaging, introduced for density matrices by Gamel and James in Ref.~\cite{Gamel}, from their density matrix formulation to operators and obtain the effective Hamiltonian in the Schrödinger picture
\begin{equation}
     \hat{H}_{\text{eff}}= \QmOp{H}_\textrm{L} + \hat{H}_\mathrm{sp}+ \hat{H}_\mathrm{pp}.
     \label{eq:H_eff_MatrixRep}
\end{equation}
$\hat{H}_{\text{eff}}$ comprises the Hamiltonian of the electromagnetic field $\QmOp{H}_\textrm{L}$, given by \Cref{eq:H_L}, the single-particle contribution $\hat{H}_\mathrm{sp}$ and the particle-particle interactions $ \hat{H}_\mathrm{pp}$.

The next step in this investigation will be to introduce the components of the effective Hamiltonian separately. We begin by defining the single-particle Hamiltonian in \Cref{sec:H_sp} and the particle-particle Hamiltonian in \Cref{sec:H_pp}.

\subsubsection{The single-particle Hamiltonian}
\label{sec:H_sp}
The single-particle Hamiltonian in \Cref{eq:H_eff_MatrixRep} describes the couplings between the internal states of the atoms. It is given by
{\medmuskip=1mu
\begin{equation}
\scalebox{0.9}{$
\begin{split}
    \hat{H}_\mathrm{sp}\! =&  \hbar \! \sum_{j\in\SetAncilla} \sum_{\alpha\in \SetRelevant} \! \int \! \dd^3  \R ~\QmOp{\Psi}_{\alpha}^\dagger(\R) 
    \biggr[ \frac{\mathcal{H}_\text{COM}^{(\alpha)}}{\hbar} +  \omega_\alpha -
    \frac{|\Omega_{j \alpha}(\R)|^2}{\omega_{jj\alpha \alpha}^{(+)}}\QmOp{n}_\alpha  + \frac{|\Lambda_{j \alpha}(\R)|^2 }{\Omega_{jj\alpha \alpha}^{(+)}}(\hat{n}_\alpha+\mathbbm{1}_\textrm{L}) ~
    \biggr]\QmOp{\Psi}_{\alpha}(\R)\\
    &- \hbar \! \sum_{j\in\SetAncilla} {\sum_{\substack{\alpha,\beta\in \SetRelevant\\\alpha\neq\beta}}} \! \int \! \dd^3 \R ~\QmOp{\Psi}_{\beta}^\dagger(\R)\biggl[ 
    \frac{\Omega_{j \beta}^*(\R) \, \Omega_{j \alpha}(\R)}{\omega_{jj\alpha \beta}^{(+)}}\hat{\beta}^\dagger  \hat{\alpha}  -\frac{\Lambda_{j \beta}^*(\R) \, \Lambda_{j \alpha}(\R)}{\Omega_{jj\alpha \beta}^{(+)}} \hat{\beta} \hat{\alpha}^\dagger
     \biggr]~\QmOp{\Psi}_{\alpha}(\R)\\
    &+ \hbar \!
    \sum_{j\in \SetAncilla} 
    \sum_{\alpha\in\SetRelevant}
    \! \int \! \dd^3 \R ~\QmOp{\Psi}_{j}^\dagger(\R) 
    \biggl[ \frac{\mathcal{H}_\text{COM}^{(j)}}{\hbar} + \omega_j +
    \frac{|\Omega_{j \alpha}(\R)|^2}{\omega_{jj\alpha \alpha}^{(+)}}(\hat{n}_\alpha+\mathbbm{1}_\textrm{L})    
   -\frac{|\Lambda_{j \alpha}(\R)|^2 }{\Omega_{jj\alpha \alpha}^{(+)}}  \QmOp{n}_\alpha
    \biggr]~\QmOp{\Psi}_{j}(\R)\\
    &+\hbar \!
    \sum_{\substack{j,k\in \SetAncilla\\ j\neq k}} 
    \sum_{\alpha\in\SetRelevant} 
    \! \int \! \dd^3 \R ~\QmOp{\Psi}_{j}^\dagger(\R) 
    \biggl[ 
    \frac{\Omega_{k \alpha}^*(\R) \, \Omega_{j \alpha}(\R)}{\omega_{jk\alpha \alpha}^{(+)}}(\hat{n}_\alpha+\mathbbm{1}_\textrm{L})    
    - \frac{\Lambda_{k \alpha}^*(\R) ~ \Lambda_{j \alpha}(\R)}{\Omega_{jk\alpha \alpha}^{(+)}}  \QmOp{n}_\alpha
    \biggr]~\QmOp{\Psi}_{k}(\R).
\end{split}$}
\label{eq:Heff_SP}
\end{equation}}
It is important to recall that the indices $\alpha,\beta \in \{\A,\B \}$ are associated with the ground and excited states, respectively. In contrast, the indices $j,k \in \{3,...,N\}$ are reserved for the ancilla states. Consequently, the first line of \Cref{eq:Heff_SP} describes the self-coupling between the ground state, or rather the excited state. In the second line, we find transitions between the ground and the excited state. The third line represents the self-coupling of each ancilla state. In the fourth line, we find transitions between all ancilla states. The co-rotating contributions to the couplings can be identified by $\Omega_{\bullet \bullet}(\R)$ and the counter-rotating contributions by $\Lambda_{\bullet \bullet}(\R)$, where the bullets correspond to the indices $j,k,\alpha$ or $\beta$.

\Cref{eq:Heff_SP} contains the inverse sum of two detunings, see \Cref{eq:Detuning}, between the same ground / excited state and two different ancilla states, given by
\begin{equation}
    \frac{1}{\omega_{jk\alpha \alpha}^{(+)}} = \frac{1}{2} \left(\frac{1}{\Delta_{j\alpha}^{(-)}} + \frac{1}{\Delta_{k\alpha}^{(-)}}\right) \text{~ and ~}     \frac{1}{\Omega_{jk\alpha \alpha}^{(+)}} = \frac{1}{2} \left(\frac{1}{\Delta_{j\alpha}^{(+)}} + \frac{1}{\Delta_{k\alpha}^{(+)}}\right).
\end{equation}
Therefore, it is obvious that $\Omega_{jj\alpha \beta}^{(+)}=\Omega_{jj\beta \alpha}^{(+)}$ and $\Omega_{jk\alpha \alpha}^{(+)}=\Omega_{kj\alpha \alpha}^{(+)}$.

\subsubsection{The particle-particle Hamiltonian}
\label{sec:H_pp}
In contrast to the single-particle Hamiltonian, the contributions of the particle-particle interaction Hamiltonian are governed by four field operators. The particle-particle interaction Hamiltonian is given by
\begin{equation}
\scalebox{0.97}{$
\begin{split}
     \hat{H}_\mathrm{pp} = & \hbar \!
    \sum_{j,k\in \SetAncilla}
    \sum_{\alpha\in \SetRelevant}
    \!  \int \dd^3 \R ~ \QmOp{\Psi}_{\alpha}^\dagger(\R) \zeta_{k\alpha j}^*(\R) \QmOp{\Psi}_{k}(\R) \cdot \!\! \int \!  \dd^3 \R^\prime~  \QmOp{\Psi}_{j}^\dagger(\R') \zeta_{j\alpha k}(\R')\QmOp{\Psi}_{\alpha}(\R') \\
      & + \hbar \!
    \sum_{j,k\in \SetAncilla}
    \sum_{\alpha\in \SetRelevant}
    \! \int \dd^3 \R ~  \QmOp{\Psi}_{\alpha}^\dagger(\R)  \eta_{k\alpha j}^*(\R) \QmOp{\Psi}_{k}(\R) \cdot \!\! \int \!  \dd^3 \R^\prime~\QmOp{\Psi}_{j}^\dagger(\R') \eta_{j\alpha k}(\R') \QmOp{\Psi}_{\alpha}(\R') \\
    &- \hbar \!
    \sum_{\alpha\in \SetRelevant} \int \!  \dd^3 \R ~\QmOp{\Psi}_{\alpha}^\dagger(\R) \mathcal{V}_\alpha(\R)  \QmOp{\Psi}_{\alpha}(\R).
\end{split}$}
\label{eq:Heff_pp}
\end{equation}
This expression exhibits two pairs of field operators with corresponding couplings, which depend on both $\R$ and $\R'$. 

The first line in \Cref{eq:Heff_pp} reveals two transitions at two different positions $\R$ and $\R'$. Each transition takes place between one ancilla state $\ket{j}$ and the ground or excited state $\ket{\alpha}$ with 
\begin{equation}
    \zeta_{j\alpha k}(\R) = \frac{\Omega_{j\alpha}(\R)}{\sqrt{\omega_{jk\alpha \alpha}^{(+)}}}
    \label{eq:zeta},
\end{equation}
which corresponds to the co-rotating contributions.

The second line in \Cref{eq:Heff_pp} contains, in complete analogy, the counter-rotating contributions
\begin{equation}
    \eta_{j\alpha k}(\R) = \frac{\Lambda_{j\alpha}(\R)}{\sqrt{\Omega_{jk\alpha \alpha}^{(+)}}}.
    \label{eq:eta}
\end{equation}

The final line reveals a modification to the self-coupling of the ground and excited states, which emerges as a consequence of the commutation of the atomic field operators in Appendix~\ref{app:PPdecomposition}. The corresponding self-coupling is given by
\begin{equation}
   \mathcal{V}_\alpha(\R)  \equiv \sum_{j,k\in \SetAncilla} \left[\frac{ |\Omega_{j \alpha}(\R)|^2}{\omega_{jk\alpha \alpha}^{(+)}}   + \frac{|\Lambda_{j \alpha}(\R)|^2}{\Omega_{jk\alpha \alpha}^{(+)}} \right] \equiv \sum_{j,k\in \SetAncilla} \left[\mathcal{V}_{jk\alpha}^{(\Omega)}(\R)  + \mathcal{V}_{jk\alpha}^{(\Lambda)}(\R) \right],
   \label{eq:V_alpha}
\end{equation}
and is composed of co- and counter-rotating contributions. 

\subsection{Energy shifts}
\label{sec:EnergyShifts}
In this section, we identify the energy shifts induced by the electromagnetic field, since the self-coupling terms of the ground state and the excited state in \Cref{eq:Heff_SP} can be expressed in terms of these. Given that we consider a multi-level atom, all levels contribute to the energy shifts. The contribution of the co-rotating terms from each atomic level $j$ is given by
\begin{equation}
    \hat{\omega}_{\text{AC},j}^{(\pm)}=\frac{|\Omega_{j\A}(\R)|^2}{\omega_{jj\A\A}^{(+)}}\hat{n}_\A \pm \frac{|\Omega_{j\B}(\R)|^2}{\omega_{jj\B\B}^{(+)}}\hat{n}_\B 
\end{equation} 
and adds up to the quantum AC-Stark shift
\begin{equation}
    \hat{\Omega}_{\text{AC}}^{(\pm)}=\sum_{j=3}^N \hat{\omega}_{\text{AC},j}^{(\pm)}.
    \label{eq:ACstark}
\end{equation}
The corresponding shift due to vacuum reads
\begin{equation}
    \Omega_\text{ACvac}^{(\pm)} = \sum_{j=3}^N \omega_{\text{ACvac},j}^{(\pm)} = \sum_{j=3}^N  \frac{|\Omega_{j\A}(\R)|^2}{\omega_{jj\A\A}^{(+)}}\pm \frac{|\Omega_{j\B}(\R)|^2}{\omega_{jj\B\B}^{(+)}}.
    \label{eq:ACstarkVac}
\end{equation}

Analogously, we find for the counter-rotating terms the Bloch-Siegert shift contribution to the lowest order \cite{Rossatto_AA} for each ancilla state denoted by
\begin{equation}
    \hat{\omega}_{\text{BS},j}^{(\pm)}=\frac{|\Lambda_{j\A}(\R)|^2}{\Omega_{jj\A\A}^{(+)}}\hat{n}_\A \pm \frac{|\Lambda_{j\B}(\R)|^2}{\Omega_{jj\B\B}^{(+)}}\hat{n}_\B 
\end{equation} 
and the quantum Bloch-Siegert shift (lowest order)
\begin{equation}
    \hat{\Omega}_{\text{BS}}^{(\pm)}=\sum_{j=3}^N \hat{\omega}_{\text{BS},j}^{(\pm)}
      \label{eq:BlochSiegert}
\end{equation}
as well as the corresponding Bloch-Siegert shift due to vacuum
\begin{equation}
    \Omega_\text{BSvac}^{(\pm)} = \sum_{j=3}^N \omega_{\text{BSvac},j}^{(\pm)} = \sum_{j=3}^N =\frac{|\Lambda_{j\A}(\R)|^2}{\Omega_{jj\A\A}^{(+)}} \pm \frac{|\Lambda_{j\B}(\R)|^2}{\Omega_{jj\B\B}^{(+)}}.
    \label{eq:BochSiegertVac}
\end{equation}
Consequently, our method of time averaging preserves the Bloch-Siegert shift, whereas the standard approach, based on the RWA, neglects this contribution. Although this energy shift is considerably smaller than the AC-Stark shift, its relevance depends on the regime under consideration \cite{Larson}.

\subsection{Identification of the couplings}
\label{sec:IdentificationCouplings}
In this section, we reduce the complexity of \Cref{eq:Heff_SP} by introducing convenient variables. The self-coupling of the ground and excited state can be written in the general form 
\begin{equation}
     \hat{\Delta}_\alpha= \mathcal{H}_0^{(\alpha)}/\hbar  -  \sum_{j=3}^N \Biggl[ \frac{|\Omega_{j\alpha}(\R)|^2}{\omega_{jj\alpha \alpha}^{(+)}} \hat{n}_\alpha - \frac{|\Lambda_{j\alpha}(\R)|^2}{\Omega_{jj\alpha \alpha}^{(+)}} (\hat{n}_\alpha+\mathbbm{1}_\textrm{L}) \Biggr].
    \label{eq:Delta_alpha}
\end{equation}
We introduced in this equation a joint variable 
\begin{equation}
    \mathcal{H}_0^{(\mathscr{l})} = \mathcal{H}_\text{COM}^{(\mathscr{l})} + \hbar \omega_\mathscr{l}
\end{equation}
for the time-independent center-of-mass component and the energy of the atomic level for the corresponding internal state $\ket{\mathscr{l}}$ with  $\mathscr{l} \in \mathcal{S}\equiv\qty{\A,\B,3,\hdots,N}$.

We use the AC-Stark shifts, \Cref{eq:ACstark,eq:ACstarkVac}, and the Bloch-Siegert shifts, \Cref{eq:BlochSiegert,eq:BochSiegertVac}, to define the self-coupling of the ground state $\ket{\A}$
\begin{equation}
   \hat{\Delta}_\A=  \mathcal{H}_0^{(\A)}/\hbar - \frac{1}{2} \bigl[\hat{\Omega}_{\text{AC}}^{(+)}+\hat{\Omega}_{\text{AC}}^{(-)} -\hat{\Omega}_{\text{BS}}^{(+)}-\hat{\Omega}_{\text{BS}}^{(-)} -\Omega_\text{BSvac}^{(+)} -\Omega_\text{BSvac}^{(-)}   \bigr] 
    \label{eq:Delta_a}
\end{equation}
and the self-coupling of the excited state $\ket{\B}$ 
\begin{equation}
    \hat{\Delta}_\B= \mathcal{H}_0^{(\B)}/\hbar - \frac{1}{2} \bigl[\hat{\Omega}_{\text{AC}}^{(+)}-\hat{\Omega}_{\text{AC}}^{(-)} -\hat{\Omega}_{\text{BS}}^{(+)}+\hat{\Omega}_{\text{BS}}^{(-)} -\Omega_\text{BSvac}^{(+)} +\Omega_\text{BSvac}^{(-)}   \bigr].
    \label{eq:Delta_b}
\end{equation}
In a similar manner, we introduce the self-coupling of the ancilla states $\ket{j}$, given by
\begin{align}
    \hat{\Delta}_j&= \mathcal{H}_0^{(j)}/\hbar + \sum_{\alpha=\A,\B} \Biggl[ \frac{|\Omega_{j\alpha}(\R)|^2}{\omega_{jj\alpha \alpha}^{(+)}} (\hat{n}_\alpha+\mathbbm{1}_\textrm{L}) - \frac{|\Lambda_{j\alpha}(\R)|^2}{\Omega_{jj\alpha \alpha}^{(+)}} \hat{n}_\alpha \Biggr] \notag \\
    &=\mathcal{H}_0^{(j)}/\hbar + \bigl[\hat{\omega}_{\text{AC},j}^{(+)}-\hat{\omega}_{\text{BS},j}^{(+)} +\omega_{\text{ACvac},j}^{(+)}  \bigr].
    \label{eq:Delta_j}
\end{align}

The transitions between the ground and excited state are mediated by the operator densities
\begin{equation}
    \hat{\Omega}_{\B\A}= - \sum_{j=3}^N \biggl[ \frac{\Omega_{j\B}^*(\R) \Omega_{j\A}(\R)}{\Omega_{jj\A\B}^{(+)}} \hat{\B}^\dagger \hat{\A} -  \frac{\Lambda_{j\B}^*(\R) \Lambda_{j\A}(\R)}{\Omega_{jj\A\B}^{(+)}} \hat{\A}^\dagger  \hat{\B}  \biggr]
    \label{eq:Rabi}
\end{equation}
with $\hat{\Omega}_{\B\A}^\dagger = \hat{\Omega}_{\A\B}$ and the transitions between different ancilla states are given by
\begin{equation}
    \hat{\chi}_{jk}= \sum_{\alpha=\A,\B} \biggl[ \frac{\Omega_{k\alpha}^*(\R) \Omega_{j\alpha}(\R)}{\omega_{jk\alpha \alpha}^{(+)}} (\hat{n}_\alpha +\mathbbm{1}_\textrm{L}) -  \frac{\Lambda_{k\alpha}^*(\R) \Lambda_{j\alpha}(\R)}{\Omega_{jk\alpha \alpha}^{(+)}} \hat{n}_\alpha  \biggr]
    \label{eq:Xi}
\end{equation}
with $\hat{\chi}_{jk}^\dagger =  \hat{\chi}_{kj}$.

\subsection{Matrix representation}
\label{sec:MatrixRepresentation}
In the following discussion we will turn to a matrix representation with respect to the atomic field operators in order to facilitate a better comparison with the first-quantized description of a Raman diffraction process. For the sake of brevity, we omit the spatial dependence of the involved operators within the matrices. We define a vector containing all field operators of the ancilla states $\QmOp{\Vect{\Psi}}_\SetAncilla^\dagger(\R) =(\QmOp{\Psi}_N^\dagger(\R),...,\QmOp{\Psi}_3^\dagger(\R))^T$ and another one, that comprises the field operators of $\ket{\A}$ and $\ket{\B}$, denoted by $\QmOp{\Vect{\Psi}}_\SetRelevant^\dagger (\R)=(\QmOp{\Psi}_\B^\dagger(\R),\QmOp{\Psi}_\A^\dagger(\R))^T$. 

With $\QmUnity_{\SetRelevant + \SetAncilla} = \QmUnity_\SetAncilla+\QmUnity_\SetRelevant$ and $\QmUnity_\gamma$ as the projector on the relevant ($\gamma = \SetRelevant$) or irrelevant ancillary ($\gamma=\SetAncilla$) internal states we express in \Cref{sec:MatrixHsp} the single-particle Hamiltonian and in \Cref{sec:MatrixHpp} the particle-particle Hamiltonian within this matrix formalism.

\subsubsection{Matrix representation of the single-particle Hamiltonian}
\label{sec:MatrixHsp}
The previously introduced vectors simplify the expression of the single-particle Hamiltonian in terms of a matrix within the subspace of the relevant states. Thus, we transform \Cref{eq:Heff_SP} into
\begin{equation}
     \hat{H}_\mathrm{sp}= \hbar\! \int \!\! \dd^3 \R  ~ \left[ \QmUnity_{\SetAncilla} \QmOp{\Vect{\Psi}}_\SetAncilla^\dagger(\R)  \QmOp{M}_\SetAncilla(\R)\QmOp{\Vect{\Psi}}_\SetAncilla(\R) \QmUnity_{\SetAncilla} + \QmUnity_{\SetRelevant} \QmOp{\Vect{\Psi}}_\SetRelevant^\dagger(\R) \QmOp{M}_\SetRelevant(\R) \QmOp{\Vect{\Psi}}_\SetRelevant(\R) \QmUnity_{\SetRelevant}  \right].
     \label{eq:H_1}
\end{equation}
\Cref{eq:H_1} reveals that in $\hat{H}_\mathrm{sp}$ the ground state and the excited state decouple from the ancilla states, forming the two-by-two Rabi-block matrix $\QmOp{M}_\SetAncilla(\R)$ and the matrix over the ancilla subspace $\QmOp{M}_\SetAncilla(\R)$.

The matrix $\QmOp{M}_\SetRelevant$ contains the single-particle contributions defined in \Cref{eq:Delta_a,eq:Delta_b,eq:Rabi} and is given by
\begin{equation}
    \QmOp{M}_\SetRelevant(\R)= 
    \begin{pmatrix}
           \hat{\Delta}_{\B}& \hat{\Omega}_{\B\A}\\
            \hat{\Omega}_{\B\A}^\dagger& \hat{\Delta}_\textrm{a}
    \end{pmatrix}.
    \label{eq:MatrixR}
  \end{equation}  
 This block matrix shows the typical Rabi dynamics. The diagonal elements $\hat{\Delta}_\alpha$ of the matrix, defined in \Cref{eq:Delta_a,eq:Delta_b}, include the AC-Stark and Bloch-Siegert shift, already introduced in \Cref{sec:EnergyShifts}. The transitions between the ground and the excited states are represented by the operator densities  $\hat{\Omega}_{\B\A}$, defined in \Cref{eq:Rabi}, and can be found on the off-diagonals of $\QmOp{M}_\SetRelevant$. In comparison to a single three-level system, the couplings in \Cref{eq:Rabi}, and self-couplings of $\ket{\A}$ and $\ket{\B}$ in \Cref{eq:Delta_alpha}, or rather the energy shifts in \Cref{eq:ACstark,eq:ACstarkVac,eq:BlochSiegert,eq:BochSiegertVac} are modified by contributions from each included ancillary level, which is indicated by the sum over $j$. Consequently, these expressions permit the investigation of more realistic systems, as in experiments not only one ancilla state can be clearly addressed by the laser, but the entire manifold is typically involved.

As the ancilla states were not eliminated, their dynamics remain accessible in the block matrix 
\begin{equation}
    \QmOp{M}_\SetAncilla(\R)= 
    \begin{pmatrix}
        \hat{\Delta}_N& \hat{\chi}_{N(N-1)}&  \hdots & \hat{\chi}_{N4} &\hat{\chi}_{N3}\\
         \hat{\chi}_{N(N-1)}^\dagger& \hat{\Delta}_{N-1}& \hat{\chi}_{(N-1)(N-2)} & &\hat{\chi}_{(N-1)3} \\
         \vdots & \ddots & \ddots& \ddots & \vdots &  &\\
        \hat{\chi}_{N4}^\dagger & \dots &   \hat{\chi}_{54}^\dagger & \hat{\Delta}_4 & \hat{\chi}_{43}\\
          \hat{\chi}_{N3}^\dagger&  \dots& \hat{\chi}_{53}^\dagger & \hat{\chi}_{43}^\dagger & \hat{\Delta}_3\\
    \end{pmatrix}.
    \label{eq:MatrixA}
\end{equation}
Similarly, as in the Rabi-block matrix, \Cref{eq:MatrixR}, the self-couplings of the ancilla states, \Cref{eq:Delta_j}, can be found on the main diagonal of $\QmOp{M}_\SetAncilla(\R)$. As the remaining entries show, transitions between all ancilla states are mediated by the couplings in \Cref{eq:Xi}.

\subsubsection{Matrix representation of the particle-particle Hamiltonian}
\label{sec:MatrixHpp}
We proceed in a similar manner to reformulate the particle-particle interactions. To do so, we have to introduce suitable matrices.

The first line in \Cref{eq:Heff_pp} corresponds to the co-rotating contributions. This term contains two transitions at two different positions $\R$ and $\R'$. Each transition takes place between one ancilla state $\ket{j}$ and the ground or excited state $\ket{\alpha}$.  Thus, the corresponding matrix connects the ancilla and the Rabi subspace and is given by
\begin{equation}
     \QmOp{\zeta}(\Vect{R})\equiv 
    \begin{pmatrix}
    \zeta_{N \B k} & \zeta_{N \A k} \\[1ex]
    \vdots & \vdots \\
     \zeta_{3 \B k} & \zeta_{3 \A k} 
    \end{pmatrix} \text{\,\,\, and \,\,\,}      \QmOp{\zeta}^\dagger(\Vect{R})\equiv 
    \begin{pmatrix}
    \zeta_{N \B j}^* & \dots &\zeta_{N \A j}^*  \\[1ex]
     \zeta_{3 \B j}^* & \dots & \zeta_{3 \A j}^*  
    \end{pmatrix}.
\end{equation}
The corresponding matrix for the counter-rotating contributions differs from its co-rotating counterpart only in the coupling strength and is defined as
\begin{equation}
     \QmOp{\eta}(\Vect{R})\equiv 
    \begin{pmatrix}
    \eta_{N \B k} & \eta_{N \A k} \\[1ex]
    \vdots & \vdots \\
     \eta_{3 \B k} & \eta_{3 \A k} 
    \end{pmatrix} \text{\,\,\, and \,\,\,}      
    \QmOp{\eta}^\dagger(\Vect{R})\equiv 
    \begin{pmatrix}
    \eta_{N \B j}^* & \dots &\eta_{N \A j}^*  \\[1ex]
     \eta_{3 \B j}^* & \dots & \eta_{3 \A j}^*  
    \end{pmatrix}.
\end{equation}
As \Cref{eq:Heff_pp} together with and \Cref{eq:zeta,eq:eta} show, the row of the first matrix in \Cref{eq:Heff_pp}, denoted by the index $k$, and the column of the second matrix, denoted by the index $j$ determine the detuning $\omega_{jk\alpha \alpha}$. It should be noted that $\omega_{jk\alpha \alpha}^{(+)}=\omega_{kj\alpha \alpha}^{(+)}$

In last line in \Cref{eq:Heff_pp}, we find the modification to the self-couplings of the ground and excited states. It is defined via  the matrix
\begin{equation}
\QmOp{\mathcal{V}}(\R)\equiv \operatorname{diag}(\mathcal{V}_\B,\mathcal{V}_\A)
\end{equation}
whose diagonal elements are specified in \Cref{eq:V_alpha}.

This leads us finally to the matrix representation of the particle-particle interactions, given by
\begin{equation}
\begin{split}
       \hat{H}_\mathrm{pp} =& \hbar\int  \dd^3 \R \, \QmUnity_{\SetRelevant} \QmOp{\Vect{\Psi}}_\SetRelevant^\dagger(\R)  \QmOp{\zeta}^\dagger(\Vect{R})  \QmOp{\Vect{\Psi}}_\SetAncilla \QmUnity_{\SetAncilla} \cdot 
      \int  \dd^3 \R' \, \QmUnity_{\SetAncilla} \QmOp{\Vect{\Psi}}_\SetAncilla^\dagger(\R')  \QmOp{\zeta}(\Vect{R'})  \QmOp{\Vect{\Psi}}_\SetRelevant(\R') \QmUnity_{\SetRelevant} \\
       +& \hbar\int  \dd^3 \R \, \QmUnity_{\SetRelevant} \QmOp{\Vect{\Psi}}_\SetRelevant^\dagger(\R)  \QmOp{\eta}^\dagger(\Vect{R})  \QmOp{\Vect{\Psi}}_\SetAncilla \QmUnity_{\SetAncilla} \cdot 
      \int  \dd^3 \R' \, \QmUnity_{\SetAncilla} \QmOp{\Vect{\Psi}}_\SetAncilla^\dagger(\R')  \QmOp{\eta}(\Vect{R'})  \QmOp{\Vect{\Psi}}_\SetRelevant(\R') \QmUnity_{\SetRelevant} \\
       -& \hbar\int  \dd^3 \R \, \QmUnity_\SetRelevant \QmOp{\Vect{\Psi}}_{\SetRelevant}^\dagger(\R)  \QmOp{\mathcal{V}}(\R)  \QmOp{\Vect{\Psi}}_{\SetRelevant}(\R) \QmUnity_{\SetRelevant}.
\end{split} 
\label{eq:H_appM}
\end{equation}
This representation clearly shows, that $\hat{H}_\mathrm{pp}$ connects the subspace of the relevant states $\mathcal{R}$ with the subspace of the ancillary states $\mathcal{A}$. This observation, will be useful in the subsequent section.

\subsection{From the effective Hamiltonian to the Heisenberg equations of motion}
\label{sec:Heisenberg}
Finally, we use the Heisenberg equation of motion, denoted by 
\begin{equation}
    \ii \hbar \frac{\dd }{\dd t} \QmOp{\Vect{\Psi}}(\R) = [\QmOp{\Vect{\Psi}}(\R), \hat{H}_\mathrm{eff}],
\end{equation}
as well as the Matrix formalism introduced in \Cref{sec:MatrixRepresentation} to present our result in the form of two coupled differential equations for the relevant Rabi subspace and the irrelevant ancilla subspace. Thus, we find for the time evolution of the Rabi subspace
\begin{equation}
\begin{split}
        \ii \hbar \frac{\dd }{\dd t} \QmOp{\Vect{\Psi}}_\SetRelevant(\R) = \hbar \QmOp{M}_\SetRelevant \QmOp{\Vect{\Psi}}_\SetRelevant(\R) 
        &+\hbar \QmOp{\zeta}^\dagger(\Vect{R}) \QmOp{\Vect{\Psi}}_\SetAncilla(\R) \cdot \int \dd^3 \R' \,\QmOp{\Vect{\Psi}}_\SetAncilla^\dagger(\R')\QmOp{\zeta}(\Vect{R}')
    \QmOp{\Vect{\Psi}}_\SetRelevant(\R') \\
    &+\hbar \QmOp{\eta}^\dagger(\Vect{R}) \QmOp{\Vect{\Psi}}_\SetAncilla(\R)  \cdot \int \dd^3 \R' \, \QmOp{\Vect{\Psi}}_\SetAncilla^\dagger(\R')\QmOp{\eta}(\Vect{R}')
    \QmOp{\Vect{\Psi}}_\SetRelevant(\R')\\
    &- \hbar \QmOp{\mathcal{V}}(\R)   \QmOp{\Vect{\Psi}}_\SetRelevant(\R) 
\end{split}
\end{equation}
and equivalently for the ancilla subspace
\begin{equation}
\begin{split}
        \ii \hbar \frac{\dd }{\dd t} \QmOp{\Vect{\Psi}}_\SetAncilla(\R) = \hbar \QmOp{M}_\SetAncilla \QmOp{\Vect{\Psi}}_\SetAncilla(\R) 
        &+ \hbar\int  \dd^3 \R \, \QmOp{\Vect{\Psi}}_\SetRelevant^\dagger(\R)  \QmOp{\zeta}^\dagger(\Vect{R})  \QmOp{\Vect{\Psi}}_\SetAncilla  \cdot \QmOp{\zeta}(\Vect{R}')
    \QmOp{\Vect{\Psi}}_\SetAncilla(\R') \\
    &+\hbar\int  \dd^3 \R \,  \QmOp{\Vect{\Psi}}_\SetRelevant^\dagger(\R)  \QmOp{\eta}^\dagger(\Vect{R})  \QmOp{\Vect{\Psi}}_\SetAncilla  \cdot \QmOp{\eta}(\Vect{R}')
    \QmOp{\Vect{\Psi}}_\SetAncilla(\R').
\end{split}
\end{equation}
As demonstrated by the given representation, the ancilla and the Rabi space are clearly decoupled at leading order. Nevertheless, the two spaces remain interconnected via the particle-particle interactions. Paulisch et al. discussed in Ref.~\cite{paulisch2014beyond} the hierarchy of adiabatic elimination with a similar model. In contrast to the approach taken in this study, we performed a block diagonalization to leading order, rather than adiabatically eliminating the ancilla states.

\section{Cavity QED Raman diffraction model}
\label{sec:Application}
In this section, we will focus on the dynamics induced by the reduced Hamiltonian $\hat{H}_{\SetRelevant}$, consisting of the Rabi-block matrix in \Cref{eq:MatrixR} and the light field Hamiltonian. Therefore, we derive in \Cref{sec:DerivationU} its corresponding time-evolution operator in the Schrödinger picture. In \Cref{sec:ApplicationSP}, we apply the tensor product of a generic atomic single-particle state with Gaussian amplitude and a generic optical state to the time evolution operator. The obtained result allows us to identify the beam splitter operator. Finally, we obtain in \Cref{sec:HongMandel} the Hong-Ou-Mandel effect for a specific input state in the optical as well as the atomic system.

\subsection{Derivation of the time evolution operator}
\label{sec:DerivationU}
In this section we will focus on the dynamics induced by the reduced Hamiltonian 
\begin{equation}
    \hat{H}_{\SetRelevant} = \hbar \int \dd^3\R \begin{pmatrix}
\hat{\Delta}_\B &  \hat{\Omega}_{\B\A} \\
\hat{\Omega}_{\B\A}^\dagger & \hat{\Delta}_\A  
\end{pmatrix}  + \QmOp{H}_\textrm{L},
\label{eq:H_2} 
\end{equation}
consisting of the Rabi-block matrix, \Cref{eq:MatrixR}, and the light field Hamiltonian \Cref{eq:H_L}. For the sake of simplicity, we will neglect the particle-particle interaction term because it does not contribute to the result for the input states considered in the following sections.

We separate the Hamiltonian in \Cref{eq:H_2} into $\hat{H}_{\SetRelevant}=\hat{{H}}_0^{(2)}+\hat{\mathcal{V}}$. The free part
\begin{equation}
    \hat{{H}}_0^{(2)} =  \hbar \int \dd^3 \R  \sum_{\alpha} \QmOp{\psi}_{\alpha}^\dagger(\R) \hat{\Delta}_{\alpha }\QmOp{\psi}_{\alpha}(\R)+ \QmOp{H}_\textrm{L}
\end{equation}
corresponds to the diagonal elements of \Cref{eq:H_2} and $\QmOp{H}_\textrm{L}$, while the interacting part
\begin{equation}
    \hat{\mathcal{V}} \equiv \hbar  \int \dd^3\R \, [\Omega(\R) \hat{\C}- \Lambda(\R) \hat{\C}^\dagger] \QmOp{\psi}_{\C}(\R) + \text{h.c.},
    \label{eq:V}
\end{equation}
corresponds to the off-diagonals of \Cref{eq:H_2}.

In this equation, we introduced the frequencies 
\begin{equation}
   \Omega(\R) \equiv -\sum_{j=3}^N \frac{\Omega^*_{j\B}(\R) \Omega_{j\A}(\R)}{\omega_{jj\A\B}^{(+)}}  \text{\,\,\, and \,\,\,}   \Lambda(\R) \equiv \sum_{j=3}^N \frac{\Lambda^*_{j\B}(\R) \Lambda_{j\A}(\R)}{\Omega_{jj\A\B}^{(+)}} 
\end{equation}
for the co-rotating, $ \Omega(\R)$, and the counter-rotating contributions, $\Lambda(\R)$, to simplify the notation.

Furthermore, we introduce the two-photon operators $\hat{\C}^\dagger \equiv \hat{\A}^\dagger \hat{\B}$ and $\hat{\C} \equiv \hat{\B}^\dagger \hat{\A}$, which describe the absorption of one photon from one field and the subsequent emission of another photon into the other field. Analogously, we establish novel atomic field operators \mbox{$\QmOp{\psi}_{\C}^\dagger(\R) \equiv \QmOp{\psi}_{\A}^\dagger(\R)\QmOp{\psi}_{\B}(\R)$} and   $\QmOp{\psi}_{\C}(\R) \equiv \QmOp{\psi}_{\B}^\dagger(\R)\QmOp{\psi}_{\A}(\R)$. It is important to acknowledge that the conventional bosonic commutation relations are not applicable to this particular set of joint operators. 

We use the delta-pulse approximation \cite{Schleich_2013,PhysRevD.100.065021} with the Hamiltonian
\begin{equation}
   \hat{H}_{\SetRelevant}(t,t')\equiv \vartheta \delta(t-t')\bigl[ \hat{H}^{(2)}_0(t) +  \mathcal{V}(t) \bigr]
   \label{eq:DeltaPulse}
\end{equation}
acting at time $t'$ with $\vartheta=\mathscr{A}/\Gamma$, where $\mathscr{A}$ denotes the pulse area and $\Gamma$ represents the strength of the pulse. 

We consider the interacting part of $\hat{H}_{\SetRelevant}$,
\begin{equation}
    \hat{U}_\text{int}(t,t_0) = \mathcal{T} \exp{-\frac{\ii}{\hbar} \vartheta \int_{t_0}^t \dd s  \, \delta(s-t') \hat{\mathcal{V}}_\text{I}(s)},
\end{equation}
in an interaction picture with respect to $\hat{H}^{(2)}_0$. The operator $\hat{\mathcal{V}}$ that mediates the interaction, is transformed into the interaction picture by 
\begin{equation}
    \hat{\mathcal{V}}_\text{I}(t)=  \hat{U}_0^\dagger(t,t_0)  \hat{\mathcal{V}}(t) \hat{U}_0(t,t_0). 
    \label{eq:V_I}
\end{equation}

The operator
\begin{equation}
    \hat{U}_0(t,t_0) = \mathcal{T} \exp{-\frac{\ii}{\hbar} \vartheta  \int_{t_0}^t \dd s \, \delta(s-t') \hat{H}^{(2)}_0(s)}.
\end{equation}
represents the reverse transformation which ensures that our results are finally calculated in the Schrödinger picture. Therefore, we split the time evolution operator in the Schrödinger picture into
\begin{equation}
     \hat{U}(t,t_0) = \hat{U}_0(t,t_0) \hat{U}_\text{int}(t,t_0).
\end{equation}

In order to evaluate the integrals over the time we take advantage of the delta-pulse and thus, employ the delta-pulse approximation. This results in
\begin{equation}
    \hat{U}(t,t_0) \overset{t' \in (t,t_0)}{\approx} \exp{-\frac{\ii}{\hbar} \vartheta\hat{H}_0^{(2)}(t')} \exp{-\frac{\ii}{\hbar} \vartheta \hat{\mathcal{V}}_\text{I}(t')} = \hat{U}_0(t')  \hat{U}_\text{int}(t')
    \label{eq:U_Magnus}
\end{equation}
Moreover we utilize for $\hat{\mathcal{V}}_\text{I}(t')$ in \Cref{eq:V_I} the fact that for an arbitrary unitary operator $\hat{S}$ the identity
\begin{equation}
   \sum_{k=0}^\infty (\hat{S}^\dagger \hat{\mathcal{O}}\hat{S})^k =\sum_{k=0}^\infty \hat{S}^\dagger \hat{\mathcal{O}}^k \hat{S}
\end{equation}
holds. Hence, the exponential of $\hat{\mathcal{V}}_\text{I}(t')$  can be decomposed into
\begin{equation}
    \hat{U}_\text{int}(t') = \hat{U}_0^\dagger(t')  \ee^{-\frac{\ii}{\hbar} \vartheta \hat{\mathcal{V}}(t')}  \hat{U}_0(t').
    \label{eq:U_IntMagnus}
\end{equation}
We plug this result into \Cref{eq:U_Magnus} and obtain
\begin{equation}
    \hat{U}(t,t_0) =  \ee^{-\frac{\ii}{\hbar} \vartheta \hat{\mathcal{V}}(t')} \hat{U}_0(t') .
    \label{eq:TimeEvolutionOperator}
\end{equation}
It is worth noting that $\hat{U}_0(t')$ in \Cref{eq:U_Magnus} cancels with $\hat{U}_0^\dagger(t')$ in \Cref{eq:U_IntMagnus}. Thus, the delta-pulse approximation and the subsequent transformations in \Cref{eq:DeltaPulse} to \Cref{eq:TimeEvolutionOperator} justify, that free propagation can be considered in the Schrödinger picture, while the interaction potential is evaluated within the Heisenberg picture of the free Hamiltonian $\hat{H}^{(2)}_0$. This is an approach frequently utilized in atom optics \cite{Hartmann_Regimes,Jenewein}.

We proceed by separating \Cref{eq:TimeEvolutionOperator} into an even and odd component. This step is presented in greater detail in Appendix~\ref{app:V}. Thus, the time evolution operator for \Cref{eq:H_2} in the Schrödinger picture can finally be expressed as
\begin{equation}
    \hat{U}(t,t_0) \overset{t' \in (t,t_0)}{\approx} \left[ \cos{\left(\vartheta \hat{\Omega}\right)} - \ii \sin{\left(\vartheta\hat{\Omega}\right)} \big(\hbar\QmOp{\Omega}\big)^{-1}\mathcal{\hat{V}}\right] \hat{U}_0(t') \equiv \hat{U},
    \label{eq:U}
\end{equation}
with the Rabi operator $\hat{\Omega}$ and
\begin{equation}
\scalebox{0.97}{$
\begin{split}
        \hat{\mathcal{V}}^2  \approx \hbar^2 \!\! \int \!\!  \dd^3\R \!  \int \!\!  \dd^3\R' 
        \Bigl\{&\Omega(\R) \Omega(\R') \hat{\C}^{2}  \QmOp{\psi}_{\C}(\R) \QmOp{\psi}_{\C}(\R') +\Omega^*(\R) \Omega^*(\R') \hat{\C}^{\dagger 2}  \QmOp{\psi}_{\C}^\dagger(\R)  \QmOp{\psi}_{\C}^\dagger(\R') \\
    + \bigl[&\Omega(\R) \Omega^*(\R') \hat{\C}\hat{\C}^\dagger + \Lambda(\R) \Lambda^*(\R')  \hat{\C}^\dagger  \hat{\C} \bigr] \QmOp{\psi}_{\C}(\R) \QmOp{\psi}_{\C}^\dagger(\R') \\
       + \bigl[& \Omega^*(\R) \Omega(\R')  \hat{\C}^\dagger  \hat{\C}  + \Lambda^*(\R) \Lambda(\R') \hat{\C}\hat{\C}^\dagger   \bigr] \QmOp{\psi}_{\C}^\dagger(\R) \QmOp{\psi}_{\C}(\R')\Bigr\}, 
    \end{split}$}
     \label{eq:V2_approx} 
\end{equation}
where we define $\hat{\mathcal{V}}^2 \equiv \big(\hbar\hat{\Omega}\big)^2$. The full expression of $\hat{\mathcal{V}}^2$ and the justification of its approximation are discussed as well in Appendix~\ref{app:V}. For improved readability, we will omit the variable $t'$ in the following discussion, but keep in mind that all time dependencies are evaluated at the pulse time $t^\prime$. 

It should be noted, that within the interaction picture we deal with the dressed states with respect to the atomic Hamiltonian, including the center-of-mass motion, and the Hamiltonian of the electromagnetic field \cite{Dalibard:85}. The delta-pulse approximation in \Cref{eq:U} and the approximation of $\mathcal{V}^2$ in \Cref{eq:V2_approx} result in the neglect of certain contributions. Thus, we cannot completely invert the interaction picture \cite{PhysRevResearch.6.043153}.

The time-evolution operator in \Cref{eq:U} is composed of the exponential $\QmOp{U}_{0}$,
which comprises the center-of-mass motion, and the trigonometric part  
\begin{equation}
    \QmOp{U}_{\mathrm{i}}\equiv\cos{\left(\vartheta \hat{\Omega}\right)} - \ii \sin{\left(\vartheta \hat{\Omega}\right)} \big(\hbar \hat{\Omega}\big)^{-1} \mathcal{\hat{V}},
    \label{eq:U_int}
\end{equation}
which corresponds to interaction within the interaction picture with respect to $\hat{H}_0^{(2)}$. 

\Cref{eq:U_int} reveals a similar structure as the time-evolution operator of the Jaynes-Cummings model \cite{schleich_quantum_2001}. In this model, the atomic eigenvectors, which are orthogonal, are chosen as basis vectors. Thus, the time evolution operator can be decomposed into an odd and even part. Written in terms of the atomic eigenvectors, the evaluation of the even terms yields a cosine-function on the diagonal of the matrix, while the evaluation of the odd terms reveals a sine-function on the off-diagonals. 

It is our objective to present the time evolution operator in \Cref{eq:U} in a form that is comparable to the time evolution operator of the Jaynes-Cummings model. Although we use in \Cref{eq:H_2} a matrix representation with the field operators as basis, the field operators themselves do not obey any orthogonality relations. Thus, the utilization of the orthogonality relations of the single-particle basis vectors is not applicable in this context, as typically employed within the framework of a first-quantized description of the atom \cite{schleich_quantum_2001}. In fact, the number of field operators increases with each power of $\hat{\mathcal{V}}$. Consequently, \Cref{eq:U} is not a matrix equation, as the atomic field operators are incorporated within the operators $\hat{\Omega}$ and $\hat{\mathcal{V}}$. Nevertheless, we use the relation $\hat{\mathcal{V}}^{2k+1} = \hat{\Omega}^{2k+1} \frac{\hat{\mathcal{V}}}{\hat{\Omega}}$, which allows us to transform the time evolution operator into a form similar to that of the Jaynes-Cummings model.

\subsection{Generic single-particle atomic and optical state}
\label{sec:ApplicationSP}
In the following, we consider the input state as a tensor product of the states of the atomic field, denoted by the subscript A, and the states of the electromagnetic field, denoted by the subscript L, each with modes a and b with the ordering
\begin{equation}
    \ket{\psi} \equiv \ket{\psi_\text{a}}_\text{A} \otimes \ket{\psi_\text{b}}_\text{A}\otimes  \ket{n_\text{a}}_\text{L} \otimes \ket{n_\text{b}}_\text{L} \equiv \ket{\psi_\textrm{a} \, \psi_\textrm{b}}_\text{A} \ket{n_\text{a} \, n_\text{b}}_\text{L}.
\end{equation}
In this section we consider the tensor product of a generic single-particle atomic state $\ket{\psi^{(1)}}_\mathrm{A}$ and a generic state of the optical field $\ket{\psi^{(w)}}_\text{L}$, denoted by 
\begin{equation}
    \ket{\psi} = \ket{\psi^{(1)}}_\mathrm{A} \otimes  \ket{\psi^{(w)}}_\text{L}. 
    \label{eq:Input}
\end{equation}

A single-particle state of the atomic field can generally be expressed as
\begin{equation}
	\ket{\psi^{(1)}}_\mathrm{A} \equiv \mathscr{N} \int \dd^3 \R \, \sum_{\alpha \in \mathcal{R}}  f_\alpha(\R) \, \QmOp{\psi}_\alpha^\dagger(\R) \ket{0 \, 0}_\mathrm{A},
    \label{eq:InputA}
\end{equation}
where $\mathscr{N}=\left( \int \dd^3 \R \, \sum_{\alpha } |f_\alpha(\R)|^2 \right)^{-1/2}$ ensures that the single-particle state is normalized. Hence, the atom is in a superposition of the two possible modes a and b with the corresponding single-particle amplitude $f_\alpha(\R)$. We focus in this article solely on amplitudes in the form of a Gaussian distribution. Thus, the amplitude is denoted by
\begin{equation}
    f_\alpha(\R) = \frac{1}{(2\pi)^{3/2} \sigma_\alpha^3} \cdot \ee^{\ii \boldsymbol{\kappa} \R} \cdot \ee^{-\frac{\R^2}{2\sigma_\alpha^2}}.
    \label{eq:GaussianAmplitude}
\end{equation}

A general state of the optical field is denoted by 
\begin{equation}
    \ket{\psi^{(w)}}_\text{L} \equiv \sum_{n_\A, n_\B=0} w_{n_\A n_\B} \ket{n_\A n_\B}_\text{L},
    \label{eq:InputL}
\end{equation}
which is a sum over Fock states in both modes, a and b, respectively, with corresponding amplitudes $w_{n_\alpha n_\beta}\equiv w_{n_\alpha} w_{n_\beta} $.

In \Cref{sec:FinalStateGeneral}, we calculate the action of the time-evolution operator on a generic single-particle atomic and a generic optical state. Subsequently, in \Cref{sec:FinalStatePlaneWaves}, we neglect the counter-propagating contributions and select plane waves as optical mode functions. 

\subsubsection{Final state: general case}
\label{sec:FinalStateGeneral}
We evaluate the action of the time evolution operator in \Cref{eq:U} in three steps. First, we apply its exponential component, $\QmOp{U}_0$, to the input state in \Cref{eq:Input} and obtain
\begin{equation}
     \QmOp{U}_0  \ket{\psi}= \mathscr{N} \int \dd^3 \R \, \sum_{\alpha} \sum_{n_\A, n_\B} \ee^{-\frac{\ii}{\hbar}\vartheta \mathcal{C}_{n_{\alpha},\alpha}(\R)}  f_\alpha(\R) \, \QmOp{\psi}_\alpha^\dagger(\R) \ket{0 \, 0}_\mathrm{A}  w_{n_\A n_\B} \ket{n_\A n_\B}_\text{L}.
     \label{eq:OutH0}
\end{equation}
This expression contains the phase factor
\begin{equation}
\begin{split}
    	\mathcal{C}_{n_{\alpha},\alpha}(\R) \equiv& \mathcal{E}_\text{COM}^{(\alpha)}(\R) + \sum_{\alpha'=\A,\B} \hbar \Omega_{\alpha'} \biggl(n_{\alpha'}+\frac{1}{2}\biggr) \\
        - &\hbar  \Biggl[\sum_{j=3}^N \frac{|\Omega_{j\alpha }(\R)|^2}{\omega_{jj\alpha \alpha}^{(+)}} n_\alpha - \sum_{j=3}^N \frac{|\Lambda_{j\alpha }(\R)|^2}{\Omega_{jj\alpha \alpha }^{(+)}}  (n_\alpha +1)+ \omega_\alpha  \Biggr],
\end{split}
\end{equation}
which encompasses the component 
\begin{equation}
    \mathcal{E}_\text{COM}^{(\alpha)}(\R) = -\frac{\hbar^2}{2M} \frac{\R^2-2\ii \sigma_\alpha^2 \boldsymbol{\kappa} \R -3\sigma_\alpha^2- \sigma_\alpha^4 \boldsymbol{\kappa}^2}{ \sigma_\alpha^4}
\end{equation}
that results from the action of the center-of-mass Hamiltonian on the Gaussian distribution. 

A more detailed discussion of the action of $\hat{H}_0^{(2)}$ on a generic single-particle state can be found in Appendix~\ref{app:COM_U}. The calculation of the specific parameters for a single-particle state with a Gaussian amplitude is performed in Appendix~\ref{app:COM_Gauss}.

We proceed by applying the cosine in \Cref{eq:U_int} to the initial state in \Cref{eq:Input}. This component of the time-evolution yields
\begin{equation}
   \begin{split}
             \cos{\left(\vartheta \hat{\Omega}\right)} \ket{\psi} = \mathscr{N} \!\!  \int \!\! \dd^3 \R  \!\! \sum_{n_\A, n_\B} \!\! \biggl\{ &\cos{\Bigl(\vartheta\sqrt{\Omega_{\B_+\A}^2(\R)+\Lambda_{\B \A_+}^2(\R)}} \Bigr) f_\A(R) \QmOp{\psi}_\A^\dagger(\R) \ket{0 \, 0}_\text{A}  \\
       +&\cos{\Bigl(\vartheta\sqrt{\Omega_{\B \A_+}^2(\R)+\Lambda_{\B_+\A}^2(\R)}} \Bigr) f_\B(R) \QmOp{\psi}_\B^\dagger(\R) \ket{0 \, 0}_\text{A}  \biggr\} \\
       \otimes & w_{n_\A n_\B} \ket{n_\A n_\B}_\text{L},
    \end{split}
    \label{eq:OutCosine}
\end{equation}
where we introduced the frequencies 
\begin{equation}
	\Omega_{\B_+\A}^2(\R) \equiv |\Omega(\R)|^2 (n_\B+1)n_\A \text{\, and \,}
     \Omega_{\B \A_+}^2(\R) \equiv |\Omega(\R)|^2 (n_\A+1) n_\B,
\end{equation}
for the co-rotating contributions and 
\begin{equation}
   \Lambda_{\B_+\A}^2(\R) \equiv |\Lambda(\R)|^2 (n_\B+1)n_\A  \text{\, and \,} \Lambda_{\B \A_+}^2(\R) \equiv |\Lambda(\R)|^2 (n_\A+1) n_\B 
\end{equation}
for the counter-rotating contributions in order to enhance the readability of \Cref{eq:OutCosine}.
The cosine represents the even part of the time-evolution operator. Consequently, it is associated with transitions that end in the initial state. 

Next, we evaluate the sine in \Cref{eq:U_int}. The sine is an odd function that mediates transitions leading to a change in the initial state. Therefore, we introduce the absolute values $\Omega(\R)\equiv |\Omega(\R)|\ee^{\ii \phi_\Omega}$ and $\Lambda(\R)\equiv |\Lambda(\R)|\ee^{\ii \phi_\Lambda}$. The action of the sine on the initial state is given by
{\medmuskip=1mu
\begin{equation}
  \begin{split}
         \sin{\left(\vartheta \hat{\Omega}\right)} \frac{\mathcal{\hat{V}}}{\hbar \hat{\Omega}} \ket{\psi} \approx & \ii \mathscr{N}  \int   \dd^3 \R \,  \sum_{n_\A, n_\B}   \\
      \Bigl\{ &\sin{\Bigl(\vartheta\Omega_{\B_+\A}(\R)} \Bigr)  \ee^{\ii \phi_\Omega}  f_\A(R) \QmOp{\psi}_\B^\dagger(\R)  \ket{0 \, 0}_\text{A} 
          w_{n_\A n_\B}  \ket{n_\A-1 \, n_\B+1}_\text{L}   \\
         -&\sin{\Bigl(\vartheta\Lambda_{\B \A_+}(\R)} \Bigr) \ee^{\ii \phi_\Lambda} f_\A(R) \QmOp{\psi}_\B^\dagger(\R)  \ket{0 \, 0}_\text{A}    w_{n_\A n_\B}  \ket{n_\A+1 \, n_\B-1}_\text{L}\\
        +  & \sin{\Bigl(\vartheta\Omega_{\B \A_+}(\R)}\Bigr) \ee^{-\ii \phi_\Omega}  f_\B(R) \QmOp{\psi}_\A^\dagger(\R) \ket{0 \, 0}_\text{A}  w_{n_\A n_\B}  \ket{n_\A+1 \, n_\B-1}_\text{L} \\
        -&\sin{\Bigl(\vartheta\Lambda_{\B_+\A}(\R)\Bigr)} \ee^{-\ii \phi_\Lambda} f_\B(R) \QmOp{\psi}_\A^\dagger(\R) \ket{0 \, 0}_\text{A} w_{n_\A n_\B}  \ket{n_\A-1 \, n_\B+1}_\text{L}  \Bigr\}. 
  \end{split}
  \label{eq:OutSine}
\end{equation}}
During the evaluation of the sine, we neglected terms that contain the product of co- and counter-rotating contributions. This is analogous to the approximation in \Cref{eq:Averaging3}.

We introduce the amplitudes
\begin{equation}
	c_{n_\alpha}(\R) \equiv  \mathscr{N}  \ee^{-\frac{\ii}{\hbar}\vartheta \mathcal{C}_{n_{\alpha},\alpha}(\R)}  f_\alpha(\R), 
    \label{eq:c_alpha}
\end{equation}
which belong to transitions which do not change the initial state. Similarly, we define amplitudes for transitions that do alter the initial states. The amplitudes for the co-rotating contributions are given by
\begin{equation}
	c_{n_\A,\Omega}(\R)  \equiv  c_{n_\A}(\R) \ee^{\ii \phi_\Omega} \text{\, and \,} c_{n_\B,\Omega}(\R)  \equiv  c_{n_\B}(\R) \ee^{-\ii \phi_\Omega},
    \label{eq:c_Omega}
\end{equation}
and by
\begin{equation}
	c_{n_\A,\Lambda}(\R)  \equiv  c_{n_\A}(\R) \ee^{\ii \phi_\Lambda} \text{\, and \,} c_{n_\B,\Lambda}(\R)  \equiv  c_{n_\B}(\R) \ee^{-\ii \phi_\Lambda} 
    \label{eq:c_Lambda}
\end{equation}
for the counter-rotating contributions.

The action of the time-evolution operator in \Cref{eq:U} on the input state, \Cref{eq:Input}, can be found by collecting the partial results in \Cref{eq:OutCosine,eq:OutSine,eq:OutH0}. Moreover, we insert the amplitudes in \Cref{eq:c_alpha,eq:c_Omega,eq:c_Lambda} and finally obtain
{\medmuskip=1mu
   \begin{equation}
	\begin{split}
		\hat{U}\ket{\psi}&= \!\! \int \!\! \dd^3 \R \!\! \sum_{n_\A, n_\B} \!\! w_{n_\A n_\B}  \biggl\{\cos{\Bigl(\vartheta\sqrt{\Omega_{\B_+\A}^2(\R)+ \Lambda_{\B \A_+}^2(\R)}} \Bigr) c_{n_\A}(\R) \QmOp{\psi}_\A^\dagger(\R) \ket{0 \, 0}_\text{A} \ket{n_\A \, n_\B}_\text{L}  \\
		& \phantom{\!\! \int \!\! \dd^3 \R \!\! \sum_{n_\A, n_\B} \!\! w_{n_\A n_\B}  \biggl\{ } +   \cos{\Bigl(\vartheta\sqrt{\Omega_{\B \A_+}^2(\R)+\Lambda_{\B_+\A}^2(\R)}} \Bigr) c_{n_\B}(\R) \QmOp{\psi}_\B^\dagger(\R) \ket{0 \, 0}_\text{A}  \ket{n_\A \, n_\B}_\text{L} \\
		&\phantom{\!\! \int \!\! \dd^3 \R \!\! \sum_{n_\A, n_\B} \!\! w_{n_\A n_\B}   \biggl\{ }-\ii \sin{\Bigl(\vartheta\Omega_{\B_+\A}(\R)} \Bigr) c_{n_\A,\Omega}(\R) \QmOp{\psi}_\B^\dagger(\R) \ket{0 \, 0}_\text{A} \ket{n_\A-1 \, n_\B+1}_\text{L}\\
        &\phantom{\!\! \int \!\! \dd^3 \R \!\! \sum_{n_\A, n_\B} \!\! w_{n_\A n_\B}   \biggl\{ }+ \ii \sin{\Bigl(\vartheta\Lambda_{\B \A_+}(\R)} \Bigr) c_{n_\A,\Lambda}(\R) \QmOp{\psi}_\B^\dagger(\R) \ket{0 \, 0}_\text{A}  \ket{n_\A+1 \, n_\B-1}_\text{L}  \\
		&\phantom{\!\! \int \!\! \dd^3 \R \!\! \sum_{n_\A, n_\B} \!\! w_{n_\A n_\B}   \biggl\{ }-\ii \sin{\Bigl(\vartheta\Omega_{\B \A_+}(\R)}\Bigr) c_{n_\B,\Omega}(\R) \QmOp{\psi}_\A^\dagger(\R) \ket{0 \, 0}_\text{A}  \ket{n_\A+1 \, n_\B-1}_\text{L}  \\
        &\phantom{\!\! \int \!\! \dd^3 \R \!\! \sum_{n_\A, n_\B} \!\! w_{n_\A n_\B}  \biggl\{ }+\ii   \sin{\Bigl(\vartheta\Lambda_{\B_+\A}(\R)\Bigr)} c_{n_\B,\Lambda}(\R) \QmOp{\psi}_\A^\dagger(\R) \ket{0 \, 0}_\text{A} \ket{n_\A-1 \, n_\B+1}_\text{L}\biggr\}. 
	\end{split}
	\label{eq:Out}
\end{equation}}
The output state is in a superposition of the initial state, \Cref{eq:Input}, denoted by the first and second term of \Cref{eq:Out}. The third and fourth term describe transitions from the atomic state $\ket{\A}$ to $\ket{\B}$. The co-rotating contributions, with $c_{n_\A,\Omega}$ are accompanied by an absorption of a photon from the a-mode, and a subsequent emission of a photon into the b-mode. The inverse photon processes can be found for the counter-rotating contributions with $c_{n_\A,\Lambda}$. The last two terms show transitions from the initial atomic state $\ket{\B}$ to $\ket{\A}$. As expected, the co- and counter-rotating processes are accompanied by the inverse photonic processes.

\subsubsection{Final state: within RWA and with plane waves optical mode functions}
\label{sec:FinalStatePlaneWaves}
In the weak coupling regime, the counter-rotating coupling factors $\Lambda_{\B \A_+}^2$ and $\Lambda_{\B_+ \A}^2$ are small compared to the co-rotating coupling factors, and thus, their corresponding states are rarely occupied. Consequently, we focus in this section solely on the co-rotating contributions. Moreover, we specify the mode functions of the electromagnetic field and choose them to be counter-propagating plane waves, i.~e. $u_\A(\R)=\ee^{\ii \mathbf{k}_\A \R}$ and $u_\B(\R)=\ee^{-\ii \mathbf{k}_\B \R}$ according to \Cref{fig:Atom}~a). Furthermore, we write the sum of the two different wave vectors as $\K=\mathbf{k}_\A+\mathbf{k}_\B$. Thus, $\Omega(\R)$ can now be decomposed into a position-independent amplitude and a position-dependent phase, given by
\begin{equation}
    \Omega(\R)=\sum_{j=3}^N \frac{\mathcal{E}_\A \mathcal{E}_\B |\Vect{\wp}_{j\B}| |\Vect{\wp}_{j\A}|}{2 \hbar^2 \omega_{jj\A\B}^{(+)}}  \ee^{\ii(\phi + \K \R)} \equiv  \sum_{j=3}^N \frac{|\Omega_{j\A}| |\Omega_{j\B}|}{\omega_{jj\A\B}^{(+)}} \ee^{\ii(\phi + \K \R)} \equiv | \Omega|\ee^{\ii(\phi + \K \R)}.
\end{equation}
The phase between the dipole transition elements is denoted by $\phi=\mathrm{arg}\{\Vect{\wp}_{j\B}^* \Vect{\wp}_{j\A}\}$ and the absolute value of the coupling  is given by $|\Omega_{j\alpha}|=\mathcal{E}_\alpha |\Vect{\wp}_{j\alpha}|/(\sqrt{2} \hbar )$. This leads to the replacement $\ee^{\ii \phi_\Omega} \equiv \ee^{\ii (\phi+\K \R)}$.

Due to the choice of the mode function the frequencies
\begin{equation}
   \Omega_{\B_+\A}^2(\R) = |\Omega|^2 (n_\B+1)n_\A \equiv  \Omega_{\B_+\A}^2 \text{\, and \,} \Omega_{\B \A_+}^2(\R) = |\Omega|^2 (n_\A+1)n_\B \equiv  \Omega_{\B \A_+}^2 
\end{equation}
lose their position-dependence, and the phase factor reduces to
\begin{equation}
    \mathcal{C}_{n_\alpha,\alpha}^{(\Omega)}(\R) = \mathcal{E}_{\mathrm{COM}}^{(\alpha)}(\R)+\hbar \left[ \sum_{\alpha'=\A,\B} \Omega_{\alpha'} \biggl( n_{\alpha'} + \frac{1}{2} \biggr) - \sum_{j=3}^N \frac{|\Omega_{j\alpha }|^2}{\omega_{jj\alpha \alpha}^{(+)}} n_\alpha + \omega_\alpha  \right].
    \label{eq:C_alpha}
\end{equation}
Moreover, we define the atomic output states 
\begin{equation}
\begin{aligned}
	C_{n_\A} \ket{1 0}_\text{A} \equiv \int \dd^3 \R \, c_{n_\A}(\R) \QmOp{\psi}_\A^\dagger(\R) \ket{0 \, 0}_\text{A}  \\
    C_{n_\B} \ket{0 1}_\text{A} \equiv \int \dd^3 \R \, c_{n_\B}(\R) \QmOp{\psi}_\B^\dagger(\R) \ket{0 \, 0}_\text{A},
    \label{eq:C_ab}
\end{aligned}    
\end{equation}
which describe processes that end in the initial state. Conversely, transitions that result in a change of the initial state are described by the output states
\begin{equation}
  \begin{aligned}
	C_{n_\A,\hbar K} \ket{0 1}_\text{A} &\equiv \int \dd^3 \R \, c_{n_\A}(\R) \ee^{\ii (\phi + \K \R)} \QmOp{\psi}_\B^\dagger(\R) \ket{0 \, 0}_\text{A}, \\ C_{n_\B,-\hbar K} \ket{1 0}_\text{A} &\equiv \int \dd^3 \R \, c_{n_\B}(\R) \ee^{-\ii (\phi + \K \R)} \QmOp{\psi}_\A^\dagger(\R) \ket{0 \, 0}_\text{A}.
    \label{eq:C_hK}
\end{aligned}  
\end{equation}
These states have gained a phase factor $\ee^{\pm \ii (\phi + \K \R)}$, that is associated with a momentum kick of $\hbar K$ in opposite directions. This momentum transfer is mediated by the absorption and subsequent emission of photons with different frequencies.

By utilizing \Cref{eq:C_alpha,eq:C_hK} and under neglect of the counter-rotating contributions, we are able to cast the output state in \Cref{eq:Out} for plane wave optical mode functions into the form
{\medmuskip=1mu
\begin{equation}
   \begin{split}
       \hat{U}\ket{\psi} \approx \!\! \sum_{n_\A, n_\B} \!\!  w_{n_\A n_\B}\!  \biggl\{& \!\! \cos{\Bigl(\vartheta\Omega_{\B_+\A}} \Bigr) C_{n_\A} \ket{1 0}_\text{A} \ket{n_\A \, n_\B}_\text{L} 
        +\cos{\Bigl(\vartheta\Omega_{\B \A_+}} \Bigr) C_{n_\B} \ket{0 1 }_\text{A} \ket{n_\A \, n_\B}_\text{L}\\
		- &\ii \sin{\Bigl(\vartheta \Omega_{\B_+\A}} \Bigr) C_{n_\A,\hbar K} \ket{0 1}_\text{A} \ket{n_\A-1 \, n_\B+1}_\text{L} \\
		- &\ii \sin{\Bigl(\vartheta \Omega_{\B \A_+}} \Bigr) C_{n_\B,\text{-}\hbar K} \ket{1 0}_\text{A} \ket{n_\A + 1 \, n_\B-1}_\text{L} \biggr\}.
   \end{split}
    \label{eq:OutPW}
\end{equation}}
This representation shows the well-known Rabi oscillations for the Raman system \cite{Moler,EnnoPHD}. The first and second terms correspond to transitions that end in the initial state. Hence, there is no change in the occupation of the states. The third and fourth terms describe transitions that alter the state. These transitions are mediated by the absorption and subsequent emission of photons with different frequencies, leading to a momentum transfer of $\pm \hbar K$ to the atom. Moreover, the occupation of the atomic and optical modes changes. Thus, the process $\ket{\A} \rightarrow \ket{\B}$ leads to the occupation $\ket{0 1}_\text{A} \ket{n_\A-1 \, n_\B+1}_\text{L}$ and the inverse process to $\ket{1 0}_\text{A} \ket{n_\A + 1 \, n_\B-1}_\text{L}$.

Finally, we employ once more the vector representation developed in \Cref{sec:MatrixRepresentation} for a comparison to the first quantized Raman model. By writing the input states in the subspace of $\mathcal{R}$, we obtain for the beam splitter operator 
\begin{equation}
     \QmOp{S}_\textrm{pw}^{(\vartheta)}  = \begin{pmatrix}
\cos(\vartheta |\Omega| \sqrt{(\hat{n}_\A+1)\hat{n}_\B})  & -\ii \sin(\vartheta |\Omega| \sqrt{(\hat{n}_\A+1)\hat{n}_\B})  \frac{ \ee^{\ii \K \R} \hat{\A}\hat{\B}^\dagger}{\sqrt{(\hat{n}_\A+1)\hat{n}_\B}}\\
-\ii \sin(\vartheta |\Omega| \sqrt{(\hat{n}_\B+1)\hat{n}_\A})  \frac{ \ee^{-\ii \K \R} \hat{\A}^\dagger \hat{\B}}{\sqrt{(\hat{n}_\B+1)\hat{n}_\A}}
& \cos(\vartheta |\Omega| \sqrt{(\hat{n}_\B+1)\hat{n}_\A})) 
\end{pmatrix}  \cdot \mathbf{v}
\label{eq:BS_operator}
\end{equation}
under the assumption of plane wave optical mode functions and without counter-rotating contributions and with the vector
\begin{equation}
    \mathbf{v} = \begin{pmatrix}
        \ee^{-\frac{\ii}{\hbar}\vartheta  \mathcal{C}_{n_\B,\B}^{(\Omega)}} \\ 
        \ee^{-\frac{\ii}{\hbar}\vartheta \mathcal{C}_{n_\A,\A}^{(\Omega)}}
    \end{pmatrix},
\end{equation}
which contains the phase factors that we will discuss at the end of this section. This result holds for single-particle atomic states with Gaussian amplitude. The diagonal of \Cref{eq:BS_operator} reveals the self-coupling, i.~e. transitions that do not change the internal state of the atom. Consequently, the number of photons remains constant. In contrast, the off-diagonals mediate transitions between the internal states. The transition from $\ket{\A}$ to $\ket{\B}$ causes an annihilation of a photon in mode a and a creation of a photon in mode b. Moreover, this process is accompanied by a momentum kick of $\hbar K$, associated with $\ee^{\ii \K \R}$. The inverse process from $\ket{\B}$ to $\ket{\A}$ annihilates a photon in the b-mode and creates another one in the a-mode, leading to a momentum kick of  $\hbar K$ in the inverse direction. This result is consistent with the findings of semi-classical models, as in Refs.~\cite{Moler,EnnoPHD}.

With the joint electromagnetic field operators
$\hat{\C}^\dagger \equiv \hat{\A}^\dagger \hat{\B}$, $\hat{\C} \equiv \hat{\B}^\dagger \hat{\A}$ and the joint number operator $\hat{n}_\C=\hat{n}_\A(\hat{n}_\B + 1)$ we rewrite \Cref{eq:BS_operator} into
\begin{equation}
     \QmOp{S}_\textrm{pw}^{(\vartheta)} = \begin{pmatrix}
\cos(\vartheta |\Omega| \sqrt{\hat{\C} \hat{\C}^\dagger})  & -\ii \sin(\vartheta |\Omega| \sqrt{\hat{\C} \hat{\C}^\dagger}) \ee^{\ii \K \R} {\sqrt{\hat{\C} \hat{\C}^\dagger}}^{-1}\hat{\C}\\
-\ii \sin(\vartheta |\Omega| \sqrt{\hat{n}_\C})  \ee^{-\ii \K \R} \sqrt{\hat{n}_\C}^{-1}\hat{\C}^\dagger 
& \cos(\vartheta |\Omega| \sqrt{\hat{n}_\C})) 
\end{pmatrix}
 \cdot \mathbf{v}
\label{eq:BS_operatorC}
\end{equation}
and transform the beam splitter operator into a form, that is similar to the corresponding operator of the Jaynes-Cummings model \cite{schleich_quantum_2001,Soukup,PhysRevResearch.4.013115}, although these studies were performed within an interaction picture.  

In contrast to Refs.~\cite{Moler,EnnoPHD,schleich_quantum_2001,Soukup}, we observe an additional phase $\ee^{-\frac{\ii}{\hbar}\vartheta \mathcal{C}_{n_\alpha,\alpha}}$ with 
\begin{equation}
    \mathcal{C}_{n_\alpha,\alpha}^{(\Omega)}(\R) = \mathcal{E}_{\mathrm{COM}}^{(\alpha)}(\R) +\hbar  \Biggl[\sum_{\alpha'=\A,\B} \Omega_{\alpha'} \biggl( n_{\alpha'} + \frac{1}{2} \biggr)-\sum_{j=3}^N \frac{|\Omega_{j\alpha }|^2}{\omega_{jj\alpha \alpha}^{(+)}} n_\alpha + \omega_\alpha  \Biggr].
\end{equation}
This factor consists of several contributions. The components $ \hbar \Omega_{\alpha'} \left( n_{\alpha'} + \frac{1}{2} \right)+\hbar \omega_\alpha $ appear, because the beam splitter operator is presented in the Schrödinger picture instead of the interaction picture. The contribution $\frac{\hbar|\Omega_{j\alpha }|^2}{\omega_{jj\alpha \alpha}^{(+)}} n_\alpha$ causes an entanglement between the optical and the atomic states and can be observed for models with large detuning, where an effective Hamiltonian has been derived \cite{schleich_quantum_2001}. Moreover, evaluating the center-of-mass motion yields the contribution, denoted by $\mathcal{E}_{\mathrm{COM}}^{(\alpha)}$ \cite{schleich_quantum_2001}, whose explicit form depends on the amplitude of the atomic input state.

\subsection{Two-particle atomic state in the spirit of Hong-Ou-Mandel}
\label{sec:HongMandel}
The famous Hong-Ou-Mandel experiment proposed \cite{HongTheory} and experimentally realized in 1987 \cite{HongExperiment}, revealed that two photons, entering the input ports of a 50:50 beam splitter separately, would always emerge together from either of the exit ports. However, the Hong-Ou-Mandel effect is merely the simplest example of interference between multiple particles \cite{Ou2007,LimBeige,Tichy_2012}.  As demonstrated in several experiments \cite{Lopes,Preiss,Li_Rydberg,Dussarrat}, atoms exhibit analogous behavior. Moreover, many-body interference has recently been realized with 10 indistinguishable atoms \cite{quensen}.  For a comprehensive review of the atomic Hong-Ou-Mandel effect we refer to \cite{KaufmannReview}. Hong-Ou-Mandel experiments based on atom interferometry have been demonstrated with Bragg diffraction \cite{Lopes,Dussarrat}. Thus, we analyze in this section a Hong-Ou-Mandel set-up for atomic Raman diffraction in the spirit of the previously mentioned pioneering experiments.

Hence, we expand our previous analysis to the case of two atoms. The input state
\begin{equation}
    \ket{\psi}_\mathrm{HM} = \ket{\psi^{(2)}}_\mathrm{A} \otimes  \ket{\psi^{(n)}}_\text{L}. 
    \label{eq:InputMandel}
\end{equation}  
is as previously in \Cref{eq:Input} a tensor product of an atomic and optical input state. However, we consider now a two-particle atomic state, given by
\begin{equation}
	\ket{\psi^{(2)}}_\mathrm{A} \equiv  \mathscr{N}_2 \int \dd^3 \R \int \dd^3 \R' \, f_{\A\B}(\R,\R') \, \big[ \QmOp{\psi}_\A^\dagger(\R) \QmOp{\psi}_\B^\dagger(\R') + \QmOp{\psi}_\B^\dagger(\R) \QmOp{\psi}_\A^\dagger(\R') \big] \ket{0 \, 0}_\mathrm{A}.
    \label{eq:InputMandelA}
\end{equation}
This state represents two atoms in different internal states, where their respective positions remain indistinguishable. Their joint amplitude is given by
\begin{equation}
    f_{\A \B}(\R,\R') = f_\A(\R) f_\B(\R')
    \label{eq:AmplitudeMandel}
\end{equation}
with $f_\alpha(\R) \equiv f_\alpha(\R')$. The normalization of the initial state is ensured by the factor \mbox{$\mathscr{N}_2 = \left(2 \int \dd^3 \R \int \dd^3 \R' [|f_\A(\R)|^2 |f_\B(\R')|^2 + |f_\A(\R')|^2 |f_\B(\R)|^2 ]\right)^{-1/2}$}. Each amplitude is, as previously in \Cref{eq:GaussianAmplitude}, chosen to be a Gaussian function. This input state does not allow one to differentiate between which atom is at which position.

The optical state 
\begin{equation}
    \ket{\psi^{(n)}}_\text{L} \equiv \sum_{n} w_{n}^2  \ket{n \, n}_\text{L}
    \label{eq:InputMandelL}
\end{equation}
is selected in a manner such that both optical modes are initially occupied by the same number of photons.

In the following discussion, the counter-rotating terms will be excluded from the analysis, because they are not the primary subject of the present discussion. Moreover, we assume  $\Omega_{j\alpha}(\R) \approx \Omega_{j\alpha}(\R')$. Thus, the atoms have to be sufficiently close to each other, to experience approximately the same electromagnetic field. Consequently, during the pulse, both atoms experience the same Rabi frequency and, thus, undergo the same AC-Stark shift. 

We will proceed in complete analogy to \Cref{sec:FinalStateGeneral} and divide the application of the time evolution operator in \Cref{eq:U} into three steps. The application of its exponential component  $\QmOp{U}_0$ to the input state reveals 
\begin{equation}
\begin{split}
         \QmOp{U}_0  \ket{\psi}_\mathrm{HM} = &\mathscr{N}_2 \int \dd^3 \R \int \dd^3 \R' \, \sum_{n} \ee^{-\frac{\ii}{\hbar}\vartheta \mathcal{C}_n(\R,\R')} \\
         \cdot & f_{\A\B}(\R,\R') \left[ \QmOp{\psi}_\A^\dagger(\R) \QmOp{\psi}_\B^\dagger(\R') + \QmOp{\psi}_\B^\dagger(\R) \QmOp{\psi}_\A^\dagger(\R') \right] \ket{0 \, 0}_\mathrm{A}  w_n^2 \ket{n \, n}_\text{L}
\end{split}
     \label{eq:OutH0Mandel}
\end{equation}
with the phase
{\medmuskip=1mu
\begin{equation}
	\mathcal{C}_n(\R,\R') \! \equiv \! \mathcal{E}_\text{COM}^{(\A)}(\R)+\mathcal{E}_\text{COM}^{(\B)}(\R') +\!\sum_{\alpha=\A,\B} \! \hbar \! \left[ \omega_\alpha - \sum_{j=3}^N \frac{|\Omega_{j\alpha }(\R)|^2}{\omega_{jj\alpha \alpha}^{(+)}} n  +\Omega_{\alpha} \biggl(n+\frac{1}{2}\biggr) \right].
\end{equation}}
As in \Cref{sec:FinalStateGeneral}, $\QmOp{U}_0$ does not modify the initial state, but only adds a phase factor. 

Subsequently, we apply the cosine of the time evolution operator in \Cref{eq:U_int} to the input state in \Cref{eq:InputMandel} and obtain
\begin{equation}
\begin{split}
      \cos{\left(\vartheta \hat{\Omega}\right)} \ket{\psi}_\mathrm{HM} = & \mathscr{N}_2 \int \dd^3 \R \int \dd^3 \R' \, \sum_{n}   \cos{\left(2\vartheta \Omega_n(\R)\right)} \\
      \cdot &f_{\A\B}(\R,\R') \left[ \QmOp{\psi}_\A^\dagger(\R) \QmOp{\psi}_\B^\dagger(\R') + \QmOp{\psi}_\B^\dagger(\R) \QmOp{\psi}_\A^\dagger(\R') \right] \ket{0 \, 0}_\mathrm{A}  w_n^2 \ket{n \, n}_\text{L}
\end{split}
    \label{eq:OutCosineMandel}
\end{equation}
with the amplitude $\Omega_n^2(\R) \equiv  |\Omega(\R)|^2 (n^2+n)$. Thus, this component of the time evolution operator does not change the occupation of the atomic or optical modes. 

The application of the sine of the time evolution operator in \Cref{eq:U_int} to the input state in \Cref{eq:InputMandel} reveals
{\medmuskip=1mu
\begin{equation}
\scalebox{0.96}{$
  \begin{split}
         \sin{\left(\vartheta \hat{\Omega}\right)} \frac{\mathcal{\hat{V}}}{\hbar \hat{\Omega}} \ket{\psi}_\mathrm{HM} =  & \mathscr{N}_2  \int \dd^3 \R \int \dd^3 \R' \,  \sum_{n} 
         \sin{\left(2\vartheta\Omega_{n}(\R) \right)} f_{\A\B}(\R,\R') w_n^2\\ 
         \cdot \Bigl[&  \ee^{\ii \phi_\Omega} \QmOp{\psi}_\B^\dagger(\R) \QmOp{\psi}_\B^\dagger(\R')  \ket{n-1 \, n+1}_\text{L}
        +  \ee^{-\ii \phi_\Omega} \QmOp{\psi}_\A^\dagger(\R) \QmOp{\psi}_\A^\dagger(\R')  \ket{n+1 \, n-1}_\text{L}\Bigr].   
  \end{split}$}
  \label{eq:OutSineMandel}
\end{equation}}
Consequently, this part of the time evolution operator changes the occupation of the atomic modes as well as the occupation of the optical modes. We find the output state in a superposition state in which both atoms can be found in the same internal state. Both atoms can be found in the internal state $\ket{\B}$, if the atom, that initially occupied the atomic mode a, changed its internal state from $\ket{\A}$ to $\ket{\B}$. This process is conducted by the absorption of one photon of the optical mode a and a subsequent emission of a photon into the optical mode b, denoted by the optical output state $\ket{n-1 \, n+1}_\text{L}$ and accompanied by a momentum kick, described by $\ee^{\ii \phi_\Omega}$. In reverse, if the atom initially in state $\ket{\B}$ changes its internal state, both atoms can be found in the internal state $\ket{\A}$. This process is associated with an absorption of one photon from mode b and an emission of one photon into mode a, denoted by the optical output state $\ket{n+1 \, n-1}_\text{L}$. As the factor $\ee^{-\ii \phi_\Omega}$ shows, this process causes a momentum kick into the opposite direction.

We conclude from \Cref{eq:OutPW}, that a 50:50 beam splitter can be produced by choosing $\vartheta \Omega_n = \pi/4$. In this particular case, the values of the sine and cosine functions are identical. Considering \Cref{eq:OutCosineMandel} and \Cref{eq:OutSineMandel}, this condition results in $\cos{\left(2\vartheta \Omega_n(\R)\right)} = 0 $ and $\sin{\left(2\vartheta\Omega_{n}(\R) \right)} = 1$. Hence, the final state for the 50:50 beam splitter is given by
{\medmuskip=1mu
\begin{equation}
   \begin{split}
       \hat{U}\ket{\psi}_\mathrm{HM} =&- \ii \, \mathscr{N}_2 \int \dd^3 \R \int \dd^3 \R' \,  \sum_{n}  \ee^{-\frac{\ii}{\hbar}\vartheta \mathcal{C}_n(\R,\R')} f_{\A\B}(\R,\R') w_{n}^2 \\
        \cdot \Bigl[& \ee^{\ii \phi_\Omega} \QmOp{\psi}_\B^\dagger(\R) \QmOp{\psi}_\B^\dagger(\R')  \ket{n-1 \, n+1}_\text{L}
        +  \ee^{-\ii \phi_\Omega} \QmOp{\psi}_\A^\dagger(\R) \QmOp{\psi}_\A^\dagger(\R') \ket{n+1 \, n-1}_\text{L}\Bigr].  
   \end{split}
    \label{eq:OutMandel}
\end{equation}}
The atomic output states
\begin{equation}
	C_{n}^{(\B)} \ket{0 2}_\text{A} \equiv  \mathscr{N}_2 \int \dd^3 \R \int \dd^3 \R' \,    \ee^{-\frac{\ii}{\hbar}\vartheta \mathcal{C}_n(\R,\R')} f_{\A\B}(\R,\R')
         \ee^{\ii \phi_\Omega} \QmOp{\psi}_\B^\dagger(\R) \QmOp{\psi}_\B^\dagger(\R')  \ket{00}_\text{A}
\end{equation}
and
\begin{equation}
C_{n}^{(\A)}  \ket{2 0}_\text{A} \equiv  \mathscr{N}_2  \int \dd^3 \R \int \dd^3 \R' \,   \ee^{-\frac{\ii}{\hbar}\vartheta \mathcal{C}_n(\R,\R')} f_{\A\B}(\R,\R')  \ee^{-\ii \phi_\Omega} \QmOp{\psi}_\A^\dagger(\R) \QmOp{\psi}_\A^\dagger(\R') \ket{00}_\text{A}
\end{equation}
allow us to cast \Cref{eq:OutMandel} into the simple form given by
\begin{equation}
       \hat{U}\ket{\psi} =- \ii  \sum_{n} w_{n}^2 \left[ C_{n}^{(\B)} \ket{0 2}_\text{A}   \ket{n-1 \, n+1}_\text{L} 
        +  C_{n}^{(\A)}  \ket{2 0}_\text{A} \ket{n+1 \, n-1}_\text{L}\right].  
    \label{eq:OutMandelShort}
\end{equation}
This expression demonstrates, that two atoms initially in a superposition of two distinct internal states can, after a 50:50 beam splitter pulse, only be found in the same internal states. Therefore, Raman diffraction may serve as an additional technique for conducting a Hong-Ou-Mandel experiment utilizing atoms.

An interesting effect occurs when we consider the atoms as a detector for an optical Hong-Ou-Mandel experiment, which corresponds to the limit $M \rightarrow \infty$. The beam splitter in the optical Hong-Ou-Mandel experiment brings two photons, which initially occupy the modes $\ket{11}_\mathrm{L}$, into the superposition $\frac{1}{\sqrt{2}}(\ket{20}_\mathrm{L}+\ket{02}_\mathrm{L}$. The atomic system realizes this mode transformation through the alteration of one atom's internal state, which can be achieved for $\vartheta \Omega_1 \equiv  2 |\Omega(\R)|^2 = \pi/4$ and the initial state $ \ket{\psi}_\mathrm{optHM} = \ket{\psi^{(2)}}_\mathrm{A} \otimes   \ket{1 \, 1}_\text{L}$. The output state for this configuration reads
\begin{equation}
       \hat{U}_\mathrm{lim}\ket{\psi}_\mathrm{optHM} = - \ii  \left[ C_{1}^{(\B)} \ket{0 2}_\text{A}   \ket{0 \, 2}_\text{L} 
        +  C_{1}^{(\A)}  \ket{2 0}_\text{A} \ket{2 \, 0}_\text{L}\right].  
    \label{eq:OutMandelShortOpt}
\end{equation}
with the phase
\begin{equation}
	\mathcal{C}_1(\R,\R') \equiv \hbar  \sum_{\alpha=\A,\B}    \left[ \omega_\alpha - \sum_{j=3}^N\frac{|\Omega_{j\alpha }(\R)|^2}{\omega_{jj\alpha \alpha}^{(+)}} + \frac{3}{2} \Omega_{\alpha}\right].
\end{equation}
Thus, the detection of the photons can be compared to the knowledge about which atom changed its internal state. As demonstrated by the expressions in \Cref{eq:OutMandelShort} and \Cref{eq:OutMandelShortOpt}, it is evident that the operation with two atoms is mandatory.

\section{Conclusions}
In this publication we introduced a method to describe the time-averaged evolution of a general operator under a harmonic Hamiltonian with two different frequency domains. We applied this method to the atomic field operators of a multiple $\Lambda$-system in which the ground and excited states are coupled to each ancilla state via the corresponding optical mode. This system treats the atoms in a more realistic manner, as the lasers typically cannot address a single ancilla state separately, but interact with all ancillas resulting in minor energy shifts.

In contrast to the usual approach, which involves neglecting the counter-rotating terms through the RWA \cite{schleich_quantum_2001} and eliminating the ancilla states adiabatically, the effective Hamiltonian resulting from our method retains both. However, the ground and excited states are decoupled from the ancilla states, thus allowing us to identify a reduced Rabi-block matrix. In this model, we identified the AC-Stark shift and, to the lowest order, the Bloch-Siegert shift \cite{Rossatto_AA}, both of which receive contributions from all ancillary states, as well as the detunings and couplings. Furthermore, the combined quantum nature of the electromagnetic field and atoms gives rise to particle-particle interactions that cause transitions between the relevant states and the ancilla states.

We closed our analysis by applying a generic optical and single-particle atomic state to the time evolution operator of the Rabi-block matrix and the light field Hamiltonian under $\delta-$pulse approximation \cite{Schleich_2013,PhysRevD.100.065021}, neglecting the rapidly oscillating cross-terms. The results revealed the expected Rabi oscillations in a Raman setting \cite{Moler,Leveque,Hartmann_Regimes,Hartmann_Third}, extended by the contributions arising from the counter-rotating terms and the ancilla manifold. Moreover, we identified the beam splitter operator, demonstrating accordance with the semi-classical Raman operator and the beam-splitter operator of the Jaynes-Cummings model \cite{schleich_quantum_2001,Soukup,PhysRevResearch.4.013115} for a low photon number and short interaction times. Subsequently, we considered a two-particle atomic state and focused on the the Hong-Ou-Mandel effect \cite{HongTheory,HongExperiment}, initially observed in a two-photon system. Recent experiments detected this effect using an atom interferometer based on Bragg diffraction \cite{Lopes,Dussarrat}. We demonstrated, that for a specific input state, similar to Bragg diffraction \cite{PhysRevA.88.053608,PhysRevLett.116.173601,PhysRevA.94.063619,Hartmann_Regimes,Jenewein}, Raman diffraction can also be utilized to observe the Hong-Ou-Mandel effect in the atomic and additionally in the optical system. 

The model in this publication is intentionally kept as general as possible. In order to apply it to a specific cavity, it is first necessary to carry out an explicit evaluation of the modes and couplings. However, this is beyond the scope of this study. The findings of this study provide a foundation for theoretical research in the field of atom interferometry, which involves the application of entanglement between atoms to enhance the sensitivity \cite{greve2022entanglement,doi:10.1126/science.adi1393,PhysRevX.15.011029,malia2022distributed}. Moreover, our formalism permits the extension of Asano et al.'s \cite{PhysRevResearch.4.013115} entanglement studies to completely second-quantized systems.

\backmatter

\bmhead{Supplementary information}
Not applicable.

\bmhead{Acknowledgements}

We thank R. Lopp, E. Giese, R. Walser and the whole QUANTUS group in Ulm as well as our partners of the QUANTUS collaboration for fruitful discussions. We would like to express our gratitude to W.~P. Schleich for his advice and support, which significantly improved this article.

\section*{Declarations}

\subsection*{Data availability}
No datasets were generated or analyzed during the current study.

\subsection*{Competing interests}
The authors declare that they have no competing interests.

\subsection*{Funding}
This work is supported by the German Aerospace Center (Deutsches Zentrum f\"ur Luft- und Raumfahrt, DLR) with funds provided by the Federal Ministry for Economic Affairs and Climate Action (Bundesministerium für Wirtschaft und Klimaschutz, BMWK) due to an enactment of the German Bundestag under Grant No. DLR 50WM2450D-2450E (QUANTUS~VI).

\subsection*{Author contributions}
S.H. did the formal analysis, found the majority of the references and wrote the original draft of the manuscript. A.F. designed the concept of the article, drew all figures, found several references and supervised the project. All authors contributed to the interpretation of the results, reviewed and edited the original draft. 

\subsection*{Ethics approval and consent to participate}
Not applicable.

\subsection{Consent for publication}
All authors read and approved the final manuscript.

\subsection*{Materials availability}
Not applicable.

\subsection*{Code availability}
Not applicable.

\noindent




\begin{appendices}

\section{The $\Lambda$-like system in the interaction picture}
\label{app:LambdaSystemIP}
We transform in this appendix the Hamiltonian of the $\Lambda$-like system, depicted in \Cref{fig:Atom}~b), into an interaction picture with respect to the Hamiltonian of the electromagnetic field $\hat{H}_\textrm{L}$ and the internal atomic Hamiltonian $\hat{H}_\textrm{int}$, resulting in the interaction picture Hamiltonian $\hat{H}^\textrm{I}\equiv \hat{H}^\textrm{I}_\textrm{A} + \hat{H}^\textrm{I}_\textrm{AL}$.
 
The atomic Hamiltonian in the interaction picture is represented by two parts. The first part corresponds to the external atomic Hamiltonian
\begin{equation}
    \hat{H}^\textrm{I}_\textrm{A}= \int \dd^3 \R \sum_{\mathscr{l} \in  \mathcal{S}}  \QmOp{\psi}_{\mathscr{l}}^\dagger(\R) \mathcal{H}_\text{COM}^{(\mathscr{l})}  \, \QmOp{\psi}_{\mathscr{l}}^{\phantom{\dagger}}(\R),
    \label{eq:H_A_I}
\end{equation}
and is assumed to be time-independent. Furthermore, we introduce the index \mbox{$\mathscr{l} \in \mathcal{S}\equiv\qty{\A,\B,3,\hdots,N}$} which encompasses all the atomic states, that is the ground state $\ket{\A}$, the excited state $\ket{\B}$ and the ancilla states $\ket{j}$. Similarly, we define the index set $\mathcal{R}\equiv\qty{\A,\B}$ containing only the (later) relevant states, that is ground and excited state. Lastly, we also introduce an index set $\mathcal{A}=\qty{3,\hdots,N}$ for the ancillary states. 

With these definitions settled, the interaction Hamiltonian \Cref{eq:InteractionHamiltonianSP} in the interaction picture takes the form
\begin{equation}
\begin{split}
    \QmOp{H}^\textrm{I}_\textrm{AL}=   \hbar \int \dd^3 \R \sum_{\alpha \in\SetRelevant} \sum_{j\in\SetAncilla} \Bigl[ &\ee^{\ii (\omega_{j\alpha}  - \Omega_{\alpha})t}  \QmOp{\psi}_j^\dagger(\R)  \Omega_{j\alpha}(\R)\hat{\alpha} \QmOp{\psi}_{\alpha}(\R) \\
         + &\ee^{-\ii (\omega_{j\alpha}+\Omega_{\alpha})t}\QmOp{\psi}_{\alpha}^\dagger(\R)  \Lambda^*_{j\alpha}(\R)\hat{\alpha} \QmOp{\psi}_{j}(\R) - \text{\text{h.c.}}  \Bigr],
\end{split}
\end{equation}
with the frequency difference $\omega_{j\alpha}=\omega_j- \omega_\alpha$ between the ancilla state $\ket{j}$ and ground or excited state $\ket{\alpha}$.

In the case of weak coupling and sufficiently small detuning, one typically proceeds with the application of the the rotating-wave approximation (RWA)~\cite{schleich_quantum_2001} in order to eliminate the fast-rotating couplings $\Lambda_{j\alpha}$ and $\Lambda^*_{j\alpha}$ with frequency differences $\omega_{j\alpha}+\Omega_\alpha$ while keeping the slow-rotating terms, coupling via $\Omega_{j\alpha}$ and $\Omega_{j\alpha}^*$ with frequency differences $\omega_{j\alpha}-\Omega_\alpha$, in the order of the two-photon detuning. Subsequently, an adiabatic elimination of the ancillary state manifold \cite{bernhardt_coherent_1981,marte_multiphoton_1992,brion_adiabatic_2007,Fewell} is typically performed to further reduce the complexity of the system.

In contrast to this approach, we will present a method that averages the atomic field operators over the interaction duration to obtain an effective Hamiltonian. We will see that this approach is in principle closely related to the RWA. In contrast to the RWA, effects from the counter-rotating couplings can be taken into account. Similar to the adiabatic elimination, the ancillary states decouple to lowest order from the dynamics of the ground and excited states. However, in contrast to adiabatic elimination, they do not disappear completely. On the contrary, they--and their dynamics--remain accessible and lead to particle-particle interactions between the ancilla state manifold and the ground and excited state.

\section{Derivation of the effective Hamiltonian in Schrödinger picture}
\label{app:DerivationHeff}
Our objective is to derive an expression for the time evolution of the atomic field operator. For this purpose, we develop in Appendix~\ref{App:Averaging} a theory based on the series expansion of the time evolution operator. Through time averaging we find an expression for the time evolution of an arbitrary operator, which is governed by an effective Hamiltonian. 

In Appendix~\ref{App:TimeAveragingAppliedToHarmonic} we determine an expression for the effective Hamiltonian for a class of so-called harmonic Hamiltonians, which are composed of a time-independent contribution and a part with time-dependent exponentials. 

Subsequently, this result is applied in Appendix~\ref{app:AveragingHamiltonian} to our specific multi-$\lambda$-model, leading to the effective Hamiltonian that governs the time evolution of the atomic field operator.

In Appendix~\ref{App:ApproximatedEvolution} we indicate that, in addition to time-averaging, the amplitude is a critical factor in determining which contributions should be retained and which can be disregarded. By inserting typical frequencies used in atom optics for transitions in alkali atoms, we estimate the size of the involved detunings and point out that the retained counter-rotating contributions are significantly smaller than the co-rotating contributions. Moreover, they are smaller than some of the co-rotating contributions that we neglected due to their fast oscillations.

The effective Hamiltonian that describes our specific multi-$\lambda$-model is determined by two commutators. Each commutator comprises atomic and light field operators. Their detailed evaluation is given in Appendix~\ref{app:evalcommutators}.

We proceed in Appendix~\ref{app:Trafo} by transforming the effective Hamiltonian back into  the Schrödinger picture. In Appendix~\ref{App:NewVariables} we separate the contributions of the effective Hamiltonian into interaction terms and self-couplings. An additional contribution contains two pairs of atomic field operators with different spatial dependencies and thus can be interpreted as particle-particle interactions.

We decompose these particle-particle interactions in Appendix~\ref{app:PPdecomposition} according to their spatial dependencies. This separation will be useful for the later comparison with the semi-classical model.

Finally, we present in the effective Hamiltonian, after several rearrangements, in the Schrödinger picture in Appendix~\ref{app:HeffSeparatePP}. With this version, we continue our analysis in \Cref{sec:DeterminationHeff}.

\subsection{Average Hamiltonian Theory for Operators}
\label{App:Averaging}
Following the derivation in~\cite{James_Avaraging,Gamel}, we first define the time average of an arbitrary operator $\hat{\mathcal{O}}(t)$  
\begin{equation}
    \overline{\hat{\mathcal{O}}}(t) \equiv \int_{-\infty}^\infty \dd t' f(t-t')  \hat{\mathcal{O}}(t'), 
    \label{eq:TimeaveragedOperator}
\end{equation}
which we denote as the convolution of the operator $\hat{\mathcal{O}}(t)$ with a real-valued function $f(t)$ that has unit area. Here, $f(t)$ smooths the operator such that high frequency-contributions are left out. \\
Moreover, the time-ordered time evolution operator $\hat{U}(t,t_0)$ fulfills the Schrödinger equation
\begin{equation}
    \ii \hbar \frac{\partial}{\partial t} \hat{U}(t,t_0) = \hat{H}(t) \hat{U}(t,t_0)
    \label{eq:SchrödingerEq}
\end{equation}
where the dynamics are governed by the Hamiltonian $\hat{H}(t)$ of the considered system. \\
To proceed, we replace $\hat{H}$ by $\lambda\hat{H}$ with the dimensionless parameter $\lambda\in[0,1]$ and expand the time evolution into a series, given by
\begin{equation}
    \hat{U}(t,t_0)=\sum_{n=0}^\infty \lambda^n \hat{U}_n(t,t_0).
    \label{eq:SeriesExpansionU}
\end{equation}
We insert the series expansion in \Cref{eq:SeriesExpansionU} into the Schrödinger equation, \Cref{eq:SchrödingerEq}, and compare coefficients in powers of $\lambda$, which results in the recursion relation
\begin{equation}
    \ii \hbar \frac{\partial}{\partial t} \hat{U}_n(t,t_0) = \hat{H}(t) \hat{U}_{n-1}(t,t_0)
    \label{eq:RecursionRelation}
\end{equation}
for $\frac{\partial}{\partial t}\hat{U}_0(t,t_0)=0$ and $n\geq1$. 

For $\lambda \rightarrow 0$ we find $\hat{U}(t,t_0)= \hat{I}$ and consequently \mbox{$\hat{U}_0(t,t_0)= \hat{I}$}. While Gamel and James \cite{Gamel} proceeded their analysis with the density matrix and its evolution, we utilize the Heisenberg equation for operators, as given by
\begin{equation}
    \hat{\mathcal{O}}(t)= \hat{U}^\dagger(t,t_0)\hat{\mathcal{O}}(t_0)\hat{U}(t,t_0).
    \label{eq:Heisenberg}
\end{equation}   

We plug \Cref{eq:Heisenberg} into \Cref{eq:TimeaveragedOperator} resulting in the time-averaged operator 
\begin{equation}
    \begin{aligned}
    \label{eq:DefinitionMap}
        \overline{\hat{\mathcal{O}}}(t) &= \overline{\hat{U}^\dagger(t,t_0) \hat{\mathcal{O}}(t_0) \hat{U}(t,t_0)}  =\sum_{m,n=0}^\infty \lambda^{m+n} \overline{\hat{U}^\dagger_m(t,t_0) \hat{\mathcal{O}}(t_0) \hat{U}_n(t,t_0)} \\
       &= \sum_{k=0}^\infty \lambda^{k} \Bigl[ \sum_{j=0}^k  \overline{\hat{U}^\dagger_{k-j}(t,t_0) \hat{\mathcal{O}}(t_0) \hat{U}_j(t,t_0)} \Bigr] \equiv    \sum_{k}^\infty \lambda^k \mathcal{E}_k[\hat{\mathcal{O}}(t_0)] \equiv \mathcal{E}[\hat{\mathcal{O}}(t_0)].
\end{aligned}
\end{equation}
In the final line, we have introduced a set of linear maps, designated as $\mathcal{E}_k$, which act on the operator $\hat{\mathcal{O}}(t_0)$ at the initial time $t_0$. The time-averaged operator $\overline{\hat{\mathcal{O}}}(t)$ is then obtained by summing over all contributions. It should be noted that, in accordance with the first definition provided on the second line of \Cref{eq:DefinitionMap}, we have that $\mathcal{E}_0=\mathcal{I}$. 

Following \cite{Gamel}, we invert \Cref{eq:DefinitionMap} and define the inverted map $ \mathcal{F}$, which we subsequently write as a series with parameter $\lambda$ resulting in
\begin{equation}
    \hat{\mathcal{O}}(t_0) = \mathcal{E}^{-1}\bigl[\overline{\hat{\mathcal{O}}}(t)\bigr] \equiv \mathcal{F}\bigl[\overline{\hat{\mathcal{O}}}(t)\bigr] \equiv \sum_{k=0}^\infty \lambda^k \mathcal{F}_k\bigl[\overline{\hat{\mathcal{O}}}(t)\bigr].
    \label{eq:InitialOperator}
\end{equation}

Now we use the fact that successive application of the map and its inverse gives the identity, i.~e. $\mathcal{F}[\mathcal{E}[\bullet]]=\mathcal{I}[\bullet]$, which leads to
\begin{equation}
    \sum_{m,n=0}^\infty \lambda^{m+n} \mathcal{F}_m[\mathcal{E}_n[\bullet]] =  \sum_{k=0}^\infty \lambda^{k} \Bigl( \sum_{j=0}^k \mathcal{F}_j[\mathcal{E}_{k-j}[\bullet]] \Bigr)=\lambda^0 \mathcal{I}[\bullet],
    \label{eq:UnityMap1}
\end{equation}
after rearranging the summations.

In the previous expression, we compare coefficients in powers of $\lambda$ up to second order, thereby arriving at
\begin{equation}
    \begin{split}
        \mathcal{F}_0=\mathcal{E}_0=\mathcal{I}, \text{\, \,}
        \mathcal{F}_1 = - \mathcal{E}_1, \text{\, \,}
        \mathcal{F}_2 = \mathcal{E}_1[\mathcal{E}_1] - \mathcal{E}_2.
    \end{split}
    \label{eq:Maps}
\end{equation}

In order to identify the effective evolution of $\overline{\hat{\mathcal{O}}}$ we differentiate \Cref{eq:DefinitionMap} and use \Cref{eq:InitialOperator}. This yields
\begin{align}
        \ii \hbar \frac{\partial}{\partial t} \overline{\hat{\mathcal{O}}}(t) = &\ii \hbar \mathcal{\dot{E}}[\hat{\mathcal{O}}(t_0)]  = \ii \hbar \mathcal{\dot{E}}[\mathcal{F}[\overline{\hat{\mathcal{O}}}(t))]= \sum_{k=0}^\infty \lambda^k \Bigl( \sum_{j=0}^k \ii \hbar \mathcal{\dot{E}}_j [\mathcal{F}_{k-j}[\overline{\hat{\mathcal{O}}}(t)]] \Bigr) \\
        \equiv &\sum_{k=0}^\infty \lambda^k \mathcal{L}_k[\overline{\hat{\mathcal{O}}}(t)]
\label{eq:TimeaveragedO}
\end{align}
where we introduced $\mathcal{L}_k$.

To proceed, we calculate $\mathcal{L}_k$ up to second order. From now on, our results deviate again from \cite{Gamel}, due to \Cref{eq:Heisenberg}. For improved readability, we will henceforth omit the time argument in both $\hat{H}(t)\equiv \hat{H}$ and $\hat{U}_n(t,t_0) \equiv \hat{U}_n$. This results in the following expressions:
\begin{equation}
    \mathcal{L}_0[\bullet] = \ii \hbar \mathcal{\dot{E}}_0[\mathcal{F}_0[\bullet]] = 0,
    \label{eq:L0}
\end{equation}
\begin{equation}
    \mathcal{L}_1[\bullet] = \ii \hbar \mathcal{\dot{E}}_1[\bullet]=-\overline{\hat{H}} \bullet + \bullet \overline{\hat{H}},
    \label{eq:L1}
\end{equation}
and for the second order
{\medmuskip=3mu
\begin{equation}
\begin{split}
        \mathcal{L}_2[\bullet] = & \ii \hbar \mathcal{\dot{E}}_2[\bullet] + \ii \hbar \mathcal{\dot{E}}_1[\mathcal{F}_1[\bullet]] \\ = &\overline{\hat{H}} \, \overline{\hat{U}_1^\dagger}\bullet +\overline{\hat{H}} \bullet \overline{\hat{U}}_1 
    - \overline{\hat{U}}_1^\dagger \bullet \overline{\hat{H}}
    -\bullet \overline{\hat{U}_1} \, \overline{ \hat{H}}  -\overline{\hat{U}_1^\dagger \hat{H} } \bullet
    -\overline{\hat{H} \bullet \hat{U}_1}
    + \overline{\hat{U}_1^\dagger \bullet \hat{H}} + \bullet \overline{\hat{H}\hat{U}_1}.
\end{split}
    \label{eq:L2}
\end{equation}}
By employing \Cref{eq:L0,eq:L1,eq:L2} we are able to calculate the effective evolution of the time averaged operator $\overline{\hat{\mathcal{O}}}$ in \Cref{eq:TimeaveragedO} up to second order, which is denoted by
\begin{equation}
    \begin{aligned}
    i\hbar\frac{\partial\overline{\hat{\mathcal{O}}}}{\partial t}=&-\left[\hat{H},\hat{\overline{\mathcal{O}}}\right]+\left(\hat{A}\overline{\hat{\mathcal{O}}}- \overline{\hat{O}} \hat{A}^\dagger \right)+\mathcal{\mathcal{D}}_2\left[\overline{\hat{\mathcal{O}}}\right]  \\
    =&-\left[\hat{H}_{\text{eff}},\hat{\overline{\mathcal{O}}}\right]+\left\{\frac{1}{2}(\hat{A}-\hat{A}^\dagger),\overline{\hat{\mathcal{O}}}\right\}+\mathcal{D}_2\left[\overline{\hat{\mathcal{O}}}\right],
    \label{eq:TimeEvolutionO}
\end{aligned}
\end{equation}
where we defined the effective Hamiltonian as
\begin{equation}
    \hat{H}_{\text{eff}}\equiv\overline{\hat{H}}-\frac{1}{2}\left(\hat{A}+\hat{A}^\dagger\right)
    \label{eq:Heff}
\end{equation}
with the operators
\begin{equation}
    \hat{A}\equiv \overline{\hat{U}_1\hat{H}}- \overline{\hat{H}} \, \overline{\hat{U}}_1, \text{\, \,}  \hat{A}^\dagger\equiv \overline{\hat{U}_1} \, \overline{\hat{H}}- \overline{\hat{H}\hat{U}_1}
    \label{eq:A}
\end{equation}

and
\begin{equation}
    \mathcal{D}_2\left[\overline{\hat{\mathcal{O}}}\right]\equiv \overline{\hat{H}} \, \overline{\hat{\mathcal{O}}} \, \overline{\hat{U}_1}   -\overline{\hat{H}\hat{\mathcal{O}}\hat{U}_1}+ \overline{\hat{U}_1^\dagger \hat{\mathcal{O}} \hat{H}} - \overline{\hat{U}_1^\dagger} \,  \overline{\hat{\mathcal{O}}} \, \overline{\hat{H}}.
    \label{eq:D2}
\end{equation}
In \Cref{eq:D2} we used the fact that $\overline{\hat{U}}_1^\dagger$ is unitary as a consequence of the Schrödinger equation of motion. Upon comparison of our results with those presented in Ref.~\cite{James_Avaraging,Gamel} we find complete agreement with the findings presented in \Cref{eq:InitialOperator,eq:UnityMap1,eq:Maps,eq:TimeaveragedO,eq:L0}. As a consequence of utilizing \Cref{eq:Heisenberg} instead of the time evolution for density matrices, $\hat{U}$ and $\hat{U}^\dagger$ are exchanged, leading to the consequential alterations in \Cref{eq:DefinitionMap,eq:L1,eq:L2,eq:TimeEvolutionO,eq:Heff,eq:A,eq:D2}.

\subsection{Time averaging applied to harmonic Hamiltonians}
\label{App:TimeAveragingAppliedToHarmonic}
We apply the method developed so far to the class of so-called harmonic Hamiltonians, which are characterized by a time-independent component $\hat{H}_0 $ and time-dependent terms oscillating with distinct frequencies $\omega_n$ and possibly $\Omega_n$ and their hermitian conjugates. This leads to Hamiltonians of the form
\begin{equation}
	\hat{H}(t)= \hat{H}_0 + \sum_{n=1}^N \bigl\{ \hat{h}_n e^{-\ii \omega_n t}  +\hat{g}_n e^{-\ii \Omega_n t} \pm \text{h.c.} \bigr\}.
    \label{eq:HarmonicHamiltonian}
\end{equation} 
This is an extension of the model presented in Refs.~\cite{James_Avaraging,Gamel} due to the introduction of a second group of frequencies. In order to determine $\hat{U}_1$ we integrate \Cref{eq:RecursionRelation} and obtain
\begin{equation}
    \hat{U}_1(t,t_0)= \frac{t-t_0}{\ii \hbar} \hat{H}_0 + \hat{V}_1(t) + \hat{V}'_1(t) - \hat{V}_1(t_0) - \hat{V}'_1(t_0)
\end{equation}
with
\begin{equation}
    \hat{V}_1(t)= \sum_{n=0}^N \frac{1}{\hbar \omega_n} \Bigl[ \hat{h}_n \ee^{-\ii \omega_n t} \mp \hat{h}_n^\dagger \ee^{\ii \omega_n t} \Bigr] \text{\quad and \quad}  \hat{V}'_1(t)= \sum_{n=0}^N \frac{1}{\hbar \Omega_n} \Bigl[ \hat{g}_n \ee^{-\ii \Omega_n t} \mp \hat{g}_n^\dagger \ee^{\ii \Omega_n t} \Bigr].
\end{equation}

This enables us to evaluate the components of the time evolution of the time-averaged operator in \Cref{eq:TimeEvolutionO}. We assume that the function $f(t)$ represents an ideal low-pass filter, which means that the frequencies $\omega_{n/m}$, $\Omega_{n/m}$ and summations thereof are filtered out as well as any combinations of $\omega_{n/m}$ and $\Omega_{n/m}$. In contrast to that, the frequencies are sufficiently close to each other such that frequency differences within same frequency domain remain, i.e., $\text{max}{\{|\omega_m-\omega_n|}\}\ll\text{min}\{\omega_m,\omega_n\}$ and $\text{max}{\{|\Omega_m-\Omega_n|}\}\ll\text{min}\{\Omega_m,\Omega_n\}$. This consideration leads to the following approximations:
\begin{align}
	\overline{\exp{ \pm \ii \omega_n t}}&\approx \overline{\exp{\pm \ii \Omega_n t}}\approx 0  \label{eq:Averaging1} \\
 \overline{\exp{\pm \ii (\omega_n+\omega_m) t}}&\approx\overline{\exp{\pm \ii (\Omega_n+\Omega_m) t}}\approx0  \label{eq:Averaging2} \\
 \overline{\exp{\pm \ii (\omega_n \pm \Omega_m) t}}&\approx \overline{\exp{\pm \ii (\omega_m \pm \Omega_n) t}}\approx0  \label{eq:Averaging3} \\
	\overline{\exp{\pm \ii (\omega_n-\omega_m) t}} &\approx\exp{\pm \ii (\omega_n-\omega_m) t}  \label{eq:Averaging4} \\
	\overline{\exp{\pm \ii (\Omega_n-\Omega_m) t}} &\approx\exp{\pm \ii (\Omega_n-\Omega_m) t}  \label{eq:Averaging5}
\end{align}
Hence, the terms in \Cref{eq:TimeEvolutionO} are given by $\hat{H}_1 \equiv  -\left[\hat{H}_{\text{eff}},\hat{\overline{O}}\right]+\left\{\frac{\hat{A}-\hat{A}^\dagger}{2},\overline{\hat{\mathcal{O}}}\right\} $
{\medmuskip=1.5mu
\begin{equation}
	\begin{split}
\hat{H}_1  \!=\! -[\hat{H}_0, \overline{\hat{\mathcal{O}}}] &\pm \sum_{n,m}^N \Biggl[ \Biggl( \frac{\hat{h}_n \hat{h}_m^\dagger}{\hbar \omega_n} - \frac{\hat{h}_m^\dagger \hat{h}_n}{\hbar \omega_m} \Biggr) \overline{\hat{\mathcal{O}}} - \overline{\hat{\mathcal{O}}}(t)  \Biggl( \frac{\hat{h}_n \hat{h}_m^\dagger}{\hbar \omega_m} - \frac{\hat{h}_m^\dagger \hat{h}_n}{\hbar \omega_n} \Biggr)\Biggr] \ee^{\ii(\omega_m-\omega_n)t} \\
&\pm \sum_{n,m}^N \Biggl[ \Biggl(\frac{\hat{g}_n \hat{g}_m^\dagger}{\hbar \Omega_n} - \frac{\hat{g}_m^\dagger \hat{g}_n}{\hbar \Omega_m} \Biggr) \overline{\hat{\mathcal{O}}} - \overline{\hat{\mathcal{O}}}  \Biggl( \frac{\hat{g}_n \hat{g}_m^\dagger}{\hbar \Omega_n} - \frac{\hat{g}_m^\dagger \hat{g}_n}{\hbar \Omega_n} \Biggr)\Biggr] \ee^{\ii(\Omega_m-\Omega_n)t} \\
\end{split}
\label{eq:Anti_Commutators}
\end{equation}}
and
\begin{equation}
\begin{split}
    	\mathcal{D}_2\left[\overline{\hat{\mathcal{O}}}\right] = \pm \sum_{n,m=1}^N \Biggl\{ &\bigg( \frac{\hat{h}_n \hat{\mathcal{O}} \hat{h}_m^\dagger + \hat{h}_m^\dagger \hat{\mathcal{O}} \hat{h}_n}{\hbar \omega_m} - \frac{\hat{h}_m^\dagger \hat{\mathcal{O}} \hat{h}_n + \hat{h}_n \hat{\mathcal{O}} \hat{h}_m^\dagger}{\hbar \omega_n} \bigg) \ee^{\ii(\omega_m-\omega_n)t} \\
	+&\bigg( \frac{\hat{g}_n \hat{\mathcal{O}} \hat{g}_m^\dagger + \hat{g}_m^\dagger \hat{\mathcal{O}} \hat{g}_n}{\hbar \Omega_m} - \frac{\hat{g}_m^\dagger \hat{\mathcal{O}} \hat{g}_n + \hat{g}_n \hat{\mathcal{O}} \hat{g}_m^\dagger}{\hbar \Omega_n} \bigg) \ee^{\ii(\Omega_m-\Omega_n)t} 
	\Biggr\}
\end{split}
 \label{eq:D2_fin}
\end{equation}   
where we have introduced
\begin{equation}
	\frac{1}{\omega_{nm}^{(\pm)}}= \frac{1}{2} \Bigg( \frac{1}{\omega_n} \pm \frac{1}{\omega_m}\Bigg) \text{\quad and \quad} \frac{1}{\Omega_{nm}^{(\pm)}}= \frac{1}{2} \Bigg( \frac{1}{\Omega_n} \pm \frac{1}{\Omega_m}\Bigg).
 \label{eq:Det1_2}
\end{equation}
Our requirements on the frequency domain directly lead to $\bigl(\omega_{nm}^{(-)}\bigr)^{-1}\ll \bigl(\omega_{nm}^{(+)}\bigr)^{-1}$ and equivalently $\bigl(\Omega_{nm}^{(-)}\bigr)^{-1} \ll\bigl(\Omega_{nm}^{(+)}\bigr)^{-1}$. We finally rewrite the evolution of the time averaged operator \Cref{eq:TimeaveragedOperator} with the help of \Cref{eq:D2_fin,eq:Det1_2} as
{\medmuskip=1.5mu
   \begin{equation}
	\begin{split}
	i\hbar\frac{\partial\overline{\hat{\mathcal{O}}}}{\partial t}\! =\!-\left[\hat{H}_{\text{eff}},\overline{\hat{\mathcal{O}}}\right] &\pm \sum_{n,m=1}^N \Bigg( \frac{\{ \{\hat{h}_m^\dagger, \hat{h}_n\},\overline{\hat{\mathcal{O}}} \}}{\hbar \omega_{nm}^{(-)}} -2 \frac{\hat{h}_m^\dagger \overline{\hat{\mathcal{O}}} \hat{h}_n + \hat{h}_n \overline{\hat{\mathcal{O}}} \hat{h}_m^\dagger}{\hbar \omega_{nm}^{(-)}}  \Bigg) \ee^{\ii (\omega_m - \omega_n)t} \\
& \pm \sum_{n,m=1}^N \Biggl( \frac{\{ \{\hat{g}_m^\dagger, \hat{g}_n\},\overline{\hat{\mathcal{O}}} \}}{\hbar \Omega_{nm}^{(-)}} -2 \frac{\hat{g}_m^\dagger \overline{\hat{\mathcal{O}}} \hat{g}_n + \hat{g}_n \overline{\hat{\mathcal{O}}} \hat{g}_m^\dagger}{\hbar \Omega_{nm}^{(-)}} \Bigg)\ee^{\ii (\Omega_m - \Omega_n)t}   
	\end{split}
 \label{eq:Evolution_fin}
\end{equation} }
with the effective Hamiltonian
\begin{equation}
    \hat{H}_{\text{eff}}= \hat{H}_0 \pm \sum_{n,m=1}^N \biggl\{ \frac{[\hat{h}_m^\dagger,\hat{h}_n]}{\hbar \omega_{nm}^{(+)}} \ee^{\ii (\omega_m-\omega_n)t} + \frac{[\hat{g}_m^\dagger,\hat{g}_n]}{\hbar \Omega_{nm}^{(+)}} \ee^{\ii (\Omega_m-\Omega_n)t} \biggr\}.
    \label{eq:Heff_gen}
\end{equation}
A comparison of our result with that presented in Ref.~\cite{Gamel} reveals, that our expression for the time evolution, as given in \Cref{eq:Evolution_fin}, exhibits an opposite sign for the commutator between $\hat{H}_{\text{eff}}$ and $\overline{\hat{\mathcal{O}}}$. The effective Hamiltonian, aside from the aforementioned discrepancy in the sign of the brace, exhibits the same form as that presented in Ref.~\cite{Gamel}. However, the operators $\hat{A}$ and $\hat{A}^\dagger$ are inverted. In the case where only a single frequency is involved, the system is solely described to second order by the effective Hamiltonian due to the fact that $\omega_{nn}^{(-)}=\Omega_{nn}^{(-)}=0$.

\subsection{Effective few level models via Averaging/low-pass filtering}
\label{sec:AveragingRaman}
In this appendix, we apply in Appendix~\ref{app:AveragingHamiltonian} the method, developed in Appendix~\ref{App:TimeAveragingAppliedToHarmonic} for harmonic Hamiltonians, to our specific $\Lambda$-like model. 

Due to the fact, that this model involves two extremely different frequency domains, we find two competing effects, which we discuss in Appendix~\ref{App:ApproximatedEvolution}. Fast oscillations, which are considered to cancel out, interplay with large detunings, which lead to small contributions. We discuss the involved detunings for typical frequencies of alkali atoms, used in atom optics. We finally address this issue, by involving solely the leading order of each frequency domain, given by the effective Hamiltonian.

\subsubsection{Effective Hamiltonian for Raman scheme}
\label{app:AveragingHamiltonian}
\begin{figure}[tb]
    \centering
    \includegraphics[width=0.3\textwidth]{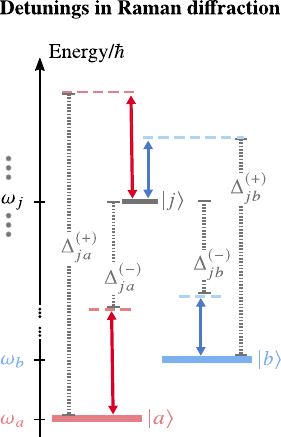}
    \caption{Schematic illustration of the frequency single-photon detunings for the laser couplings indicated in red and blue in the $\Lambda$-scheme consisting of $\ket{\A},\ket{\B}$ and $\ket{j}$. We assume an analogous situation for $\ket{j}$ with $j=3,\hdots,N$. As before, the magnitude of the frequencies and detunings in the illustration are not to scale.}
    \label{fig:frequency-detunings}
\end{figure}

In order to employ our method of time-averaging, it is necessary to separate the Hamiltonian $\hat{H}^\textrm{I}$ into two distinct parts: the term constant in time, represented by $\hat{H}^\textrm{I}_\textrm{A}\neq \hat{H}^\textrm{I}_\textrm{A}(t)$, and the time-dependent component $\hat{H}^\textrm{I}_\textrm{AL}=\hat{H}^\textrm{I}_\textrm{AL}(t)$. The latter one consists of the co-rotating part oscillating with single photon detuning $\Delta_{j\alpha}^{(-)}$ and the counter-rotating part oscillating with single photon detuning $\Delta_{j\alpha}^{(+)}$, whereas
\begin{equation}
    \Delta_{j\alpha}^{(\pm)}\equiv \Omega_\alpha \pm \omega_{j\alpha}.
    \label{eq:detuning}
\end{equation}
In the present study, we focus on a model applicable to e.g., atomic Raman diffraction~\cite{Moler,Hartmann_Regimes,Hartmann_Third} with commonly used atoms such as Sodium~\cite{PhysRevLett.67.181}, Cesium~\cite{Leveque} or Rubidium~\cite{Malossi,Perrin}. In all these cases, the typical frequency of the ancilla state (manifold), $\omega_j$, is much larger than that of the ground or excited state, $\omega_\alpha$, or more precisely we often have for the frequency separation $\omega_{j\alpha} \propto 10^2$~THz which is an optical frequency. The frequency of the light field $\Omega_\alpha$ is also of the same order with $10^2$~THz but $\Omega_\alpha<\omega_j$. Therefore, it can be concluded that $\omega_j$ dominates the expression in \Cref{eq:detuning}, and consequently $\Delta_{j\alpha}^{(+)}>0$ and $\Delta_{j\alpha}^{(-)}<0$. The magnitude of all involved frequency scales is tabulated in~\Cref{fig:frequency-detunings} together with a schematic of the single-photon detunings.

Next we define the coupling Hamiltonians
\begin{equation}
    \hat{h}_{j\alpha}\!\equiv \!\hbar \!\! \int \!\! \dd^3 \R \, \QmOp{\psi}_j^\dagger(\R) \, \Omega_{j \alpha}(\R) \, \hat{\alpha} \, \QmOp{\psi}_{\alpha}(\R)
    \text{~and~}  
    \hat{g}_{j\alpha}\! \equiv \!\hbar \!\! \int \!\! \dd^3 \R \, \QmOp{\psi}_{\alpha}^\dagger(\R)  \Lambda^*_{j\alpha}(\R)\hat{\alpha} \QmOp{\psi}_{j}(\R),
    \label{eq14}
\end{equation}
which enable us to cast the Hamiltonian into the concise form $\hat{H}^\textrm{I} = \hat{H}^\textrm{I}_\textrm{A} +\hat{H}^\textrm{I}_\textrm{AL}(t)$ with a time-independent part $\QmOp{H}_\text{A}^\text{I}$, \Cref{eq:H_A_I}, and an oscillatory time-dependent part $\QmOp{H}_\text{AL}^\text{I}(t)$. The oscillatory contribution is given by
\begin{equation}
    \hat{H}^\textrm{I}_\textrm{AL}(t) = \sum_{\alpha\in\SetRelevant} \sum_{j\in\mathcal{A}} \Bigl[ \hat{h}_{j\alpha} \ee^{-\ii \Delta_{j\alpha}^{(-)} t} + \hat{g}_{j\alpha}  \ee^{-\ii \Delta_{j\alpha}^{(+)}t} - \text{h.c.} \Bigl]
    \label{eq:interaction-picture-hamiltonian-generator-decomposed}.
\end{equation}

Our next goal is to extract by time-averaging an effective Hamiltonian valid at low frequencies compared to optical oscillation frequencies. This Hamiltonian should still incorporate the corrections resulting from fast oscillations.

When we apply this method of Hamiltonian time-averaging, developed in Appendices~\ref{App:Averaging} and \ref{App:TimeAveragingAppliedToHarmonic}, to the interaction picture Hamiltonian $\hat{H}^\textrm{I}$ and consider typical parameters for a $\Lambda$-system for atomic Raman diffraction with hyperfine states of ${}^{87}\text{Rb}$ on the $D_2$-line, we obtain 
\begin{equation}
    \ii \hbar \frac{\partial}{\partial t}\QmOp{\Psi}_\mathscr{l}(\R)  = -\left[\hat{H}_{\text{eff}}^\textrm{I},\QmOp{\Psi}_\mathscr{l}(\R)   \right]
    \label{eq:TimeAveragedPsi}
\end{equation}
for the dynamical equation of the time-averaged field operator ${\QmOp{\Psi}_\mathscr{l}(\R)  = \overline{\QmOp{\psi}_\mathscr{l}}(\R)}\approx\QmOp{\psi}_\mathscr{l}$ in the interaction picture\footnote{In the case of atomic Raman diffraction, the expressions in~\Cref{eq:Averaging1,eq:Averaging2,eq:Averaging3} are typically in the order of THz and thus average to zero very fast in the spirit of the Riemann-Lebesque lemma~\cite{Burgarth2022,Momcilovic2022}, while the terms in \Cref{eq:Averaging4,eq:Averaging5} still provide relevant time-averages since their detuning is only in the order of a few GHz.}.

Similar to the average Hamiltonian theory in Ref.~\cite{Gamel} for the density matrix, we obtain a Heisenberg-like equation of Lindblad type, but now for each of the averaged atomic field operators. It consists of an effective Hamiltonian part, an anti-commutator part and an effective dissipative part with dissipator $\mathcal{D}_2$, whose explicit forms can be found in Appendix~\ref{App:TimeAveragingAppliedToHarmonic}. Since the Heisenberg equation for the field operator is formulated in the interaction picture with respect to the internal energy levels and the light field only, the effective field operator $\QmOp{\Psi}_\mathscr{l}$ and the operators in \Cref{eq:TimeAveragedPsi} are still the Schrödinger field operators.

A comparison with \Cref{eq:HarmonicHamiltonian} reveals, that our specific model is more complex due to an additional summation index. Thus, we replace the index $n$ in \Cref{eq:HarmonicHamiltonian} by $n \equiv j\alpha$. Moreover, the frequencies are redefined to $\omega_n \equiv \Delta_{j\alpha}^{(-)}$ and $\Omega_n \equiv \Delta_{j\alpha}^{(+)}$. As \Cref{eq:Heff_gen} states, a second index is required to construct the effective Hamiltonian. Therefore, we additionally define $m=k\beta$ and find for the effective Hamiltonian of our multi-$\lambda$-scheme
    \begin{equation}
    \hat{H}_{\text{eff}}^\textrm{I} = \hat{H}^\textrm{I}_\textrm{A} - \sum_{\alpha,\beta\in\SetRelevant} \sum_{j,k\in\mathcal{A}} \Biggl\{ \frac{[\hat{h}_{k \beta}^\dagger, \hat{h}_{j\alpha}]}{\hbar \omega_{jk\alpha \beta}^{(+)}} \ee^{\ii(\Delta_{k \beta}^{(-)} - \Delta_{j \alpha}^{(-)})t}  + \frac{[\hat{g}_{k \beta}^\dagger, \hat{g}_{j\alpha}]}{\hbar \Omega_{jk\alpha \beta}^{(+)}} \ee^{\ii(\Delta_{k \beta}^{(+)} - \Delta_{j \alpha}^{(+)})t} \Biggr\}.
    \label{eq:H_eff_Raman}
\end{equation}
It should be noted, that the h.c.-term in \Cref{eq:interaction-picture-hamiltonian-generator-decomposed} has a negative sign. Thus, the sign between the constant and time-dependent contribution in \Cref{eq:H_eff_Raman} is negative as well.

In this case, the inverse sum or rather difference between two detunings are given by
\begin{equation}
    \frac{1}{\omega_{jk\alpha \beta}^{(\pm)}} =\frac{1}{2}\Biggl( \frac{1}{\Delta_{j\alpha}^{(-)}} \pm \frac{1}{\Delta_{k\beta}^{(-)}} \Biggr)
    \quad \text{and} \quad
    \frac{1}{\Omega_{jk\alpha \beta}^{(\pm)}} =\frac{1}{2} \Biggl( \frac{1}{\Delta_{j\alpha}^{(+)}} \pm \frac{1}{\Delta_{k\beta}^{(+)}} \Biggr)
    \label{eq:InverseDetunings}
\end{equation}
in order to simplify the denominators. 

This effective Hamiltonian incorporates the principal couplings of the co-rotating and counter-rotating contributions, with the respective detunings $\omega_{jk\alpha\beta}^{(+)}$ and $\Omega_{jk\alpha\beta}^{(+)}$, where the relative magnitudes can be found in~\Cref{tab:frequencies}. 

\begin{table}[h]
\caption{Relevant frequency scales~\cite{Ufrecht2019} and their approximate magnitudes in typical Raman diffraction experiments in terms of the notation of the main text.}\label{tab:frequencies}
\begin{tabular}{@{}llll@{}}
\toprule \toprule
\textbf{Variable} & \textbf{Quantity 2}  & \textbf{Frequency [Hz]}  \\
\midrule
$\Omega_\alpha$   & Laser frequency & $10^{15}$  \\
$\Omega_\A-\Omega_\B$ & Laser frequency difference & $10^{9}$ \\
$\omega_{j\alpha}$ & Transition frequency $\ket{\alpha} \leftrightarrow \ket{j}$ & $10^{15}$  \\
$\omega_{\A \B}$ & Transition frequency $\ket{\A} \leftrightarrow \ket{\B}$ & $10^{9}$  \\
$\Delta_{j \alpha}^{(-)} = \Omega_\alpha - \omega_{j\alpha}$ & Single-photon detuning $(\Omega)$ & $10^{11}$  \\
$\Delta_{j \alpha}^{(+)} = \Omega_\alpha + \omega_{j\alpha}$ & Single-photon detuning $(\Lambda)$ & $10^{15}$  \\
$\delta_{\alpha \beta}^{(+)} = \Omega_\beta- \Omega_\alpha + (\omega_{\beta}-\omega_{\alpha})$ &  & $10^{4}$ $(\alpha \neq \beta)$  \\
$\delta_{\alpha \beta}^{(-)} = \Omega_\beta- \Omega_\alpha - (\omega_{\beta}-\omega_{\alpha})$ &  & $10^{9}$ $(\alpha \neq \beta)$  \\
$\omega_{jk\alpha \beta}^{(+)}$ & & $10^{11}$ $(\alpha \neq \beta)$ \\
$\omega_{jk\alpha \beta}^{(-)}$ & & $10^{13}$ $(\alpha \neq \beta)$ \\
$\Omega_{jk\alpha \beta}^{(+)}$ & & $10^{14}$ $(\alpha \neq \beta)$ \\
$\Omega_{jk\alpha \beta}^{(-)}$ & & $10^{19}$ $(\alpha \neq \beta)$ \\
\botrule \toprule
\end{tabular}
\end{table}

\subsubsection{Approximated evolution of the field operator}
\label{App:ApproximatedEvolution}
In this publication, we consider for the evolution of the time-averaged operator in \Cref{eq:Evolution_fin} only the dominant term for both frequency domains with detuning $\omega_{nm}^{(+)}$ and $\Omega_{nm}^{(+)}$, given by the effective Hamiltonian in \Cref{eq:Heff_gen}, and exclude contributions with  $\omega_{jk\alpha\beta}^{(-)}$ and $\Omega_{jk\alpha\beta}^{(-)}$ due to their rapid oscillation.

 In a typical Raman diffraction setup, c.f.~\Cref{fig:frequency-detunings}, we have $\bigl(\omega_{jk\alpha\beta}^{(-)}\bigr)^{-1} \ll \bigl(\omega_{jk\alpha\beta}^{(+)}\bigr)^{-1}$ and \mbox{$\bigl(\Omega_{jk\alpha\beta}^{(-)}\bigr)^{-1} \ll \bigl(\Omega_{jk\alpha\beta}^{(+)}\bigr)^{-1}$}. We note that the contributions from the counter-rotating couplings are comparatively small due to their large detuning. This can result in scenarios where \mbox{$\bigl(\Omega_{jk\alpha\beta}^{(+)}\bigr)^{-1} < \bigl(\omega_{jk\alpha\beta}^{(-)}\bigr)^{-1}$}. In principle, terms with significant detunings can be disregarded. Concurrently, the time-averaging process enables the neglect of terms characterized by rapid oscillations. Nevertheless, there exists an interplay between the detuning and the oscillatory terms, which determines the relevance of each contribution. Thus, we restrict ourselves to presenting only the leading order expressions from co- and counter-rotating couplings.

We provide in this section a rough estimation of the values of the involved frequencies and detunings. For the detuning of the co-rotating terms we find
\begin{align}
    \frac{1}{\omega_{jk\alpha \beta}^{(+)}}	= & \frac{1}{2} \Biggl(\frac{1}{\Omega_\alpha-(\omega_j-\omega_\alpha)}+\frac{1}{\Omega_\beta-(\omega_k-\omega_\beta) } \Biggr) \notag \\
    = & \frac{1}{2} \frac{\overbrace{\Omega_\alpha-\omega_j+\omega_\alpha}^{\approx 10^{11}  \text{Hz}}+\overbrace{\Omega_\beta-\omega_k+\omega_\beta}^{\approx 10^{11} \text{Hz}  }}{\underbrace{[\Omega_\alpha-(\omega_j-\omega_\alpha)]}_{\approx 10^{11} \text{Hz}}\underbrace{[\Omega_\beta-(\omega_k-\omega_\beta)]}_{\approx 10^{11} \text{Hz}}} 
    \approx 10^{-11} \text{Hz}
\end{align}
and similarly for the detuning of the counter-rotating terms
\begin{align}
	\frac{1}{\Omega_{jk\alpha \beta}^{(+)}}	= &\frac{1}{2} \Biggl(\frac{1}{\Omega_\alpha+(\omega_j-\omega_\alpha)}+\frac{1}{\Omega_\beta+(\omega_k-\omega_\beta) } \Biggr) \notag \\ 
    =& \frac{1}{2} \frac{\overbrace{\Omega_\alpha+\omega_j-\omega_\alpha}^{\approx 10^{15} \text{Hz}}+\overbrace{\Omega_\beta+\omega_k-\omega_\beta}^{\approx 10^{15} \text{Hz}}}{[\underbrace{\Omega_\alpha}_{\approx 10^{15 \text{Hz}}}+\underbrace{(\omega_j-\omega_\alpha)}_{\approx 10^{15} \text{Hz}}][\underbrace{\Omega_\beta}_{\approx 10^{15} \text{Hz}}+\underbrace{(\omega_k-\omega_\beta)}_{\approx 10^{15} \text{Hz}}]} 
	\approx  10^{-15} \text{Hz}.
\end{align}

When we consider the difference of the inverses of two detunings, it becomes evident that the difference between the laser frequencies is a significant factor. Accordingly, we reorganize the terms compared to the previous expression, resulting for the co-rotating contributions in
\begin{align}
	\frac{1}{\omega_{jk\alpha \beta}^{(-)}}	= &\frac{1}{2} \Biggl(\frac{1}{\Omega_\alpha-(\omega_j-\omega_\alpha)}-\frac{1}{\Omega_\beta-(\omega_k-\omega_\beta) } \Biggr) \notag \\
    =& \frac{1}{2} \frac{\overbrace{\Omega_\beta-\Omega_\alpha}^{\approx 10^{9}  \text{Hz}}+\overbrace{\omega_j-\omega_\alpha}^{\approx 10^{15} \text{Hz}  }+ \overbrace{\omega_\beta-\omega_k}^{\approx -10^{15}  \text{Hz}}}{\underbrace{[\Omega_\alpha-(\omega_j-\omega_\alpha)]}_{\approx 10^{11} \text{Hz}}\underbrace{[\Omega_\beta-(\omega_k-\omega_\beta)]}_{\approx 10^{11} \text{Hz}}}  \approx 10^{-13} \text{Hz}
\end{align}
and in
\begin{align}
	\frac{1}{\Omega_{jk\alpha \beta}^{(-)}}	= &\frac{1}{2} \Biggl(\frac{1}{\Omega_\alpha+(\omega_j-\omega_\alpha)}-\frac{1}{\Omega_\beta+(\omega_k-\omega_\beta) } \Biggr) \notag \\
	=& \frac{1}{2} \frac{\overbrace{\Omega_\beta-\Omega_\alpha}^{\approx 10^{9}  \text{Hz}}+\overbrace{-\omega_j+\omega_\alpha}^{\approx -10^{15} \text{Hz}  }+ \overbrace{-\omega_\beta+\omega_k}^{\approx 10^{15}  \text{Hz}}}{\underbrace{[\Omega_\alpha-(\omega_j-\omega_\alpha)]}_{\approx 10^{15} \text{Hz}}\underbrace{[\Omega_\beta-(\omega_k-\omega_\beta)]}_{\approx 10^{15} \text{Hz}}} \approx 10^{-21} \text{Hz}
\end{align}
for the counter-rotating contributions. Hence, a fully consistent calculation would necessitate the consideration of the magnitude of the various contributions including the detuning $\omega_{jk\alpha \beta}^{(+)}$.

\subsection{Evaluation of commutators}
\label{app:evalcommutators}
In order to compute the effective Hamiltonian in \Cref{eq:H_eff_Raman}, we evaluate its inherent commutators. By inserting \Cref{eq14}, the commutator between $\hat{h}_{{k \beta}^\dagger}^{(\dagger)}$ and $\hat{h}_{j\alpha}$ can explicitly be written as
\begin{equation}
\begin{split}
        [\hat{h}_{k \beta}^\dagger, \hat{h}_{j\alpha}] \!	= \! \hbar^2 \!\! \int \!\! \dd^3 \R  \!\! \int \!\! \dd^3 \R' \, \Omega_{k \beta}^*(\R) \Omega_{j\alpha}(\R') \Bigl\{ &\QmOp{\psi}_{\beta}^\dagger(\R) \QmOp{\psi}_{k}(\R) \QmOp{\psi}_{j}^\dagger(\R') \QmOp{\psi}_{\alpha}(\R') \hat{\beta}^\dagger \hat{\alpha}   \\
	- &\QmOp{\psi}_{j}^\dagger(\R') \QmOp{\psi}_{\alpha}(\R')  \QmOp{\psi}_{\beta}^\dagger(\R) \QmOp{\psi}_{k}(\R) \hat{\alpha} \hat{\beta}^\dagger  \Bigr\}.
\end{split}
\end{equation}

In the next step we use the commutation relations of the atomic field operators in \Cref{eq:commutation-relations-matter-field} and emphasize, that the indices of the ground and excited states are represented by $\alpha,\beta \in \mathcal{R}$, respectively, and that $j,k \in \mathcal{A}$ are the indices designated for the ancilla states. In the first term we commute $\QmOp{\psi}_{k}(\R)$ and $ \QmOp{\psi}_{j}^\dagger(\R')$, in the second term we commute $\QmOp{\psi}_{\alpha}(\R')$ and $\QmOp{\psi}_{\beta}^\dagger(\R)$, which results in 
\begin{equation}
\begin{split}
  [\hat{h}_{k \beta}^\dagger, \hat{h}_{j\alpha}] 	  = &\hbar^2 \!\!\int \!\! \dd^3 \R  \!\! \int \!\! \dd^3 \R' \, \Omega_{k \beta}^*(\R) \Omega_{j\alpha}(\R')\\
  \Bigl\{& \QmOp{\psi}_{\beta}^\dagger(\R) \Bigl[\QmOp{\psi}_{j}^\dagger(\R')\QmOp{\psi}_{k}(\R) + \delta_{jk} \delta(\R\!-\!\R') \Bigr] \QmOp{\psi}_{\alpha}(\R') \hat{\beta}^\dagger \hat{\alpha}   \\
	-& \QmOp{\psi}_{j}^\dagger(\R') \Bigl[ \QmOp{\psi}_{\beta}^\dagger(\R) \QmOp{\psi}_{\alpha}(\R') + \delta_{\alpha \beta} \delta(\R\!-\!\R') \Bigr] \QmOp{\psi}_{k}(\R) \hat{\alpha} \hat{\beta}^\dagger  \Bigr\}. \\
  \end{split}
  \label{eq:evalcommutator}
\end{equation}

We rearrange the equation in such a way, that we first collect the terms containing a Delta function. The remaining terms contain four atomic field operators. Since the creation and annihilation field operators commute among themselves, likewise the annihilating atomic field operators, it is obvious, that those remaining terms differ in the ordering of the light field operators only. Thus, we can combine these two terms with a commutator of the light field operator. We rearrange \Cref{eq:evalcommutator} to
{\medmuskip=1mu
\begin{equation}
\scalebox{0.97}{$
\begin{split}
    [\hat{h}_{k \beta}^\dagger, \hat{h}_{j\alpha}] \!=\!\hbar^2 \!\! \int \!\! \dd^3 \R \!\! \int \!\!\dd^3 \R' \, \Omega_{k \beta}^*(\R) \Omega_{j\alpha}(\R') \Bigl\{ &\QmOp{\psi}_{\beta}^\dagger(\R)    \QmOp{\psi}_{\alpha}(\R') \hat{\beta}^\dagger \hat{\alpha} \delta_{jk}  \delta(\R\!-\!\R') \\
    - &\QmOp{\psi}_{j}^\dagger(\R') \QmOp{\psi}_{k}(\R) \hat{\alpha} \hat{\beta}^\dagger \delta_{\alpha \beta} \delta(\R\!-\!\R')  \\
	+& \QmOp{\psi}_{\beta}^\dagger(\R)   \QmOp{\psi}_{j}^\dagger(\R')  \QmOp{\psi}_{\alpha}(\R')  \QmOp{\psi}_{k}(\R) [\hat{\beta}^\dagger, \hat{\alpha}]  \Bigr\} \\
\end{split}$}
\end{equation}}

Finally, we use the implicit bosonic commutation relations for the annihilation and creation operators of the light field to evaluate the remaining commutator, resulting in
{\medmuskip=1mu
\begin{equation}
\scalebox{0.97}{$
\begin{split}
     [\hat{h}_{k \beta}^\dagger, \hat{h}_{j\alpha}] 	
     \!=\! \hbar^2 \!\! \int \!\! \dd^3 \R \!\! \int \!\! \dd^3 \R' \, \Omega_{k \beta}^*(\R) \Omega_{j\alpha}(\R') \Bigl\{ &\QmOp{\psi}_{\beta}^\dagger(\R) \QmOp{\psi}_{\alpha}(\R') \hat{\beta}^\dagger \hat{\alpha} \, \delta_{jk} \delta(\R-\R')  \\
    - &\QmOp{\psi}_{j}^\dagger(\R') \QmOp{\psi}_{k}(\R)  \hat{\alpha} \hat{\beta}^\dagger \, \delta_{\alpha \beta } \delta(\R-\R') \\
     -&\QmOp{\psi}_{\beta}^\dagger(\R)   \QmOp{\psi}_{j}^\dagger(\R')  \QmOp{\psi}_{\alpha}(\R')  \QmOp{\psi}_{k}(\R) \, \delta_{\alpha \beta} \Bigr\}.
\end{split}$}
\label{eq:CommutatorCo}
\end{equation}}

In addition to the terms that contain a combination of atomic and light field operators the result reveals a term which describes the particle-particle interactions of ground and ancilla states. The latter is induced by the quantization of the light field.

Similarly, the commutator for the operators $\hat{g}_{\bullet \bullet}^{(\dagger)}$ yields
{\medmuskip=1mu
\begin{equation}
\begin{split}
     [\hat{g}_{k \beta}^\dagger, \hat{g}_{j\alpha}] \!=\! \hbar^2 \!\! \int\!\!  \dd^3 \R \!\! \int \!\! \dd^3 \R' \, \Lambda_{k \beta}(\R) \Lambda_{j\alpha}^*(\R') \Bigl\{ &\QmOp{\psi}_{k}^\dagger(\R) \QmOp{\psi}_{j}(\R') \hat{\beta}^\dagger \hat{\alpha} \, \delta_{\alpha \beta} \delta(\R-\R')  \\
    -  &\QmOp{\psi}_{\alpha}^\dagger(\R') \QmOp{\psi}_{\beta}(\R)  \hat{\alpha} \hat{\beta}^\dagger \, \delta_{kj} \delta(\R-\R') \\
    -&  \QmOp{\psi}_{k}^\dagger(\R)   \QmOp{\psi}_{\alpha}^\dagger(\R')  \QmOp{\psi}_{j}(\R')  \QmOp{\psi}_{\beta}(\R) \, \delta_{\alpha \beta} \Bigr\}
\end{split}
\label{eq:CommutatorCounter}
\end{equation}}
where we used \Cref{eq14}, respectively.

\subsection{Effective Hamiltonian in the Schrödinger picture}
\label{app:H_effSP}
In Appendix~\ref{app:Trafo} we insert the expression for the commutators, \Cref{eq:CommutatorCo,eq:CommutatorCounter}, in \Cref{eq:H_eff_Raman} to find the concrete expression of the effective Hamiltonian for the considered $\Lambda$-like model. Moreover, we reverse the interaction picture and find the effective Hamiltonian in the Schrödinger picture.

Subsequently, we separate in Appendix~\ref{App:NewVariables} the contributions in the effective Hamiltonians into transitions between the different atomic states and self-couplings. Finally, we use this decomposition in Appendix~\ref{app:finH_effSP} to cast the effective Hamiltonian into a form that facilitates its physical interpretation.

\subsubsection{Transformation of the effective Hamiltonian from Heisenberg to Schrödinger picture}
\label{app:Trafo}
Fundamentally, the commutators between operators of the form $\hat{h}_{j\alpha}$ and $\hat{h}_{k\beta}^\dagger$ in \Cref{eq:H_eff_Raman} originate from the co-rotating couplings, while the commutators with operators of the form $\hat{g}_{j\alpha}$ and $\hat{g}_{k\beta}^\dagger$ stem from the counter-rotating couplings. 

We insert the evaluation of both commutators, \Cref{eq:CommutatorCo,eq:CommutatorCounter}, into \Cref{eq:H_eff_Raman}  and obtain for the effective Hamiltonian
    \begin{equation} \label{effectivehamiltonian-interim-result}
    \begin{split}
    \hat{H}_\text{eff}^\textrm{I} = \hat{H}_\textrm{A}^\textrm{I}-&\hbar \int \dd^3 \R \sum_{j\in\SetAncilla} \sum_{\alpha\in\SetRelevant}  \Biggl\{  \sum_{\beta\in\SetRelevant} \biggl[ \frac{\Omega_{j \beta}^*(\R) \, \Omega_{j \alpha}(\R)}{\omega_{jj\alpha \beta}^{(+)}} \QmOp{\Psi}_{\beta}^\dagger(\R) \QmOp{\Psi}_{\alpha}(\R)  \hat{\beta}^\dagger  \hat{\alpha} \ee^{\ii \delta_{\alpha \beta}^{(+)} t} \\
    -  &\frac{\Lambda_{j \beta}(\R) \, \Lambda_{j \alpha}^*(\R)}{\Omega_{jj\alpha \beta}^{(+)}} \QmOp{\Psi}_{\alpha}^\dagger(\R) \QmOp{\Psi}_{\beta}(\R)  \hat{\alpha} \hat{\beta}^\dagger \ee^{\ii \delta_{\alpha \beta}^{(-)} t} \biggr] \\
    -\sum_{k\in\SetAncilla}  \biggl[&\frac{\Omega_{k \alpha}^*(\R) \, \Omega_{j \alpha}(\R)}{\omega_{jk\alpha \alpha}^{(+)}} \QmOp{\Psi}_{j}^\dagger(\R) \QmOp{\Psi}_{k}(\R)  \hat{\alpha} \hat{\alpha}^\dagger \\
    + \int \dd^3 \R' & \frac{\Omega_{k \alpha}^*(\R) \, \Omega_{j \alpha}(\R')}{\omega_{jk\alpha \alpha}^{(+)}} \QmOp{\Psi}_{\alpha}^\dagger(\R) \QmOp{\Psi}_{j}^\dagger(\R') \QmOp{\Psi}_{\alpha}(\R')  \QmOp{\Psi}_{k}(\R)  \biggr] \ee^{\ii(\omega_j-\omega_k)t} \\
    +\sum_{k\in\SetAncilla}  \biggl[&\frac{\Lambda_{k \alpha}(\R) \, \Lambda_{j \alpha}^*(\R)}{\Omega_{jk\alpha \alpha}^{(+)}} \QmOp{\Psi}_{k}^\dagger(\R) \QmOp{\Psi}_{j}(\R)  \hat{\alpha}^\dagger \hat{\alpha} \\
    - \int \dd^3 \R' &\frac{\Lambda_{k \alpha}(\R) \, \Lambda_{j \alpha}^*(\R')}{\Omega_{jk\alpha \alpha}^{(+)}} \QmOp{\Psi}_{k}^\dagger(\R) \QmOp{\Psi}_{\alpha}^\dagger(\R') \QmOp{\Psi}_{j}(\R')  \QmOp{\Psi}_{\alpha}(\R)  \biggr] \ee^{-\ii(\omega_j-\omega_k)t}  \Biggr\}
    \end{split}
    \end{equation}
The time-independent part $\hat{H}_\textrm{A}^\textrm{I}$ incorporates the center-of mass motion, while the second term contains the co-rotating and the third term the counter-rotating contributions to the transitions between the ground and excited state. The fourth term of \Cref{effectivehamiltonian-interim-result} features first the co-rotating contributions to transitions between two ancilla states $\ket{j}$ and $\ket{k}$, which are induced by photons with the same frequency. This contribution appears only due to the fact that we are dealing with an ancilla manifold and not a single ancillary state. The fifth term of \Cref{effectivehamiltonian-interim-result} represents particle-particle interactions. We note that while the initial formulation of the model did not include particle-particle interactions, it is in fact the combination of the non-commutative nature of the quantized light field and the averaging procedure which gives rise to effective particle-particle interactions. The final two contributions of \Cref{effectivehamiltonian-interim-result} arise in the same ways as in the third and fourth line, but from the counter-rotating couplings.
    
In \Cref{effectivehamiltonian-interim-result} we used the two-photon detunings 
\begin{align}
    \Delta_{j\beta}^{(-)}-\Delta_{j\alpha}^{(-)}&=\Omega_\beta - \Omega_\alpha +(\omega_\beta - \omega_\alpha) \equiv \delta_{\alpha \beta}^{(+)},\\
    \Delta_{j\beta}^{(+)}-\Delta_{j\alpha}^{(+)}&=\Omega_\beta - \Omega_\alpha +(\omega_\alpha - \omega_\beta) \equiv \delta_{\alpha \beta}^{(-)}.
\end{align}
In a typical Raman diffraction scheme, the laser frequencies are adjusted in such a way that their frequency difference corresponds to the frequency difference between the ground state and the excited state, plus the recoil frequency of the atom when it undergoes a momentum change of $\hbar K$. Our model relies on the definition $\Omega_\A > \Omega_\B$, while $\omega_\B > \omega_\A$, therefore, the frequency differences in $\delta_{\alpha \beta}^{(+)}$ have an opposite sign when compared to $\delta_{\alpha \beta}^{(-)}$. Thus, $\delta_{\alpha \beta}^{(+)}$ is usually in the order of the recoil frequency, i.~e. several KHz, while $\delta_{\alpha \beta}^{(-)}$ is in the order of GHz. Similarly, the residual time-dependencies with frequencies $\omega_j-\omega_k$, which correspond to the frequency spacing between ancilla states, are typically on the order of a few hundred $\mathrm{MHz}$ (see Refs.~\cite{Steck2019_Cs,Steck2019_Na,Steck2021_85Rb,Steck2021_87Rb} for data on the Cesium, Sodium, and Rubidium $D_2$-lines). These time-dependencies will disappear once the interaction picture is reversed.

Subsequent to the implementation of the averaging procedure, we invert the interaction picture in order to eliminate the residual time dependence. Thereafter, the contributions are reorganized, resulting in the effective Hamiltonian 
{\medmuskip=1mu
\begin{equation}
\scalebox{0.95}{$
    \begin{split}
    \hat{H}_\text{eff} = \hat{h}_0 - \hbar \!\! \int \!\! \dd^3 \R \!\sum_{j\in\SetAncilla} \sum_{\alpha\in\SetRelevant}  \Biggl\{  \sum_{\beta\in\SetRelevant} \biggl[ &\frac{\Omega_{j \beta}^*(\R) \, \Omega_{j \alpha}(\R)}{\omega_{jj\alpha \beta}^{(+)}} \QmOp{\Psi}_{\beta}^\dagger(\R) \QmOp{\Psi}_{\alpha}(\R)  \hat{\beta}^\dagger  \hat{\alpha} \\
    -  &\frac{\Lambda_{j \beta}(\R) \, \Lambda_{j \alpha}^*(\R)}{\Omega_{jj\alpha \beta}^{(+)}} \QmOp{\Psi}_{\alpha}^\dagger(\R) \QmOp{\Psi}_{\beta}(\R)  \hat{\alpha} \hat{\beta}^\dagger  \biggr] \\
    -\sum_{k\in\SetAncilla}  \biggl[&\frac{\Omega_{k \alpha}^*(\R) \, \Omega_{j \alpha}(\R)}{\omega_{jk\alpha \alpha}^{(+)}} \QmOp{\Psi}_{j}^\dagger(\R) \QmOp{\Psi}_{k}(\R) (\hat{n}_\alpha+\mathbbm{1}_\textrm{L})  \\
    -&\frac{\Lambda_{k \alpha}(\R) \, \Lambda_{j \alpha}^*(\R)}{\Omega_{jk\alpha \alpha}^{(+)}} \QmOp{\Psi}_{k}^\dagger(\R) \QmOp{\Psi}_{j}(\R)  \QmOp{n}_\alpha   \\
    -\int \dd^3 \R' \biggl(& \frac{\Omega_{k \alpha}^*(\R) \, \Omega_{j \alpha}(\R')}{\omega_{jk\alpha \alpha}^{(+)}} \QmOp{\Psi}_{\alpha}^\dagger(\R) \QmOp{\Psi}_{j}^\dagger(\R') \QmOp{\Psi}_{\alpha}(\R')  \QmOp{\Psi}_{k}(\R) \\
    + &\frac{\Lambda_{k \alpha}(\R) \, \Lambda_{j \alpha}^*(\R')}{\Omega_{jk\alpha \alpha}^{(+)}} \QmOp{\Psi}_{k}^\dagger(\R) \QmOp{\Psi}_{\alpha}^\dagger(\R') \QmOp{\Psi}_{j}(\R')  \QmOp{\Psi}_{\alpha}(\R) \biggr) \biggr]  \Biggr\} 
    \end{split}$}
    \label{eq:Heff_SP_time}
    \end{equation}}
in the Schrödinger picture. In this expression, the Hamiltonian 
\begin{equation}
    \hat{h}_0= \QmOp{H}_\textrm{A}+\QmOp{H}_\textrm{L}
\end{equation}
entails the Hamiltonian of the atomic and electromagnetic field.

We rearrange the field operators and exchange indices in the last summation of each line. Additionally, we exchange $\R \longleftrightarrow \R'$ in the last summation, resulting in
{\medmuskip=1mu
\begin{equation}
\scalebox{0.89}{$
\begin{split}
    \hat{H}_\text{eff} &= \QmOp{h}_0 -  \hbar \!\!\!\sum_{\alpha,\beta\in \SetRelevant} \!\! \int \!\! \dd^3 \R \, \QmOp{\Psi}_{\beta}^\dagger(\R) \! \left[ \!\sum_{j\in\SetAncilla} \! \frac{\Omega_{j \beta}^*(\R) \, \Omega_{j \alpha}(\R)}{\omega_{jj\alpha \beta}^{(+)}}\hat{\beta}^\dagger  \hat{\alpha}  - \! \sum_{j\in\SetAncilla} \! \frac{\Lambda_{j \beta}^*(\R) \, \Lambda_{j \alpha}(\R)}{\Omega_{jj\beta \alpha }^{(+)}}\hat{\beta} \hat{\alpha}^\dagger \right] \! \QmOp{\Psi}_{\alpha}(\R)\\
    &+ \hbar \!\!\!
    \sum_{j,k\in \SetAncilla} \!\!
    \int \!\! \dd^3 \R \, \QmOp{\Psi}_{j}^\dagger(\R) \!
    \left[ \! \sum_{\alpha\in\SetRelevant} \!
    \frac{\Omega_{k \alpha}^*(\R) \, \Omega_{j \alpha}(\R)}{\omega_{jk\alpha \alpha}^{(+)}}(\hat{n}_\alpha+\mathbbm{1}_\textrm{L}) - \! \sum_{\alpha\in\SetRelevant} \! \frac{\Lambda_{k \alpha}^*(\R)~\Lambda_{j \alpha}(\R) }{\Omega_{kj\alpha \alpha}^{(+)}} \QmOp{n}_\alpha \right] \!\QmOp{\Psi}_{k}(\R) \\
    &+ \hbar \!\!
    \sum_{j,k\in \SetAncilla} \!
    \sum_{\alpha\in \SetRelevant} \!\!
    \int\!\! \dd^3 \R \!\! \int \!\!\dd^3 \R^\prime\,\QmOp{\Psi}_{\alpha}^\dagger(\R) \QmOp{\Psi}_{j}^\dagger(\R') \!
     \left[
    \mathcal{V}^{(\Omega)}_{jk\alpha}(\R,\R') + \mathcal{V}^{(\Lambda)}_{jk\alpha}(\R,\R') 
    \right] \!
    \QmOp{\Psi}_{\alpha}(\R')  \QmOp{\Psi}_{k}(\R),
\end{split}$}
\label{eq:Heff_SP_time2}
\end{equation}}
where we define 
\begin{equation}
     \mathcal{V}^{(\Omega)}_{jk\alpha}(\R,\R') \equiv \frac{ \Omega_{k \alpha}^*(\R) \, \Omega_{j \alpha}(\R')}{\omega_{jk\alpha \alpha}^{(+)}}   \text{\quad and \quad}    \mathcal{V}^{(\Lambda)}_{jk\alpha}(\R,\R') \equiv \frac{\Lambda^*_{k \alpha}(\R) \, \Lambda_{j \alpha}(\R')}{\Omega_{jk\alpha \alpha}^{(+)}}
     \label{eq:PotentialApp}
\end{equation}
for the co-rotating and counter-rotating contributions, respectively. 

Due to the definitions in \Cref{eq:InverseDetunings}, it is obvious that $\Omega_{jj\alpha \beta}^{(+)}=\Omega_{jj\beta \alpha}^{(+)}$ and $\Omega_{jk\alpha \alpha}^{(+)}=\Omega_{kj\alpha \alpha}^{(+)}$. In the following, we use only $\Omega_{jk\alpha \alpha}^{(+)}$ and $\Omega_{jj\alpha \beta}^{(+)}$ instead of the equivalent expressions.

\subsubsection{Separation into transitions and self-couplings}
\label{App:NewVariables}
The effective Hamiltonian in \Cref{eq:Heff_SP_time2} comprises three summations. Separating this summation enables the differentiation between self-couplings and transitions between two states. 

The initial two terms of \Cref{eq:Heff_SP_time2} are characterized by field operators with index $\alpha$ and $\beta$. Thus, these terms encompass solely the ground state $\ket{\A}$ and the excited state $\ket{\B}$. By explicitly writing the sum over $\beta$ we obtain
\begin{equation}
\scalebox{0.94}{$
\begin{split}
   &\sum_{j=3}^N \sum_{\alpha, \beta=\A,\B} \QmOp{\psi}_{\beta}^\dagger(\R)  \frac{\Omega_{j\beta}^*(\R) \, \Omega_{j \alpha}(\R)}{\omega_{jj\alpha \beta}^{(+)}}    \hat{\beta}^\dagger  \hat{\alpha} \QmOp{\psi}_{\alpha}(\R)\\
    &= \!\!  \sum_{j=3}^N \sum_{\alpha=\A,\B} \!\! \QmOp{\psi}_{\alpha}^\dagger(\R) \frac{|\Omega_{j \alpha}(\R)|^2}{\omega_{jj\alpha \alpha}^{(+)}}   \hat{n}_\alpha  \QmOp{\psi}_{\alpha}(\R) \! +\! \biggl[\QmOp{\psi}_{\B}^\dagger(\R) \!\sum_{j=3}^N \! \frac{\Omega_{j \B}^*(\R) \, \Omega_{j \A}(\R)}{\omega_{jj\A\B}^{(+)}} \hat{\B}^\dagger \hat{\A} \QmOp{\psi}_{\A}(\R) \!+\!\text{h.c.} \biggr],
    \end{split}
    \label{eq:DecompRabiCo}$}
\end{equation}
where the first term denotes the self-coupling of $\ket{\A}$ and $\ket{\B}$, and the second term describes transitions between those two states for the co-rotating contributions. 

Analogously, we are able to separate counter-rotating contributions for $\ket{\A}$ and $\ket{\B}$ into self-couplings and transitions, given by
\begin{equation}
\scalebox{0.91}{$
\begin{split}
   \sum_{j=3}^N  \! \sum_{\alpha, \beta=\A,\B} \!\!\!\! \QmOp{\psi}_{\beta}^\dagger(\R) \frac{\Lambda_{j\beta}^*(\R) \, \Lambda_{j \alpha}(\R)}{\Omega_{jj\alpha \beta}^{(+)}}   \QmOp
   {\psi}_{\alpha}(\R) \hat{\beta} \hat{\alpha}^\dagger \! = \! &\sum_{j=3}^N \! \sum_{\alpha=\A,\B} \!\! \QmOp{\psi}_{\alpha}^\dagger(\R) \frac{|\Lambda_{j \alpha}(\R)|^2}{\Omega_{jj\alpha \alpha}^{(+)}} (\hat{n}_\alpha\!+\! \mathbbm{1}_\textrm{L}) \QmOp{\psi}_{\alpha}(\R) \! \\ 
   +  \biggl[&\sum_{j=3}^N \QmOp{\psi}_{\B}^\dagger(\R) \frac{\Lambda_{j \B}^*(\R) \, \Lambda_{j \A}(\R)}{\Omega_{jj\A\B}^{(+)}} \hat{\B} \hat{\A}^\dagger  \QmOp{\psi}_{\A}(\R) \! + \!\text{h.c.} \biggr].
    \end{split}$}
    \label{eq:DecompRabiCounter}
\end{equation}

A similar procedure can be applied to the ancilla states, which can be found in line two of \Cref{eq:Heff_SP_time2}. We rearrange the expression in such a way that we separate the sum into contributions where $j=k$, and into contributions where $j\neq k$. Thus, the co-rotating part can be expressed as
{\medmuskip=1mu
\begin{equation}
\hspace*{-0.35cm}
\scalebox{0.91}{$
\begin{split}
   &\sum_{j,k=3}^N \! \sum_{\alpha=\A,\B} \!\!\! \QmOp{\psi}_{j}^\dagger(\R) \frac{\Omega_{k\alpha}^*(\R) \, \Omega_{j \alpha}(\R)}{\omega_{jk\alpha \alpha}^{(+)}} (\hat{n}_\alpha+\mathbbm{1}_\textrm{L}) \QmOp{\psi}_{k}(\R)\\
   &= \! \sum_{j=3}^N \!\sum_{\alpha=\A,\B} \!\!  \QmOp{\psi}_{j}^\dagger(\R) \frac{|\Omega_{j \alpha}(\R)|^2}{\omega_{jj\alpha \alpha}^{(+)}} \!(\hat{n}_\alpha+\mathbbm{1}_\textrm{L}) \QmOp{\psi}_{j}(\R) 
   + \! \sum_{j\neq k}^N  \! \sum_{\alpha=\A,\B} \!\! \QmOp{\psi}_{j}^\dagger(\R) \frac{\Omega_{k \alpha}^*(\R) \, \Omega_{j \alpha}(\R)}{\omega_{jk\alpha \alpha}^{(+)}} (\hat{n}_\alpha +\mathbbm{1}_\textrm{L}) \QmOp{\psi}_{k}(\R). \\
       \end{split}$}
       \label{eq:DecompAncillaCo}
\end{equation}}
The first line represents the self-coupling of the ancilla states, while the second term denotes transitions between the involved ancilla states. 

In an equal manner, we separate the counter-rotating contributions to the ancilla states into self-coupling and transitions, given by 
{\medmuskip=1mu
\begin{equation}
\scalebox{0.91}{$
\begin{split}
   \sum_{j,k=3}^N \!\sum_{\alpha=\A,\B} \!\!\! \QmOp{\psi}_{j}^\dagger(\R) \frac{\Lambda_{k\alpha}^*(\R) \, \Lambda_{j \alpha}(\R)}{\Omega_{jk\alpha \alpha}^{(+)}} \hat{n}_\alpha \QmOp{\psi}_{k}(\R)  
  =& \sum_{j=3}^N \! \sum_{\alpha=\A,\B} \!\! \QmOp{\psi}_{j}^\dagger(\R) \frac{|\Lambda_{j \alpha}(\R)|^2}{\Omega_{jj\alpha \alpha}^{(+)}} \hat{n}_\alpha \QmOp{\psi}_{j}(\R)\\ + \! &\sum_{j\neq k}^N  \sum_{\alpha=\A,\B} \!\! \QmOp{\psi}_{j}^\dagger(\R) \frac{\Lambda_{k \alpha}^*(\R) \, \Lambda_{j \alpha}(\R)}{\Omega_{jk\alpha \alpha}^{(+)}} \hat{n}_\alpha \QmOp{\psi}_{k}(\R). 
       \end{split}$}
       \label{eq:DecompAncillaCounter}
\end{equation}}

Moreover, the particle-particle interactions, i.e. the last line of \Cref{eq:Heff_SP_time2}, are governed by a potential that depends on two different positions $\R$ and $\R'$. 

In the following section, we employ these rearrangements and new variables to cast the effective Hamiltonian into a form that facilitates its physical interpretation.

\subsubsection{Rearranged effective Hamiltonian in the Schrödinger picture}
\label{app:finH_effSP}
Next, we use the decompositions in \Cref{eq:DecompRabiCo,eq:DecompRabiCounter,eq:DecompAncillaCo,eq:DecompAncillaCounter} and the potentials in \Cref{eq:PotentialApp} and transform the effective Hamiltonian in \Cref{eq:Heff_SP_time2} into a more comprehensive form given by
{\medmuskip=1mu
\begin{equation}
\scalebox{0.93}{$
\begin{split}
    \hat{H}_\text{eff} &= \QmOp{h}_0 - \hbar \! \sum_{j\in\SetAncilla} \! \sum_{\alpha\in \SetRelevant} \!\int \!\!\dd^3 \R \, \QmOp{\Psi}_{\alpha}^\dagger(\R) 
    \biggr[ 
    \frac{|\Omega_{j \alpha}(\R)|^2}{\omega_{jj\alpha \alpha}^{(+)}}\QmOp{n}_\alpha  -\frac{|\Lambda_{j \alpha}(\R)|^2 }{\Omega_{jj\alpha \alpha}^{(+)}}(\hat{n}_\alpha+\mathbbm{1}_\textrm{L}) ~
    \biggr]\QmOp{\Psi}_{\alpha}(\R)\\
    &- \hbar \! \sum_{j\in\SetAncilla} \! {\sum_{\substack{\alpha,\beta\in \SetRelevant\\\alpha\neq\beta}}} \! \int \!\! \dd^3 \R \,\QmOp{\Psi}_{\beta}^\dagger(\R) \biggl[ 
    \frac{\Omega_{j \beta}^*(\R) \, \Omega_{j \alpha}(\R)}{\omega_{jj\alpha \beta}^{(+)}}\hat{\beta}^\dagger  \hat{\alpha}  -\frac{\Lambda_{j \beta}^*(\R) \, \Lambda_{j \alpha}(\R)}{\Omega_{jj\alpha \beta}^{(+)}}\hat{\beta} \hat{\alpha}^\dagger
     \biggr]\QmOp{\Psi}_{\alpha}(\R)\\
    &+ \hbar \!
    \sum_{j\in \SetAncilla} \!
    \sum_{\alpha\in\SetRelevant} \!
    \int \!\! \dd^3 \R \; \QmOp{\Psi}_{j}^\dagger(\R) 
    \biggl[ 
    \frac{|\Omega_{j \alpha}(\R)|^2}{\omega_{jj\alpha \alpha}^{(+)}}(\hat{n}_\alpha+\mathbbm{1}_\textrm{L})    
    -\frac{|\Lambda_{j \alpha}(\R)|^2 }{\Omega_{jj\alpha \alpha}^{(+)}}  \QmOp{n}_\alpha
    \biggr]\QmOp{\Psi}_{j}(\R) \\
    &+ \hbar \!\!
    \sum_{\substack{j,k\in \SetAncilla\\ j\neq k}} \!
    \sum_{\alpha\in\SetRelevant} \!
    \int \!\! \dd^3 \R \, \QmOp{\Psi}_{j}^\dagger(\R) 
    \biggl[ 
    \frac{\Omega_{k \alpha}^*(\R) \, \Omega_{j \alpha}(\R)}{\omega_{jk\alpha \alpha}^{(+)}}(\hat{n}_\alpha+\mathbbm{1}_\textrm{L})    
    - \frac{\Lambda_{k \alpha}^*(\R) ~ \Lambda_{j \alpha}(\R)}{\Omega_{kj\alpha \alpha}^{(+)}}  \QmOp{n}_\alpha
    \biggr]  \QmOp{\Psi}_{k}(\R) \\
    &+\hbar \!\!
    \sum_{j,k\in \SetAncilla}\!
    \sum_{\alpha\in \SetRelevant}\!
    \int \!\! \dd^3 \R \! \int \!\! \dd^3 \R^\prime~\QmOp{\Psi}_{\alpha}^\dagger(\R) \QmOp{\Psi}_{j}^\dagger(\R') \mathcal{V}_{jk\alpha}(\R,\R^\prime) 
    \QmOp{\Psi}_{\alpha}(\R')  \QmOp{\Psi}_{k}(\R),
\end{split}$}
\label{eq:Heff_long}
\end{equation}}
with $\mathcal{V}_{jk\alpha}(\R,\R^\prime)= \mathcal{V}^{(\Omega)}_{jk\alpha}(\R,\R^\prime) +    \mathcal{V}^{(\Lambda)}_{jk\alpha}(\R,\R^\prime)$. We find in this equation for each line the co-rotating contribution on the left, indicated by $\Omega_{\bullet \bullet}(\R)$, 
and the counter-rotating contributions, indicated by $\Lambda_{\bullet \bullet}(\R)$, on the right, where the bullets represent the corresponding indices $\alpha,\beta, k, j$. The first line represents the self-coupling of the ground state and the excited state, while the subsequent line describes the transitions between these states. Equivalently, the third line reveals the self-coupling of the ancilla states and the subsequent line shows transitions between them. The last line contains the particle-particle interactions.

\subsection{Spatial separation of the particle-particle interactions}
\label{app:PPdecomposition}
In this section we will rearrange the particle-particle interactions, i.e. the last line in \Cref{eq:Heff_long}, denoted by
\begin{equation}
   \hat{H}_\mathrm{pp} \equiv  \hbar
    \sum_{j,k\in \SetAncilla}
    \sum_{\alpha\in \SetRelevant}
    \int \dd^3 \R ~ \int \dd^3 \R^\prime~\QmOp{\Psi}_{\alpha}^\dagger(\R) \QmOp{\Psi}_{j}^\dagger(\R') 
   \mathcal{V}_{jk\alpha}(\R,\R^\prime)
    \QmOp{\Psi}_{\alpha}(\R')  \QmOp{\Psi}_{k}(\R).
    \label{eq:H_ppApp}
\end{equation} 
First, we focus solely on the involved operators and the potentials. The following discussion is independent of the specific form of the potential. Therefore, we utilize $V^{\bullet}_{jk\alpha}(\R,\R^\prime)$, where the bullet may represent the index $(\Omega)$ or $(\Lambda)$. 

Due to $\alpha \in \mathcal{R}$ and $j,k \in \mathcal{A}$, their corresponding field operators commute. Thus, the particle-particle interactions can be reformulated to
\begin{equation}
\begin{split}
        \QmOp{\Psi}_{\alpha}^\dagger(\R) \QmOp{\Psi}_{j}^\dagger(\R') V^{\bullet}_{jk\alpha}(\R,\R^\prime)
\QmOp{\Psi}_{\alpha}(\R')  \QmOp{\Psi}_{k}(\R) =  &\QmOp{\Psi}_{\alpha}^\dagger(\R) V^{\bullet}_{jk\alpha}(\R,\R^\prime)
\bigl[\QmOp{\Psi}_{k}(\R) \QmOp{\Psi}_{j}^\dagger(\R') \\
- &\delta(\R-\R') \delta_{jk}\bigr] \QmOp{\Psi}_{\alpha}(\R').
\end{split}
\label{eq:PP_IA}
\end{equation}

First, we focus on the first term of \Cref{eq:PP_IA}, which contains as before four field operators. However, the field operators depending on $\R$ can now be moved to the left-hand side of the potential, while the field operators depending on $\R'$, can be moved to the right-hand side. Consequently, for $\bullet=(\Omega)$, \Cref{eq:PP_IA} can be further decomposed into
\begin{equation}
 \begin{split}
          &\QmOp{\Psi}_{\alpha}^\dagger(\R) \QmOp{\Psi}_{k}(\R)  V^{(\Omega)}_{jk\alpha}(\R,\R^\prime)
\QmOp{\Psi}_{j}^\dagger(\R') \QmOp{\Psi}_{\alpha}(\R') \\&= \QmOp{\Psi}_{\alpha}^\dagger(\R) \frac{\Omega_{k\alpha}^*(\R)}{\sqrt{\omega_{jk\alpha \alpha}^{(+)}}} \QmOp{\Psi}_{k}(\R) \cdot \QmOp{\Psi}_{j}^\dagger(\R') \frac{\Omega_{j\alpha}(\R')}{\sqrt{\omega_{jk\alpha \alpha}^{(+)}}}\QmOp{\Psi}_{\alpha}(\R'). 
 \end{split}
\label{eq:pp_IA_4}
\end{equation}
The counter-rotating contributions with $V^{(\Lambda)}_{jk\alpha}(\R,\R^\prime)$ have to be decomposed in an equal manner.

Next, we address the second term of \Cref{eq:PP_IA} and add the integrals and summations from \Cref{eq:H_ppApp}. This expression depends only on two field operators. We evaluate the integration over the Delta function, which leads to a modification of the self-couplings of ground and excited state, denoted by
\begin{equation}
\begin{split}
    &\hbar \sum_{j,k\in \SetAncilla}
    \sum_{\alpha\in \SetRelevant} \int \dd^3 \R ~ \int \dd^3 \R^\prime~ \QmOp{\Psi}_{\alpha}^\dagger(\R) \mathcal{V}^{\bullet}_{jk\alpha}(\R,\R^\prime) \delta(\R-\R') \delta_{jk}\QmOp{\Psi}_{\alpha}(\R') \\
    &= \hbar \sum_{j\in \SetAncilla}
    \sum_{\alpha\in \SetRelevant} \int \dd^3 \R ~\QmOp{\Psi}_{\alpha}^\dagger(\R) \mathcal{V}^{\bullet}_{jj\alpha}(\R) \QmOp{\Psi}_{\alpha}(\R).
\end{split}
\end{equation}
The Kronecker delta enables the simplification of the potentials in \Cref{eq:PotentialApp} into
\begin{equation}
    \mathcal{V}^{(\Omega)}_{jj\alpha}(\R) = \frac{|\Omega_{j\alpha}(\R)|^2}{\omega_{jj\alpha \alpha}^{(+)}}
    \quad \text{and} \quad 
    \mathcal{V}^{(\Lambda)}_{jj\alpha}(\R) = \frac{|\Lambda_{j\alpha}(\R)|^2}{\Omega_{jj\alpha \alpha}^{(+)}}
\end{equation}
As \Cref{eq:H_ppApp} indicates, the dipole coupling contains the sum of the potentials in $\Omega$ and $\Lambda$. Consequently, the total modification of the self-couplings reads
\begin{equation}
    \hbar \sum_{j\in \SetAncilla}
    \sum_{\alpha\in \SetRelevant} \int \dd^3 \R ~\QmOp{\Psi}_{\alpha}^\dagger(\R) \sum_{j\in \SetAncilla} \bigl[\mathcal{V}^{(\Omega)}_{jj\alpha}(\R) + \mathcal{V}^{(\Lambda)}_{jj\alpha}(\R) \bigr] \QmOp{\Psi}_{\alpha}(\R).
    \label{eq:pp_IA_2}
\end{equation}
Moreover, we combine the potentials into the joint expression
\begin{equation}
    \mathcal{V}_\alpha (\R) = \sum_{j\in \SetAncilla} \bigl[\mathcal{V}^{(\Omega)}_{jj\alpha}(\R) + \mathcal{V}^{(\Lambda)}_{jj\alpha}(\R) \bigr] 
\end{equation}
to enhance the readability of the equations.

\subsection{Effective Hamiltonian in Schrödinger picture with separated particle-particle interactions}
\label{app:HeffSeparatePP}
We conclude Appenix~\ref{app:DerivationHeff} with the presentation of the effective Hamiltonian in the Schrödinger picture, which has been derived through a series of rearrangements. This may offer a valuable foundation for the subsequent discussion. By combining \Cref{eq:Heff_long}, \Cref{eq:pp_IA_4} and \Cref{eq:pp_IA_2} we are able to express the effective Hamiltonian as
{\medmuskip=1mu
\begin{equation}
\scalebox{0.93}{$
\begin{split}
    \hat{H}_\text{eff} &= \QmOp{h}_0  - \hbar \! \sum_{j\in\SetAncilla} \! \sum_{\alpha\in \SetRelevant} \!\! \int \!\! \dd^3 \R ~\QmOp{\Psi}_{\alpha}^\dagger(\R) 
    \biggr[ 
    \frac{|\Omega_{j \alpha}(\R)|^2}{\omega_{jj\alpha \alpha}^{(+)}}\QmOp{n}_\alpha  -\frac{|\Lambda_{j \alpha}(\R)|^2 }{\Omega_{jj\alpha \alpha}^{(+)}}(\hat{n}_\alpha+\mathbbm{1}_\textrm{L}) ~
    \biggr]\QmOp{\Psi}_{\alpha}(\R)\\
    &- \hbar \! \sum_{j\in\SetAncilla} \! {\sum_{\substack{\alpha,\beta\in \SetRelevant\\\alpha\neq\beta}}} \!\! \int \!\! \dd^3 \R ~\QmOp{\Psi}_{\beta}^\dagger(\R)\biggl[ 
    \frac{\Omega_{j \beta}^*(\R) \, \Omega_{j \alpha}(\R)}{\omega_{jj\alpha \beta}^{(+)}}\hat{\beta}^\dagger  \hat{\alpha}  -\frac{\Lambda_{j \beta}^*(\R) \, \Lambda_{j \alpha}(\R) }{\Omega_{jj\alpha \beta}^{(+)}}\hat{\beta} \hat{\alpha}^\dagger
     \biggr] \QmOp{\Psi}_{\alpha}(\R)\\
    &+ \hbar \!
    \sum_{j\in \SetAncilla} \!
    \sum_{\alpha\in\SetRelevant}
    \!\! \int \!\! \dd^3 \R ~\QmOp{\Psi}_{j}^\dagger(\R) 
    \biggl[ 
    \frac{|\Omega_{j \alpha}(\R)|^2}{\omega_{jj\alpha \alpha}^{(+)}}(\hat{n}_\alpha+\mathbbm{1}_\textrm{L})    
    -\frac{|\Lambda_{j \alpha}(\R)|^2 }{\Omega_{jj\alpha \alpha}^{(+)}}  \QmOp{n}_\alpha
    \biggr] \QmOp{\Psi}_{j}(\R) \\
    &+\hbar \!\!
    \sum_{\substack{j,k\in \SetAncilla\\ j\neq k}}  \!
    \sum_{\alpha\in\SetRelevant} 
    \!\! \int \!\! \dd^3 \R ~\QmOp{\Psi}_{j}^\dagger(\R) 
    \biggl[ 
    \frac{\Omega_{k \alpha}^*(\R) \, \Omega_{j \alpha}(\R)}{\omega_{jk\alpha \alpha}^{(+)}}(\hat{n}_\alpha+\mathbbm{1}_\textrm{L})    
    - \frac{\Lambda_{k \alpha}^*(\R) ~\Lambda_{j \alpha}(\R)}{\Omega_{jk\alpha \alpha}^{(+)}}  \QmOp{n}_\alpha
    \biggr]\QmOp{\Psi}_{k}(\R) \\
    &+\hbar \!
    \sum_{j,k\in \SetAncilla} \!
    \sum_{\alpha\in \SetRelevant}
    \!\! \int \!\! \dd^3 \R ~ \QmOp{\Psi}_{\alpha}^\dagger(\R) \zeta_{k\alpha j}^*(\R) \QmOp{\Psi}_{k}(\R) \cdot \int \dd^3 \R^\prime~  \QmOp{\Psi}_{j}^\dagger(\R') \zeta_{j\alpha k}(\R')\QmOp{\Psi}_{\alpha}(\R') \\
      &+ \hbar \!
    \sum_{j,k\in \SetAncilla} \!
    \sum_{\alpha\in \SetRelevant}
    \!\! \int \!\! \dd^3 \R ~  \QmOp{\Psi}_{\alpha}^\dagger(\R)  \eta_{k\alpha j}^*(\R) \QmOp{\Psi}_{k}(\R) \cdot \int \dd^3 \R^\prime~\QmOp{\Psi}_{j}^\dagger(\R') \eta_{j\alpha k}(\R') \QmOp{\Psi}_{\alpha}(\R') \\
    &- \hbar \!
    \sum_{\alpha\in \SetRelevant} \int \dd^3 \R ~\QmOp{\Psi}_{\alpha}^\dagger(\R) \mathcal{V}_\alpha(\R)  \QmOp{\Psi}_{\alpha}(\R), 
\end{split}$}
\label{eq:Heff_long_sum}
\end{equation}}
where we used the variables
\begin{equation}
    \zeta_{j\alpha k}(\R) = \frac{\Omega_{j\alpha}(\R)}{\sqrt{\omega_{jk\alpha \alpha}^{(+)}}} \text{\,~ and ~\,}  \eta_{j\alpha k}(\R) = \frac{\Lambda_{j\alpha}(\R)}{\sqrt{\Omega_{jk\alpha \alpha}^{(+)}}}.
\end{equation}
For a thorough examination of \Cref{eq:Heff_long_sum} and the subsequent development of this article, we direct the reader back to \Cref{sec:DeterminationHeff}.

\section{Separation of the time evolution operator into even and odd components}
\label{app:V} 
The light-matter interaction within the time-evolution operator is mediated by the operator
\begin{equation}
     \hat{\mathcal{V}}= \hbar \int \dd^3\R  [\Omega(\R) \hat{\C}- \Lambda(\R) \hat{\C}^\dagger] \QmOp{\psi}_{\C}(\R) + \text{h.c.},
 \end{equation}
with the joint light field operators $\hat{\C}^\dagger \equiv \hat{\A}^\dagger \hat{\B}$ and $\hat{\C} \equiv \hat{\B}^\dagger \hat{\A}$ which contain one creation and one annihilation operator of the different optical fields. However, the conventional bosonic commutation relations do not apply to these novel operators, leading to the following result: $[\hat{\C},\hat{\C}^\dagger]=\hat{n}_\B-\hat{n}_\A$.
A similar approach leads to the new atomic field operators  \mbox{$\QmOp{\psi}_{\C}^\dagger(\R) \equiv \QmOp{\psi}_{\A}^\dagger(\R)\QmOp{\psi}_{\B}(\R)$} and   $\QmOp{\psi}_{\C}(\R) \equiv \QmOp{\psi}_{\B}^\dagger(\R)\QmOp{\psi}_{\A}(\R)$ with the commutation relation
 $[\QmOp{\psi}_{\C}(\R),\QmOp{\psi}_{\C}^\dagger(\R')]= [\QmOp{\psi}_{\B}^\dagger(\R) \QmOp{\psi}_{\B}(\R') - \QmOp{\psi}_{\A}^\dagger(\R') \QmOp{\psi}_{\A}(\R)] \delta(\R-\R')$. Moreover, we introduced the frequencies
\begin{equation}
   \Omega(\R) \equiv -\sum_{j=3}^n \frac{\Omega^*_{j\B}(\R) \Omega_{j\A}(\R)}{\Omega_{jj\A\B}^{(+)}}  \text{\,~ and ~\,}   \Lambda(\R) \equiv \sum_{j=3}^n \frac{\Lambda^*_{j\B}(\R) \Lambda_{j\A}(\R)}{\Omega_{jj\A\B}^{(+)}} 
\end{equation}
for shorter notation. 

The analysis of the exponential containing $\hat{\mathcal{V}}$ in \Cref{eq:TimeEvolutionOperator} can be facilitated by separating the exponential into odd and even contributions, which yields
\begin{equation}
    \ee^{- \frac{\ii \vartheta}{\hbar }\hat{\mathcal{V}}} \! = \! \sum_{k=0}^\infty \frac{(-1)^k(\frac{\vartheta}{\hbar })^{2k}}{(2k)!} \hat{\mathcal{V}}^{2k} - \ii \! \sum_{k=0}^\infty \frac{(-1)^k(\frac{\vartheta}{\hbar})^{2k+1}}{(2k+1)!} \hat{\mathcal{V}}^{2k+1}.
    \label{eq:Exponential_Interaction}
\end{equation}

In order to evaluate the expression in \Cref{eq:Exponential_Interaction} we have to determine the square of $\hat{\mathcal{V}}$, which is given by
{\medmuskip=1mu
\begin{equation}
\scalebox{0.895}{$
\begin{split}
    &\hat{\mathcal{V}}^2= \hbar^2 \int \dd^3\R  \int \dd^3\R' \bigl\{ \\
    \cdot &\bigl[ \Omega(\R) \Omega(\R') \hat{\C}^{2} + \Lambda(\R) \Lambda(\R') \hat{\C}^{\dagger 2} - \Omega(\R) \Lambda(\R') \hat{\C}\hat{\C}^\dagger - \Lambda(\R) \Omega(\R')  \hat{\C}^\dagger  \hat{\C} \bigr] \QmOp{\psi}_{\C}(\R) \QmOp{\psi}_{\C}(\R')  \\
    +& \bigl[ \Omega(\R) \Omega^*(\R') \hat{\C}\hat{\C}^\dagger + \Lambda(\R) \Lambda^*(\R')  \hat{\C}^\dagger  \hat{\C}  - \Omega(\R) \Lambda^*(\R') \hat{\C}^{2} - \Lambda(\R) \Omega^*(\R') \hat{\C}^{\dagger 2} \bigr] \QmOp{\psi}_{\C}(\R) \QmOp{\psi}_{\C}^\dagger(\R') \\
    +& \bigl[ \Omega^*(\R) \Omega(\R')  \hat{\C}^\dagger  \hat{\C} + \Lambda^*(\R) \Lambda(\R') \hat{\C} \hat{\C}^\dagger - \Omega^*(\R) \Lambda(\R') \hat{\C}^{\dagger 2} - \Lambda^*(\R) \Omega(\R') \hat{\C}^{2} \bigr] \QmOp{\psi}^\dagger_{\C}(\R) \QmOp{\psi}_{\C}(\R') \\
    +&\bigl[ \Omega^*(\R) \Omega^*(\R') \hat{\C}^{\dagger 2} + \Lambda^*(\R) \Lambda^*(\R') \hat{\C}^{2}  - \Omega^*(\R) \Lambda(\R')  \hat{\C}^\dagger  \hat{\C}  - \Lambda^*(\R) \Omega^*(\R') \hat{\C} \hat{\C}^\dagger \bigr] \QmOp{\psi}_{\C}^\dagger(\R) \QmOp{\psi}_{\C}^\dagger(\R')  \! \bigr\} 
\end{split}$}
 \end{equation}}
In an interaction picture with respect to the internal Hamiltonian and the light field we observe that $\hat{\C}^{\dagger 2}$ and $\hat{\C}^{2}$ oscillate with the difference of the laser frequencies $\pm(\Omega_\A-\Omega_\B)$, which is usually in the order of GHz. $\QmOp{\psi}_{\C}^\dagger(\R) \QmOp{\psi}_{\C}^\dagger(\R')$ and $\QmOp{\psi}_{\C}(\R) \QmOp{\psi}_{\C}(\R')$ oscillate with the difference of the atomic level frequencies $\mp(\omega_\B-\omega_\A)$, which is in the order of GHz as well. Other terms do not oscillate in this interaction picture. 

We neglect terms that oscillate in the GHz range and keep constant terms or terms that oscillate with \mbox{$\pm(\Omega_\A-\Omega_\B) \pm (\omega_\B-\omega_\A)$}, which is approximately in the kHz range. This approximation finally leads to 
{\medmuskip=1mu
  \begin{equation}
  \scalebox{0.97}{$
  \begin{split}
      \hat{\mathcal{V}}^2 \approx \hbar^2 \!\!\! \int \!\! \dd^3\R  \int \!\!\dd^3\R' \biggl\{\!&\Omega(\R) \Omega(\R') \hat{\C}^{2} \QmOp{\psi}_{\C}(\R) \QmOp{\psi}_{\C}(\R') +\Omega^*(\R) \Omega^*(\R') \hat{\C}^{\dagger 2} \QmOp{\psi}_{\C}^\dagger(\R) \QmOp{\psi}_{\C}^\dagger(\R') \\
    + \bigl[&\Omega(\R) \Omega^*(\R') \hat{\C}\hat{\C}^\dagger + \Lambda(\R) \Lambda^*(\R')  \hat{\C}^\dagger  \hat{\C} \bigr] \QmOp{\psi}_{\C}(\R) \QmOp{\psi}_{\C}^\dagger(\R') \\
    + \bigl[&\Omega^*(\R) \Omega(\R')  \hat{\C}^\dagger  \hat{\C}  + \Lambda^*(\R) \Lambda(\R') \hat{\C}\hat{\C}^\dagger   \bigr] \QmOp{\psi}_{\C}^\dagger(\R) \QmOp{\psi}_{\C}(\R') \biggr\}
      \end{split}$}
       \label{eq:V2_approxApp}
 \end{equation}}
in the Schrödinger picture, with $\hbar^2 \hat{\Omega}^2 \equiv \hat{\mathcal{V}}^2 $. We emphasize that  $\hat{\Omega} \neq \hat{\mathcal{V}}$, due to the approximation in \Cref{eq:V2_approxApp}. Therefore, we reformulate the argument of the sine in \Cref{eq:Exponential_Interaction} into
\begin{equation}
    \hat{\mathcal{V}}^{2k+1} =  \hat{\mathcal{V}}^{2k} \cdot  \hat{\mathcal{V}} = (\hbar \hat{\Omega})^{2k} \cdot \hat{\mathcal{V}},
\end{equation}
where we used the identity $\hbar^2 \hat{\Omega}^2 = \hat{\mathcal{V}}^2$. Thereafter, we expand the expression in terms of $ \hat{\Omega}$ and obtain 
\begin{equation}
    \hat{\mathcal{V}}^{2k+1} = (\hbar \hat{\Omega})^{2k+1} (\hbar \hat{\Omega})^{-1}  \hat{\mathcal{V}}.
    \label{eq:V_2k+1}
\end{equation}

We utilize \Cref{eq:V_2k+1} and rewrite the even and odd parts of \Cref{eq:Exponential_Interaction} as cosine and sine, respectively, resulting in
 \begin{equation}
    \ee^{- \frac{\ii \vartheta}{\hbar}\hat{\mathcal{V}}}  = \cos{\left(\vartheta \hat{\Omega}\right)} - \ii \sin{\left(\vartheta \hat{\Omega}\right)} (\hbar \hat{\Omega})^{-1}  \hat{\mathcal{V}}.
\end{equation}    

Consequently, the time-evolution operator in \Cref{eq:TimeEvolutionOperator} can be reformulated into
\begin{equation}
        \hat{U}(t,t_0) \overset{t' \in (t,t_0)}{\approx} \left[  \cos{\left(\vartheta \hat{\Omega}\right)} - \ii \sin{\left(\vartheta \hat{\Omega}\right)} (\hbar \hat{\Omega})^{-1}  \hat{\mathcal{V}} \right] \exp{-\frac{\ii}{\hbar} \vartheta \hat{H}_0^{(2)}(t')}.
    \label{eq:TimeEvolutionOperatorApp}
\end{equation}
The trigonometric part serves as the mediator for the interaction, while the exponential contains the center-of-mass motion and the energy shifts.

\section{Evaluation of the center-of-mass component of the time-evolution operator}
\label{app:EvalCOM}
The application of the time evolution operator to the input state necessitates the evaluation of its center-of-mass component. This calculation is somewhat cumbersome. To address this challenge, we provide a comprehensive presentation of the aforementioned calculation in Appendix~\ref{app:COM_U} for a generic single-particle atomic state. In Appendix~\ref{app:COM_Gauss} we consider a Gaussian amplitude of the input state and evaluate the center-of-mass component under this assumption. 

\subsection{The center-of-mass component for a generic single-particle atomic state}
\label{app:COM_U}
$\hat{H}_0^{(2)}$ in \Cref{eq:TimeEvolutionOperatorApp} contains number operators, which can readily be applied to the input state. However, it also comprises the center-of-mass component, given by
\begin{equation}
   \hat{H}_\mathrm{COM} = -\int \dd^3 \R \, \sum_\alpha \QmOp{\psi}^\dagger_\alpha(\R) \frac{\hbar^2 \nabla^2_{\R}}{2M} \QmOp{\psi}_\alpha(\R), 
   \label{eq:H_COMapp}
\end{equation}
that requires a more intricate approach. We present in this section the action of the center-of-mass motion on a generic single-particle state of the matter field. The state
\begin{equation}
	\ket{\psi^{(1)} }_\mathrm{A} \equiv \mathscr{N} \int \dd^3 \R \, \sum_\alpha f_\alpha(\R) \, \QmOp{\psi}_\alpha^\dagger(\R) \ket{0 \, 0}_\mathrm{A},
    \label{eq:InputGenApp}
\end{equation}
describes a single particle in mode $\alpha$ of the atomic field. This excitation of the atomic field is denoted by the field operator $ \QmOp{\psi}_\alpha^\dagger(\R)$ and the corresponding single-particle amplitude $f_\alpha(\R)$. The factor $\mathscr{N}$ ensures the normalization of the state.

We apply the center-of-mass Hamiltonian in \Cref{eq:H_COMapp} to the initial state in \Cref{eq:InputGenApp}, as given by
\begin{equation}
    \hat{H}_\mathrm{COM}\ket{\psi^{(1)} }_\mathrm{A}= \hat{H}_\mathrm{COM} \,\, \mathscr{N} \int \dd^3 \R \, \sum_\alpha f_\alpha(\R) \, \QmOp{\psi}_\alpha^\dagger(\R) \ket{0 \, 0}_\mathrm{A}.
\end{equation}
Subsequently, we make use of the commutation relation $[\hat{H}_\mathrm{COM},\QmOp{\psi}_\alpha^\dagger(\R')]=-\frac{\hbar}{2M} \nabla^2_{\R'} \QmOp{\psi}_\alpha^\dagger(\R')$ and obtain
\begin{equation}
   \hat{H}_\mathrm{COM}\ket{\psi^{(1)} }_\mathrm{A}= - \mathscr{N} \int \dd^3 \R \sum_\alpha  \frac{\hbar^2}{2M}  f_\alpha(\R)  \nabla^2 \QmOp{\psi}_{\alpha}^\dagger(\R)   \ket{0 \, 0}. 
\end{equation}
Finally, we employ Green's second identity, but disregard the contributions originating from the boundaries. This is valid for wave functions that sufficiently decrease at the boundaries and results in
\begin{equation}
     \hat{H}_\mathrm{COM}\ket{\psi^{(1)} }_\mathrm{A}= - \mathscr{N} \int \dd^3 \R \sum_\alpha  \frac{\hbar^2}{2M}  \QmOp{\psi}_{\alpha}^\dagger(\R)  \, \nabla^2 f_\alpha(\R) \ket{0 \, 0}. 
        \label{eq:COM_transformFinalApp}
\end{equation}
The specification of $f_\alpha(\R)$ enables the subsequent evaluation of \Cref{eq:COM_transformFinalApp} in \Cref{sec:ApplicationSP}. 

It is important to acknowledge that the type of optical state does not influence this transformation.

\subsection{The center-of-mass component for Gaussian single-particle states}
\label{app:COM_Gauss}
The expression $\nabla^2 f_\alpha(\R)$ in \Cref{eq:COM_transformFinalApp} can finally be evaluated when the amplitude of the input state is specified. Therefore, we choose the Gaussian function 
\begin{equation}
    f_\alpha(\R) \equiv \frac{1}{(2\pi)^{3/2} \sigma_\alpha^3} \cdot \ee^{\ii \boldsymbol{\kappa} \R} \cdot \ee^{-\frac{\R^2}{2\sigma_\alpha^2}}.
\end{equation}
This choice allows us to evaluate the Laplacian acting on the single-particle amplitude in \Cref{eq:COM_transformFinalApp}, resulting in
\begin{equation}
 {\nabla}^2	f_\alpha(\R)=\frac{1}{\sigma_\alpha^4} \left( \R^2-2\ii \sigma_\alpha^2 \boldsymbol{\kappa} \R -3\sigma_\alpha^2- \sigma_\alpha^4 \boldsymbol{\kappa}^2 \right) \cdot f_\alpha (\R) 
 \equiv  - \frac{2M}{\hbar^2}\mathcal{E}_\text{COM}^{(\alpha)}(\R) f_\alpha(\R).
\end{equation}
Hence, \Cref{eq:COM_transformFinalApp} for Gaussian single-particle states reads
\begin{equation}
    \hat{H}_\mathrm{COM}\ket{\psi^{(1)} }_\mathrm{A} = \mathscr{N} \int \dd^3 \R \sum_\alpha \mathcal{E}_\text{COM}^{(\alpha)} \,  f_\alpha (\R) \QmOp{\psi}_{\alpha}^\dagger(\R)  \,  \ket{0 \, 0}. 
        \label{eq:COM_Gauss}
\end{equation}
With the results of this section, the subsequent application of the time evolution operator to specific input states is now possible. 

Finally, we will discuss the change of the initial amplitude due to the center-of-mass Hamiltonian. The components of $\hat{H}_0^{(2)}$ commute. Consequently, they can be applied separately to the initial state, as 
\begin{equation}
    \ee^{-\frac{\ii}{\hbar} \vartheta \hat{H}_\mathrm{COM}(t')}\ket{\psi^{(1)} }_\mathrm{A} =  \mathscr{N} \int \dd^3 \R \, \sum_\alpha \tilde{f}_\alpha(R)  \, \QmOp{\psi}_\alpha^\dagger(\R) \ket{0 \, 0}_\mathrm{A}
\end{equation}
Thus, the center-of-mass component of $\hat{H}_0^{(2)}$ modifies the amplitude $f_\alpha(R)$ of the initial state in  \Cref{eq:InputGenApp} to
\begin{equation}
    \tilde{f}_\alpha(R)= \frac{1}{(2\pi)^{3/2} \sigma_\alpha^3} \cdot \ee^{\ii \boldsymbol{\kappa} \R} \cdot \ee^{-\frac{\R^2}{2\sigma_\alpha^2}}  \cdot \ee^{\frac{i \vartheta \hbar}{2M}  (\frac{\R^2 -3\sigma_\alpha^2- \sigma_\alpha^4 \boldsymbol{\kappa}^2}{ \sigma_\alpha^4})} \cdot \ee^{\frac{ \vartheta}{M} \frac{ \hbar \boldsymbol{\kappa} \R }{ \sigma_\alpha^2}}.
\end{equation}
The expression for the absolute value of this amplitude is again a Gaussian function given by
\begin{align}
    |\tilde{f}_\alpha(\R)|= &\frac{1}{(2\pi)^{3/2} \sigma_\alpha^3} \cdot \ee^{-\frac{\R^2}{2\sigma_\alpha^2}} \cdot \ee^{\frac{ \vartheta}{M} \frac{\hbar \boldsymbol{\kappa} \R }{ \sigma_\alpha^2}} = \frac{1}{(2\pi)^{3/2} \sigma_\alpha^3} \cdot \ee^{-\frac{\R^2}{2\sigma_\alpha^2}} \cdot \ee^{\frac{ \frac{ \vartheta \hbar \boldsymbol{\kappa} \R }{2M} }{ 2\sigma_\alpha^2}} \notag \\
    = &\frac{1}{(2\pi)^{3/2} \sigma_\alpha^3} \cdot \ee^{-\frac{1}{2\sigma_\alpha^2} \left(\R - \frac{\vartheta \hbar \boldsymbol{\kappa}}{M} \right)^2} \cdot \ee^{\frac{1}{2\sigma_\alpha^2}\left( \frac{\vartheta \hbar \boldsymbol{\kappa}}{M} \right)^2}.
\end{align}
This expression shows that the action of the center-of-mass Hamiltonian induces a translation of the Gaussian function by an amount equivalent to $\frac{\vartheta \hbar \boldsymbol{\kappa}}{M}$, which corresponds to a recoil, and simultaneously modifies the amplitude with the momentum $\hbar \kappa$ of the initial wave function.

\end{appendices}


\bibliography{sn-bibliography}

\end{document}